\documentclass[aps,prb,twocolumn,showpacs,amsmath,amssymb,superscriptaddress,floatfix]{revtex4-2}
\usepackage[english]{babel}
\usepackage{blindtext}
\usepackage[utf8]{inputenc}
\usepackage{graphicx}
\usepackage{array} 
\usepackage{bm}
\usepackage{latexsym}
\usepackage{physics}
\usepackage{amsfonts}
\usepackage{bbold}
\usepackage{scalerel}
\usepackage{appendix}
\usepackage[colorlinks = true,
linkcolor = blue,
urlcolor  = blue,
citecolor = blue,
anchorcolor = blue]{hyperref}
\usepackage{fixltx2e}
\usepackage[usenames,dvipsnames]{color}
\makeatletter
\newcommand*{\rom}[1]{\expandafter\@slowromancap\romannumeral #1@}
\newcommand{\vebm}[1]{
	\ifcat\noexpand#1\relax
	{\bm #1}
	\else
	{\bf #1}
	\fi
}
\makeatother
\begin{document}
	
	\author{Kirill Alpin}
	\thanks{These two authors contributed equally.}
	\affiliation{Max Planck Institute for Solid State Research, Heisenbergstrasse 1, D-70569 Stuttgart, Germany}
	\author{Moritz M. Hirschmann}
	\thanks{These two authors contributed equally.}
	\affiliation{Max Planck Institute for Solid State Research, Heisenbergstrasse 1, D-70569 Stuttgart, Germany}
	\author{Niclas Heinsdorf}
	\affiliation{Max Planck Institute for Solid State Research, Heisenbergstrasse 1, D-70569 Stuttgart, Germany}
	\affiliation{Stewart Blusson Quantum Matter Institute, University of British Columbia, Vancouver, BC V6T 1Z4, Canada}
	\author{Andreas Leonhardt}
	\affiliation{Max Planck Institute for Solid State Research, Heisenbergstrasse 1, D-70569 Stuttgart, Germany}
	\author{Wan Yee Yau}
	\affiliation{Max Planck Institute for Solid State Research, Heisenbergstrasse 1, D-70569 Stuttgart, Germany}
	\affiliation{Institute for Theoretical Physics III, University of Stuttgart, D-70550 Stuttgart, Germany}
	\author{Xianxin Wu}
	\affiliation{Max Planck Institute for Solid State Research, Heisenbergstrasse 1, D-70569 Stuttgart, Germany}
	\affiliation{Institute for Theoretical Physics, Chinese Academy of Sciences, Beijing, China}
	\author{Andreas P. Schnyder}
	\affiliation{Max Planck Institute for Solid State Research, Heisenbergstrasse 1, D-70569 Stuttgart, Germany}
	
	\title{Classification of chiral band crossings: local and global constraints}
	\title{Fundamental laws of chiral band crossings: local constraints, \\ global constraints, and topological phase diagrams}
	
	\date{\today}
	\begin{abstract}
		We derive two fundamental laws of chiral band crossings:
		(i) a local constraint relating the Chern number to phase
		jumps of rotation eigenvalues; and (ii) a global constraint
		determining the number of chiral crossings on rotation axes.
		Together with the fermion doubling theorem, these laws 
		describe all conditions that a network of chiral
		band crossing must satisfy.
		We apply the fundamental laws to prove the existence of enforced
		double Weyl points, nodal planes, and generic Weyl points, among others.
		In addition, we show that chiral space-group symmetries cannot
		stabilize nodal lines with finite Chern numbers.
		Combining the local constraint with explicit low-energy models, we determine the generic 
		topological phase diagrams of all multi-fold crossings. Remarkably, we
		find a four-fold crossing with Chern number 5, which exceeds the previously 
		conceived maximum Chern number of~4. We identify
		BaAsPt as a suitable material with this four-fold crossing exhibiting Chern number~5 near the Fermi energy.
	\end{abstract}
	\maketitle
	\section{Introduction}
	
	Recent theoretical and experimental efforts have
	uncovered a huge number of materials with chiral band crossings near the Fermi level~\cite{chiu_rev_mod_phys_16,armitage_rev_mod_phys_18, bernevig2018recent, bernevig2022progress,top_semimetals_review_19,top_semi_RMP_21}.
	These include Weyl semimetals~\cite{Wan2011Weyl, xu2011chern, soluyanov2015type, Weyl_semimetal_review_felser}, where two bands cross at isolated points, 
	nodal plane materials with two-fold degeneracies on  planes~\cite{chang2018topological, yu2019circumventing, xiao2020experimental, wilde2021symmetry, huber2022network},
	as well as materials with multifold band crossings~\cite{ bradlyn2016beyond,tang2017multiple, schroter2019chiral, cano2019multifold},
	such as threefold, fourfold, or sixfold Weyl points.
	Common to these band crossings is that their little groups are chiral (i.e., do not contain inversion or mirror symmetries), such that 
	they act as sources or sinks of Berry flux with quantized Chern numbers. 
	As a consequence, chiral Weyl materials exhibit a number of unusual phenomena,
	e.g., the chiral anomaly~\cite{ong_review_chiral_anomaly,HOSUR2013857,Burkov_2015}, large negative magnetoresistance~\cite{magneto_resistance_Cd3As2}, 
	or a quantized circular photogalvanic effect~\cite{quantized_CPGE_nat_comm_moore,chiral_optical_response_PRB_flicker,CPGE_in_CoSi_nat_commun}, which might be utilized 
	for the development of new devices and technologies~\cite{liu2019topological, burkov_nat_mat_review_16}.  
	In combination with magnetic order, chiral band crossings are highly tunable via an external magnetic field~\cite{schoop_mag_weyl_sci_adv_18,hirschmann_PRB_22,wilde2021symmetry}, a property that is vital for future applications.

	Importantly, the Chern numbers of chiral band crossings, as well as their multiplicities and arrangements in the Brillouin zone (BZ) is not only constrained by crystallographic symmetries, but must obey further topological conditions that originate from the periodicity of the BZ.
	For example, screw rotations lead to symmetry-enforced chiral crossings, due to a nontrivial winding of the symmetry eigenvalues along the BZ torus~\cite{zhao_schnyder_PRB_16}. 
	In addition, there are in a band structure, in general, also chiral crossings at high-symmetry points, due to higher-dimensional irreducible representations of the little groups, and moreover there can exist accidental crossings at generic positions in the BZ. 
	All of these crossings carry a Chern number, forming an interrelated network of
	band topologies~\cite{huber2022network}. The Chern numbers of this topological network 
	are restricted by the fermion doubling theorem~\cite{NIELSEN1983389}, which dictates
	that the sum of the Chern numbers over the entire BZ must add up to zero.
	The values of the individual Chern numbers, in turn, are constrained by the crystalline symmetries and, moreover, each chiral crossing has symmetry related partners with the same Chern number. Hence, there is a complicated interplay between crystallography and topology that determines the possible network of band crossings   that can exist in a given space group (SG).

	These conditions due to symmetry and topology can be used 
	to systematically classify networks of chiral band crossings.
	For example, Refs.~\cite{watanabe2016gaplessness, Young2015enforcedCrossings, Takahashi2017lines, Young2017Enforced, Zhang2018hexagonal, Chan2019Trig, Leonhardt2021Orthorhombic,  Hirschmann2021Tetragonals, Yu2021Encyclopedia, xie2021kramers}
	classified symmetry-enforced topological features of periodic band structures, Refs.~\cite{bradlyn2016beyond, cano2019multifold,tang_wan_PRB_21, tang_wan_PRB_22, Knoll2022MagPointGroups}
	studied band topologies of low-energy models, and 
	Refs.~\cite{Chen2012MultiWeyl, tsirkin2017compositeWeyl, gonzalez2021chiralonscrew} 
	investigated the connection between rotation eigenvalues and Chern numbers of Weyl points.
	While these works uncovered a number of interesting band topologies, 
	a deeper understanding for why these arise is lacking, and a fundamental theory
	of chiral band crossings still needs to be developed. 
	
	In this article, we close this gap by deriving two fundamental laws of chiral band crossings (Secs.~\ref{sec_two} and~\ref{Sec_ProofsMaintext}),
	namely, (i) a local constraint, Eq.~\eqref{Eq_ChargeFromRotEigVal}, which relates
	the Chern number of a chiral crossing to the phase jumps of the rotation eigenvalues;
	and (ii) a global constraint, Eq.~\eqref{Eq_PhaseConstraint},
	which specifies the number and types of band crossings existing on a rotation axis.
	Together with the famous fermion doubling theorem~\cite{NIELSEN1983389}, these two fundamental
	laws describe all the conditions that a chiral topological network must satisfy. 
	To demonstrate the usefulness of these concepts, we present in Sec.~\ref{sec_four} several
	applications of the local and global constraints. For example, we use the fundamental theorems
	to prove the existence of enforced double Weyl points on a twofold rotation axis away from time-reversal invariant momenta (TRIMs) (Sec.~\ref{SubSec_doubleWeyl}), we derive the existence of  topological nodal planes (Sec.~\ref{sec_enforced_nodal_planes}), and we 
	deduce the existence of Weyl points
	at generic positions in the BZ (Sec.~\ref{sec_global_constraint_multifold_degen}). 
	In addition, we show that chiral nodal lines with finite Chern numbers cannot be
	stabilized by space-group symmetries, but only by internal (artificial) symmetries. We present a low-energy model that realizes such a chiral nodal line (Sec.~\ref{SubSubSec_ChiralNL}). 
	
	We complement these fundamental considerations by an explicit construction
	of low-energy models of all multi-fold crossings (Sec.~\ref{gen_clas_hamiltonians}). Combining
	these with the non-Abelian local constraint, we determine the generic topological phase
	diagrams of these multi-fold crossings. 
	Remarkably, we find that there exist four-fold crossings with Chern number 5 (Sec.~\ref{4foldcrossingssection}),
	which exceeds the previously conceived maximum Chern number of 4 for multi-fold 
	crossings~\cite{Schroeter2020MaximalChern,yao2020MaximalChern}.
	We perform a database search for materials with multi-fold crossings
	exhibiting Chern number 5 and identify BaAsPt in SG 198 as a suitable compound (Sec.~\ref{Sec_BaAsPt}).
	We also briefly discuss the materials NbO$_2$ and TaO$_2$
	in SG 80, which realize double Weyl points at TRIMs and away from TRIMs (Sec.~\ref{Subsec_NbO2TaO2}).
	In  Sec.~\ref{sec_conclusions} we conclude with a discussion and provide directions
	for future work. Technical details are presented in several appendices.

	\section{Two symmetry constraints on chiral crossings}
	\label{sec_two}
	
	A chiral band crossing point, commonly referred to as a Weyl point, acts as a monopole of Berry curvature $\Omega({\bf k})$. 
	Each Weyl point can be characterized by a topological charge, the chirality, which is given by the Chern number $\nu$ calculated on a closed manifold of gapped bands surrounding the crossing point.
	Previously, it has been found that a crossing at ${\bf k}_c$ is always topologically charged, if the little group $G_{{\bf k}_c}$ is chiral, i.e., if there are neither inversion nor mirror symmetries \cite{chang2018topological}.
	Without fine-tuning or additional symmetries the charge of a Weyl point is $\nu = \pm 1$.
	
	If one considers one or more rotation symmetries, the Chern numbers $\nu$ of all chiral crossing in the BZ as well as their multiplicity are subject to local and global constraints, respectively.
	In this section we formulate these  two constraints on the existence and the topological charge of chiral crossings, generalizing previous works~\cite{Chen2012MultiWeyl, tsirkin2017compositeWeyl, gonzalez2021chiralonscrew}. 
	The proofs are then given in Sec.~\ref{Sec_ProofsMaintext}. 
	
	\subsection{The local constraint}
	\label{sub_sec_local_constraint}
	
	We find a simple relation how the charge $\nu_{b,c_b}$ of a crossing $c_b$ between the bands numbered by $b$ and $b+1$ is related to the change of complex phase $\Delta \varphi_{b,c_b}$ of an $n$-fold rotation eigenvalue $\lambda_b$.
	Here and in what follows, we sort the bands by their energy, i.e., 
	$ E_{b+1}({\bf k}) > E_b({\bf k}) \; \; \forall {\bf k} $.
	For a given band $b$  the eigenvalue $\lambda_b({\bf k})$ is generally a function of the crystal momentum ${\bf k}$, which is restricted to the rotation axis.
	The eigenvalue may, but does not need to jump at each crossing on the axis, yielding $\Delta \varphi_{b,c_b} =\arg \left( \lambda_b({\bf k}_{c_b} + \epsilon {\bf \hat{z}}) / \lambda_b({\bf k}_{c_b} - \epsilon {\bf \hat{z}}) \right)$ with the unit vector along the rotation axis ${\bf \hat{z}}$ and $\epsilon > 0$ is sent to 0, see Fig.~\ref{fig_IntegrationPathAndWinding}(a).

	With these definitions we will show that
	\begin{align}\label{Eq_ChargeFromRotEigVal}
		\nu_{b,c_b} = \Delta \varphi_{b,c_b} \frac{n}{2 \pi} \mod n,
	\end{align}
	where the complex phase is only determined up to the order $n$ of the rotation axis. 
	Equation~(\ref{Eq_ChargeFromRotEigVal}) includes previous results obtained for low-energy models subject to one rotation symmetry and a time-reversal symmetry \cite{tsirkin2017compositeWeyl,Chen2012MultiWeyl}, and agrees with the expression derived by classifying equivariant line bundles \cite{gonzalez2021chiralonscrew}. 
	If one recalls that at generic low-symmetry positions in the vicinity of the rotation axis the number of singly-charged Weyl points is also restricted by symmetry to be equal $n$, one finds that larger $|\nu_{b,c_b}| > n/2$ would actually be fine-tuned. 
	Therefore, although the relation is only valid $\mod n$, it is expected that real systems are restricted to crossings with $|\nu_{b,c_b}| < n/2$. 
	Equation~(\ref{Eq_ChargeFromRotEigVal}) is valid even with time-reversal symmetry or other crystalline symmetries as long as the crossing $c_b$ is point-like, e.g., also if time-reversal enables a gapless crossings between equal rotation eigenvalues.
	With this insight a recently discovered type of unusual twofold double Weyl point, which occurs on a twofold instead of a fourfold or sixfold rotation axis away from time-reversal invariant momenta (TRIMs), can be understood, see Sec.~\ref{SubSec_doubleWeyl}. 
	But a caveat is in order here: If time-reversal and screw symmetries appear together not only can equal eigenvalues be paired but in several cases this enforces nodal planes, in which case Eq.~(\ref{Eq_ChargeFromRotEigVal}) does not apply. 
	Nevertheless, we will see that Eq.~(\ref{Eq_ChargeFromRotEigVal}) is a central tool to identify the topology of nodal planes. 
	Since our results can be applied to more than one rotation symmetry at a time, it provides a handle to study higher-fold crossings, where more than two bands intersect. 
	In such crossings every band $b$ is subject to Eq.~(\ref{Eq_ChargeFromRotEigVal}) for each  rotation symmetry. We will see in Sec.~\ref{SubSec_MultipleScrew} that this not only explains the observed topological charges, but results in more than one possible configuration of topological charges. 
	
	\subsection{The global constraint}
	\label{sub_sec_global_constraint}
	
	The second implication of rotation (and mirror) symmetries that governs the qualitative band topology of topological semimetals, is a global constraint on the number and type of required crossings $c_b$ per band $b$. 
	Generally, two types of crystalline symmetries can be distinguished, those with symmorphic operations, which leave at least one point in space invariant, and those with nonsymmorphic operations that leave no point invariant, e.g., screw rotations and glide mirror operations. 
	Since the BZ is periodic, nonsymmorphic symmetries lead to an exchange of bands, due to the ${\bf k}$-dependence of their eigenvalues $\lambda_b({\bf k})$, which implies the existence of at least one band crossing on a nonsymmorphic rotation axis~\cite{zhao_schnyder_PRB_16}.
	Conversely, for bands along a symmorphic rotation axis it must be possible to undo all band crossings via pair annihilation, due to the BZ periodicity.
	These constraints can be formalized with complex phase differences $\Delta \varphi_{b,c_b}$.
	We consider an $n$-fold symmetry comprising a translation $(a,b,\tfrac{m}{n})$, e.g., for a rotation this corresponds to $\{C^z_n | a,b,\tfrac{m}{n} \}$ in Seitz notation.
	If a band $b$ is not part of a multifold crossings, one finds for each of the rotation axes
	\begin{align}\label{Eq_PhaseConstraint}
		\sum_{b,c_b} \Delta \varphi_{b,c_b} 
		= 
		- b \cdot 2 \pi \frac{m}{n} \mod 2 \pi,
	\end{align}
	which only depends on the band index $b$, the translation part $\frac{m}{n}$, and the phase difference $\Delta \varphi_{b,c_b}$ for crossings $c_b$ between the bands $b$ and $b+1$. 
	If there is a multifold crossing for band $b$, a similar relation has to be considered, where $c_b$ comprises crossing to higher and lower bands, see Eq.~(\ref{Eq_EigvalChanges}). 
	Equation~(\ref{Eq_PhaseConstraint}) constrains the complex phase that must be accumulated as one moves through the BZ, up to multiples of $2\pi$. 
	If the right side of Eq.~(\ref{Eq_PhaseConstraint}) is non-zero up to $2\pi$, it is clear that there must be at least one crossing, which contributes to the summation on the left side. 
	We note that a glide mirror symmetry can be treated analogously, by considering crossings on any path within the mirror plane that crosses the entire BZ, such that it is closed due the periodicity in ${\bf k}$. 
	The usefulness of this formalization for rotation symmetric systems becomes evident in conjunction with our first result, Eq.~(\ref{Eq_ChargeFromRotEigVal}), which relates each $\Delta \varphi_{b,c_b} \neq 0$ to a topological charge. 
	Therefore, Eq.~(\ref{Eq_PhaseConstraint}) states that the \emph{total chirality} of all crossings on a rotation axis is given by the band index and the translation part of the screw, up to multiples of the order of the rotation axis.
	This implies that accidental crossings on the rotation axis may change the total charge only by multiples of the order of the rotation axis, which is reminiscent of what happens at generic positions, where a symmetry imposes certain multiplicities of topological crossings. 
	
	\section{Derivation of the two constraints}\label{Sec_ProofsMaintext}
	
	In this section we derive the local and global constraints, which were discussed in the previous section.
	For pedagogical reasons, we first present the proof for 
	nondegenerate bands in Sec.~\ref{Sec_ChernAndEigVal}, and then generalize it to degenerate bands in Sec.~\ref{Sec_ChernAndEigValNonAbelian}. 
	In Sec.~\ref{sec_sew_matrix_anti_unitary} we discuss properties of the sewing matrix with anti-unitary symmetries.
	The global constraint is derived in Sec.~\ref{Sec_GlobalBandTopology}. 
	
	\subsection{Abelian Chern numbers and eigenvalue jumps}\label{Sec_ChernAndEigVal}

\begin{figure}
		\centering
		\includegraphics[width = 0.30 \textwidth]{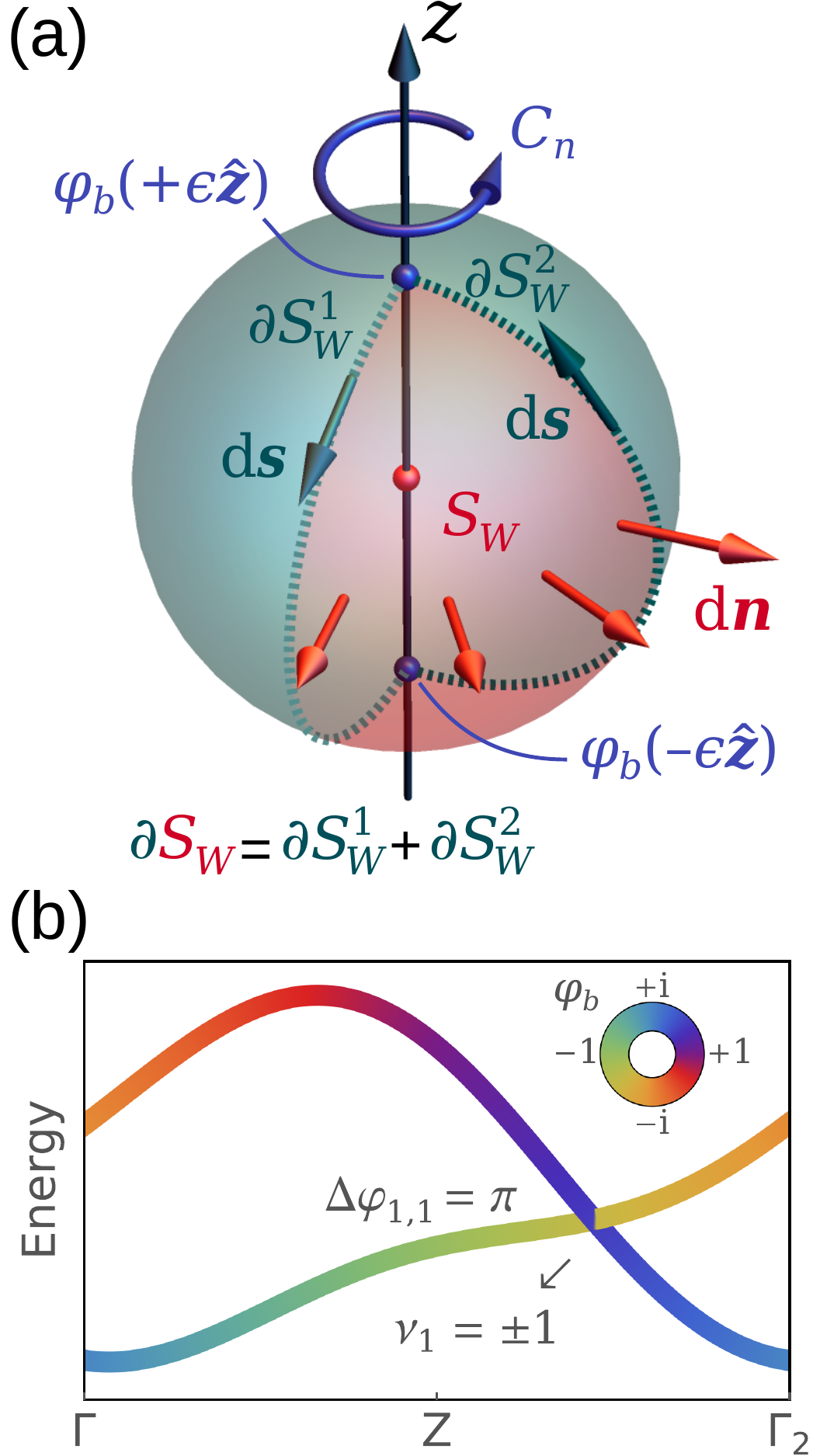}
		\caption{\label{fig_IntegrationPathAndWinding}
			(a) \emph{Local constraint}: Integration area to determine the Chern number around a point crossing (red sphere). 
			In the presence of a $C_n$ rotation the full sphere of radius $\epsilon$ (green + red) comprises $n$ symmetry-related copies of the spherical wedge $S_W$ (red).
			By Stokes theorem the flux of Berry curvature ${\bf \Omega}$ through  $S_W$ is equal to the line integral of the Berry connection ${\bf A}$ along its boundary $\partial S_W$ (dashed lines).
			The value of the Berry phase depends on $\Delta \varphi_{b,c_b} = \varphi(+\epsilon {\bf \hat{z}}) - \varphi(-\epsilon {\bf \hat{z}})$.
			(b) \emph{Global constraint}: The phase $\varphi_b$ of screw rotation symmetry eigenvalues enforces a band crossing in band $b$ with chirality $\nu_b$. 
			The points $\Gamma (0,0,0)$ and $\Gamma_2 (0,0,2\pi)$ are related by a reciprocal lattice vector. 
		}
	\end{figure}

	In the following we  derive the constraint, Eq.~(\ref{Eq_ChargeFromRotEigVal}), on the Chern number $\nu$ of a crossing $c_b$ in band $b$, which is protected by an $n$-fold rotation symmetry $C_n$.
	A related proof is given in Ref.~\cite{gonzalez2021chiralonscrew}, where the Picard group of complex line bundles is computed over a sphere subject to a cyclic group action. 
	To give self-contained proofs, we calculate the Chern number by generalizing the formalism used in Ref.~\cite{Hughes2011ChernFromSymMethod, Fang2012ChernFromSymmetry, Alexandradinata2014WilsonLoops, Bouhon2017EigvalForDirac} to spherical integration surfaces. 
	The Chern number for a nondegenerate band is defined using the flux of Berry curvature ${\bf \Omega}({\bf k})$ through a surface $S$ enclosing the crossing as
	\begin{align} \label{Eq_DefChernnumber}
		\nu
		=
		\frac{1}{2 \pi}\int_{S} \dd {\bf n} \cdot {\bf \Omega}({\bf k}),
	\end{align}
	where the surface $S$ is assumed to be a sphere in reciprocal space, without loss of generality, and ${\bf n}$ is the vector normal to the sphere, see Fig.~\ref{fig_IntegrationPathAndWinding}(a). 
	For ease of presentation we have excluded here the case of bands that are degenerate also away from the crossing point, for which a non-Abelian Berry curvature must be considered, see Sec.~\ref{Sec_ChernAndEigValNonAbelian}. 
	To calculate the Chern number, Eq.~(\ref{Eq_DefChernnumber}), we split the sphere $S$ into $n$ spherical wedges $S_W$, which are related by the $C_n$ rotation symmetry.
	The Abelian Berry curvature transforms as a vector under rotations, i.e.,  $D(C_n){\bf \Omega}({\bf k}) = {\bf \Omega}(D(C_n) {\bf k})$ \cite{Fang2012ChernFromSymmetry}, where $D(C_n)$ is the spatial representation of the rotation $C_n$.
	Noting that the scalar product ${\bf n} \cdot {\bf \Omega}({\bf k})$ is left invariant under the introduction of the orthogonal matrix $D(C_n){\bf \Omega}$, one obtains
	\begin{align}
		\nu
		=
		\frac{n}{2 \pi}  \int_{S_W} \dd {\bf n} \cdot {\bf \Omega}({\bf k}).
	\end{align}
	Further, by using that the curvature, ${\bf \Omega}({\bf k}) = \nabla \times {\bf A}$, is the rotation of the Berry connection $({\bf A}({\bf k}))_{ab} = i \matrixel{u_a({\bf k})}{\nabla}{u_b({\bf k})}$, where $\ket{u_b({\bf k})}$ is the orbital part of a Bloch eigenfunction of the considered Hamiltonian. 
	For a sufficiently small $S$ the only relevant divergence of the Berry curvature ${\bf \Omega}({\bf k})$ occurs at the crossing $c_b$, i.e., ${\bf \Omega}({\bf k})$ has continuous derivatives on $S$. 
	We can thus apply Stokes theorem and find 
	\begin{align}\label{Eq_ChernFromBerryPhase}
		\nu
		=
		\frac{n}{2 \pi}  \int_{\partial S_W} \dd {\bf s} \cdot {\bf A}({\bf k}) \mod n.
	\end{align}
	We note, that the integration in Eq.~(\ref{Eq_ChernFromBerryPhase}) corresponds to the Berry phase. Hence, Stokes theorem holds up to multiples of $2\pi$, which, when taking the factor $\frac{n}{2 \pi}$ into account, amounts to an equation valid $\mod n$.
	In other words, the $U(1)$ gauge freedom of eigenstates implies that the integration of ${\bf A}({\bf k})$ in Eq.~(\ref{Eq_ChernFromBerryPhase}) can be changed by any integer multiple of $2 \pi$, whereas the Chern number $\nu$ is gauge-invariant. 
	When we want to determine the value of the Chern number, the corresponding gauge choice is not known, and  thus Eq.~(\ref{Eq_ChernFromBerryPhase}) holds modulo $n$. 
	
	The closed path $\partial S_W$ can be split into two open and $C_n$-symmetry related paths $\partial S_W^1$ and $\partial S_W^2$, i.e., $\partial S_W = \partial S_W^1 + \partial S_W^2$. 
	But since a non-zero Berry flux through the surface $S_W$ implies that no single-valued phase convention can be found on the full edge $\partial S_W$, we need to account for a mismatch in the phase convention.
	For this purpose we consider the sewing matrix $\mathcal{B}$, which is defined as \cite{Fang2012ChernFromSymmetry}
	\begin{align}\label{Eq_SewingMatrix}
		[\mathcal{B}_{C_n}({\bf k})]_{ab} 
		=
		\matrixel{u_a(D(C_n){\bf k})}{U(C_n)}{u_b({{\bf k}})},
	\end{align}
	where $U(C_n)$ describes the action of the rotation on the eigenstates of the Hamiltonian $\ket{u_b({\bf k})}$. 
	For nondegenerate bands the sewing matrix is simply a complex phase factor $[\mathcal{B}_{C_n}({\bf k})]_{ab} =  \delta_{ab} \mathrm{e}^{i \phi_b({\bf k}) }$. 
	Specifically, at symmetry invariant momenta ${\bf k}_{\text{inv}}$ with $D(C_n) {\bf k}_{\text{inv}} = {\bf k}_{\text{inv}}$ the sewing matrix, Eq.~(\ref{Eq_SewingMatrix}), reduces to  the symmetry eigenvalue $\lambda_b = \mathrm{e}^{i \varphi_b({\bf k}_{\text{inv}}) }$ of $U(C_n)$ for band $b$, i.e., $\phi_b({\bf k}_{\text{inv}}) = \varphi_b({\bf k}_{\text{inv}})$. 
	More generally, at ${\bf k}_{\text{inv}}$ the sewing matrix becomes a diagonal matrix for an appropriate basis within a degenerate subspace.
	The Berry connection is then given by\,\cite{Fang2012ChernFromSymmetry}
	\begin{align}\label{Eq_symmetrytransformA}
		{\bf A}(D(C_n) {\bf k}) &= D(C_n)\big[\mathcal{B}_{C_n} ({\bf k}){\bf A}({\bf k})\mathcal{B}^{-1}_{C_n}({\bf k}) \nonumber\\&+ i \mathcal{B}_{C_n}({\bf k}) \nabla \mathcal{B}^{-1}_{C_n}({\bf k}) \big],
	\end{align}
	see appendix~\ref{app_symmetrytransfromA} for details.
	For nondegenerate bands $i \mathcal{B}_{C_n}({\bf k}) \nabla \mathcal{B}^{-1}_{C_n}({\bf k}) = \nabla \phi_b({\bf k}) $ and $\mathcal{B}_{C_n}({\bf k}){\bf A}({\bf k})\mathcal{B}^{-1}_{C_n}({\bf k})={\bf A}({\bf k})$.
	Using the fact that the path $\partial S_W^2$ corresponds to the rotated path $\partial S_W^1 $ but traversed in the reversed direction, we perform an integral substitution with $ {\bf k} =(C_n){\bf k}'$ in the line integral over  $\partial S_W^2 $, which turns the integration path $\partial S_W^2 $ into $- \partial S_W^1 $. 
	The integral substitution has a unity Jacobian determinant, such that the term ${\bf A}({\bf k})$ cancels leaving only the sewing matrix term.
	Using Eq.~(\ref{Eq_ChernFromBerryPhase}) we complete the proof of Eq.~(\ref{Eq_ChargeFromRotEigVal}),  
	\begin{align}
		\nu
		&=
		\frac{n}{2 \pi} \left( \int_{\partial S_W^1} \kern-1em \dd {\bf s} \cdot {\bf A}({\bf k}) + \int_{\partial S_W^2} \kern-1em \dd {\bf s} \cdot {\bf A}({\bf k}) \right)\kern-0.5em  && \text{mod } n \label{Eq_berry_connection_integration}
		\\
		&=
		-\frac{n}{2 \pi}   \int_{\partial S_W^1} \dd {\bf s} \cdot \nabla \phi_b({\bf k}) && \text{mod } n
		\\
		&=
		\frac{n}{2 \pi}  \left( \varphi({\bf k}_{c_b} + \epsilon {\bf \hat{z}}) -  \varphi({\bf k}_{c_b} - \epsilon {\bf \hat{z}}) \right) && \text{mod } n
		\label{Eq_identifyingPhases}\\
		&=
		\frac{n}{2 \pi}  \Delta\varphi && \text{mod } n,
	\end{align}
	where ${\bf k}_{c_b} + \epsilon {\bf \hat{z}}$ and ${\bf k}_{c_b} - \epsilon {\bf \hat{z}}$ are the north and south pole of the original sphere, respectively, see Fig.~\ref{fig_IntegrationPathAndWinding}(a).
	The difference in complex phases $ \Delta\varphi_{b,c_b}$ of the enclosed crossing is only meaningful up to multiplies of $2\pi$, which is consistent with the equality up to $\mod n$. 
	A comment on the used gauge is in order. 
	Here, we have used the cell-periodic part of the Bloch functions in the calculation of the Chern number and Berry phase \cite{vanderbilt2018berry}, the ${\bf k}$-dependence of the phases $ \varphi_b({\bf k})$ originates only from the exchange of symmetry eigenvalues and the wave functions $u_{b{\bf k}}({\bf r})$ that correspond to $\ket{u_b({\bf k})}$ are periodic in ${\bf r}$ in agreement with the crystal lattice. 
	In the next section, we will consider Bloch functions $\psi_{b{\bf k}}({\bf r}) = \mathrm{e}^{i {\bf k} \cdot {\bf r} } u_{b{\bf k}}({\bf r})$, which capture the periodicity of the Brillouin zone, i.e., $\psi_{b{\bf k} + {\bf K}}({\bf r}) = \psi_{b{\bf k}}({\bf r}) $ for all reciprocal lattice vectors ${\bf K}$.
	The symmetry action for the periodic gauge $\psi_{b{\bf k}}({\bf r})$ captures the global symmetry constraints on the band structure, because the symmetry eigenvalues of nonsymmorphic symmetries gain a phase factor that represents the translation part of the screw and glide symmetry operations.
	Nevertheless, Eq.~(\ref{Eq_ChargeFromRotEigVal}) holds independently of the gauge choice, because all symmetry eigenvalues on a rotation axis obtain the same additional ${\bf k}$-dependence in $\varphi_b({\bf k})$. 
	In other words, practically we think of $\Delta\varphi_{b,c_b}$ in the limit of $\epsilon \rightarrow 0$ [see, e.g., Eq.~(\ref{Eq_identifyingPhases})], whereby $\Delta\varphi_{b,c_b}$ becomes the same for both conventions.
	
	\subsection{Non-abelian Chern numbers \\ and eigenvalue jumps}\label{Sec_ChernAndEigValNonAbelian}
	
	For bands with degeneracies on $S$, for example pairs of bands forming a nodal plane, Eq.\,(\ref{Eq_ChargeFromRotEigVal}) is not applicable, since Chern numbers can either become undefined or assume non-integer values. But in these cases, a non-abelian Chern number can still be defined~\cite{soluyanov2012smooth}
	\begin{align}\label{Eq_nonabelianOmega}
		\nu_{b_1 b_2}
		=
		\frac{1}{2 \pi}  \int_{S} \dd {\bf n} \cdot \text{tr\,}{\bf \Omega}^{b_1 b_2}({\bf k}),
	\end{align}
	where the trace runs over band indices $b$ with $b_1 \leq b \leq b_2$ and the non-abelian Berry curvature and connection\,\cite{xiao2010berry} are
	\begin{align}\label{Eq_nonabelianA}
		{\bf \Omega}^{b_1 b_2} &= \grad_{{\bf k}} \cross {\bf A} - i {\bf A} \cross {\bf A},
		\\
		{\bf A}^{n,m} &= i\braket{n}{\grad_{{\bf k}}|m}, \label{eq_nonabelianberryconnection}
	\end{align}
	respectively, with $b_1 \leq n,m \leq b_2$. The band index range $b_1,...,b_2$ must be chosen such that these bands have a non-zero bandgap to bands $b_1-1$ and $b_2+1$ on the surface $S$.
	
	A similar equation as Eq.\,(\ref{Eq_ChargeFromRotEigVal}) can be derived for non-abelian Chern numbers. Using Eq.\,(\ref{Eq_nonabelianOmega}) and (\ref{Eq_nonabelianA}) we have
	\begin{align}
		\nu_{b_1 b_2}
		&=
		\frac{1}{2 \pi}  \int_{S} \dd {\bf n} \cdot \text{tr}(\grad_{{\bf k}} \cross {\bf A} - i {\bf A} \cross {\bf A})\nonumber\\
		&=\frac{1}{2 \pi}  \int_{S} \dd {\bf n} \cdot \text{tr}(\grad_{{\bf k}} \cross {\bf A})\nonumber\\
		&=\frac{n}{2 \pi}  \int_{\partial S_W} \dd {\bf s} \cdot \text{tr}{\bf A}({\bf k}) \mod n ,
	\end{align}
	where we used the fact that $\text{tr}({\bf A} \cross {\bf A})=0$ since $\text{tr}(A_iA_j-A_jA_i)=0$. In going from the second to the third line we reduced the integration area using symmetry and applied Stokes theorem just like in the proof for the abelian case. Splitting $\partial S_W$ into $\partial S^1_W$ and $\partial S^2_W$ and mapping the latter to the former with Eq.\,(\ref{Eq_symmetrytransformA}) we get
	\begin{align}
		\nu_{b_1 b_2}
		&=\biggl[\frac{n}{2 \pi}  \int_{\partial S^1_W} \dd {\bf s} \cdot \text{tr}{\bf A}({\bf k}) \nonumber\\
		&- \frac{n}{2 \pi} \int_{\partial S^1_W} \dd {\bf s} \cdot \text{tr}(\mathcal{B}_{C_n}({\bf k}){\bf A}({\bf k})\mathcal{B}^{-1}_{C_n}({\bf k})) \nonumber\\
		&-i\frac{n}{2 \pi} \int_{\partial S^1_W} \dd {\bf s} \cdot \text{tr}(\mathcal{B}_{C_n}({\bf k}) \nabla \mathcal{B}^{-1}_{C_n}({\bf k}))\biggr]&&\mod n \nonumber\\
		&=-i\frac{n}{2 \pi} \int_{\partial S^1_W} \dd {\bf s} \cdot \text{tr}(\mathcal{B}_{C_n}({\bf k}) \nabla \mathcal{B}^{-1}_{C_n}({\bf k}))&&\mod n,\label{Eq_nonabeliansinglepath}
	\end{align}
	where we used $\text{tr}(\mathcal{B}_{C_n}({\bf k}){\bf A}({\bf k})\mathcal{B}^{-1}_{C_n}({\bf k})) = \text{tr}{\bf A}({\bf k})$. Using Jacobi's formula we have
	\begin{align}
		\frac{\dd}{\dd\lambda} \log\det\mathcal{B}^{-1}_{C_n}(\lambda) &= \frac{1}{\det\mathcal{B}^{-1}_{C_n}(\lambda)} \frac{\dd \det\mathcal{B}^{-1}_{C_n}(\lambda)}{\dd\lambda} \nonumber\\
		&=\text{tr}\left(\mathcal{B}_{C_n}(\lambda)\frac{\dd \mathcal{B}^{-1}_{C_n}(\lambda)}{\dd \lambda}\right).
	\end{align}
	Combining this with Eq.\,(\ref{Eq_nonabeliansinglepath}) we obtain
	\begin{align}
		\nu_{b_1 b_2}
		&=-i\frac{n}{2 \pi} \int_{\partial S^1_W} \dd {\bf s} \cdot \grad_{{\bf k}}\log \text{det}\mathcal{B}^{-1}_{C_n}({\bf k}) && \mod n \nonumber\\
		&=
		-i\frac{n}{2 \pi}  [ \log \text{det}\mathcal{B}_{C_n}({\bf k}_{c_b} + \epsilon {\bf \hat{z}}) \nonumber\\
		&-  \log \text{det}\mathcal{B}_{C_n}({\bf k}_{c_b} - \epsilon {\bf \hat{z}}) ] && \mod n. \label{Eq_nonabeliansewingjump}
	\end{align}
	This is equivalent to
	\begin{align}
		\nu_{b_1 b_2}
		&=\frac{n}{2 \pi}  \sum_{b=b_1}^{b_2}\Delta\varphi_{b,c_b} && \text{mod } n, \label{Eq_nonabeliansymjump}
	\end{align}
	when the bands $b_1,...,b_2$ are nondegenerate, consistent with the abelian case, although here this is also true if the bands are degenerate somewhere on the sphere except at the poles. When they are degenerate at the poles, one must either resort to using Eq.\,(\ref{Eq_nonabeliansewingjump}) or choose an eigenbasis in the degenerate subspace, such that the sewing matrix $\mathcal{B}_{C_n}$ is diagonal and use Eq.\,(\ref{Eq_nonabeliansymjump}).
	
	\subsection{Sewing matrices of anti-unitary symmetries}
	\label{sec_sew_matrix_anti_unitary}
	In this section we derive similar expressions for the anti-unitary symmetries as in section \ref{Sec_ChernAndEigVal}.
	Applying these to generic crossings,
	we find that single band Chern numbers of crossings with time-reversal symmetry have even (odd) Chern numbers without (with) SOC. In the following, $W$ is either just the time-reversal symmetry $W=T$ with $R=1$ or $W$ is a combination of time-reversal and crystalline symmetry. We start with the derivation of the sewing matrix $\alpha$ for degenerate bands
	\begin{align}
		WH(\vebm{k})W^{-1}&=H(-R\vebm{k}) \\
		\implies WH(\vebm{k}) \ket{u(\vebm{k})}&= H(-R\vebm{k}) W \ket{u(\vebm{k})} \\
		\implies E(\vebm{k}) W\ket{u(\vebm{k})}&= H(-R\vebm{k}) W \ket{u(\vebm{k})}
	\end{align}
	so $W \ket{u(\vebm{k})}$ must be an eigenstate of $H(-R\vebm{k})$. Therefore 
	\begin{align}
		\alpha(\vebm{k})\ket{u(-R\vebm{k})}=W\ket{u(\vebm{k})},\label{AppEq_Tsymmetry_apply}
	\end{align} 
	which leads to the sewing matrix for anti-unitary symmetries
	\begin{align}
		\alpha(\vebm{k}) = \bra{u(-R\vebm{k})}W\ket{u(\vebm{k})}.\label{Eq_antiunitsewing}
	\end{align}
	The Berry connection transforms under $W$ like so
	\begin{align}
		A(\vebm{k}) &= i \bra{u(\vebm{k})} \partial_\vebm{k} \ket{u(\vebm{k})} \nonumber\\
		&= - i \bra{u(\vebm{k})} W^\dagger \partial_\vebm{k} W \ket{u(\vebm{k})} \nonumber\\
		&= - i \alpha^*(\vebm{k}) \bra{u(-R\vebm{k})} \partial_\vebm{k} \alpha(\vebm{k}) \ket{u(-R\vebm{k})} \nonumber\\
		&= i R^{-1}\bra{u(-R\vebm{k})} \partial_{-R\vebm{k}} \ket{u(-R\vebm{k})} - i \alpha^*(\vebm{k}) \partial_\vebm{k} \alpha(\vebm{k}) \nonumber\\
		&= R^{-1}A(-R\vebm{k}) - i \alpha^*(\vebm{k}) \partial_\vebm{k} \alpha(\vebm{k}),\label{AppEq_timerev_A}
	\end{align}
	where we used
	\begin{align}
		&\bra{u(\vebm{k})} W^\dagger \partial_\vebm{k} W \ket{u(\vebm{k})}   \nonumber\\
		=&\lim_{\epsilon\to 0} \frac{1}{\epsilon} (\bra{u(\vebm{k})} W^\dagger W \ket{u(\vebm{k}+\epsilon)} - \bra{u(\vebm{k})} W^\dagger W \ket{u(\vebm{k})}) \nonumber\\
		=&\lim_{\epsilon\to 0} \frac{1}{\epsilon} (\bra{u(\vebm{k}+\epsilon)}\ket{u(\vebm{k})} - \bra{u(\vebm{k})}\ket{u(\vebm{k})})\nonumber\\
		=&\braket{\partial_\vebm{k} u(\vebm{k})}{u(\vebm{k})} =-\bra{u(\vebm{k})}\partial_\vebm{k} \ket{u(\vebm{k})}
	\end{align}
	together with the anti-unitarity of $W$
	\begin{align}
		\bra{\Psi}W^\dag W\ket{\Phi} = \bra{W \Psi} \ket{W\Phi}
		= \bra{\Phi} \ket{\Psi}
	\end{align}
	and
	\begin{align}
		\braket{u(\vebm{k})}{u(\vebm{k})} &= 1 \\
		\implies \partial_\vebm{k} \braket{u(\vebm{k})}{u(\vebm{k})} &= 0 \\
		\implies \braket{\partial_\vebm{k} u(\vebm{k})}{u(\vebm{k})} &= -\braket{u(\vebm{k})}{\partial_\vebm{k} u(\vebm{k})}.
	\end{align}
	
	\subsubsection{Chern number constraints from $C_nT$ symmetries\label{subsq_chernnumberconstCnT}}
	
	Using Eq.(\ref{AppEq_timerev_A}), we can derive expressions similar to Eq.(\ref{Eq_ChargeFromRotEigVal}) for $C_4T$
	\begin{align}
		\nu_{c_b}&=\frac{2}{\pi}
		\left[\phi(\vebm{S})-\phi(\vebm{N}) \right] \mod 4\nonumber\\
		&=-\frac{2}{\pi}\Delta\phi_{b,c_b}\mod 4\label{Eq_constrantiunitaryC4T} ,
	\end{align}
	and for $C_6T$
	\begin{align}
		\nu_{c_b}=-\frac{3}{\pi}\Delta\phi_{b,c_b}\mod 3,
		\label{Eq_constrantiunitaryC6T}
	\end{align}
	with $\alpha(\vebm{k})=e^{-i\phi(\vebm{k})}$. The constraint for $C_6T$ is only defined mod $3$ instead of mod $6$, since $C_6T$ relates the Berry curvature of wedges spanning $1/3$ of a sphere to each other, instead of $1/6$ wedges. The main difference of Eq.\,(\ref{Eq_constrantiunitaryC4T}) and (\ref{Eq_constrantiunitaryC6T}) to Eq.\,(\ref{Eq_ChargeFromRotEigVal}) is that $\Delta\phi_{b,c_b}$ is no longer the change of a symmetry eigenvalue but the phase change of the anti-unitary symmetry sewing matrix (\ref{Eq_antiunitsewing}).
	
	$C_2T$ relates the Berry curvature of the upper to the lower hemisphere, so to derive a local constraint we need to consider a path $S_\text{equator}$ on the equator
	\begin{align}
		\nu_{c_b}&=\frac{1}{\pi}\int_{S_\text{equator}} \vebm{A}(\vebm{k}) \dd \vebm{k} \mod 2\\
		&=\frac{1}{\pi}\int_{S_\text{equator}} C_2\vebm{A}(\vebm{k}) \dd \vebm{k} \nonumber\\
		&- \frac{1}{\pi}\int_{S_\text{equator}} i \alpha^*(\vebm{k}) \partial_\vebm{k} \alpha(\vebm{k}) \dd \vebm{k} \mod 2\\
		&=- \frac{1}{\pi}\int_{S_\text{equator}} \vebm{A}(\vebm{k}) \dd \vebm{k} \nonumber\\
		&- \frac{1}{\pi}\int_{S_\text{equator}} i \alpha^*(\vebm{k}) \partial_\vebm{k} \alpha(\vebm{k}) \dd \vebm{k} \mod 2\\
		\implies \nu_{c_b}&= - \frac{1}{2\pi}\int_{S_\text{equator}} \partial_\vebm{k} \phi(\vebm{k}) \dd \vebm{k} \mod 2\label{Eq_constrantiunitaryC2T}
	\end{align}
	where we used Eq.\,(\ref{AppEq_timerev_A}), ${\bf A}(-C_2{\bf k})={\bf A}({\bf k})$ and $(C_2{\bf v})\dd \vebm{k}=-{\bf v}\dd  \vebm{k}$ with any vector ${\bf v}$ on the equator. So $\nu_{c_b}$ is even (odd) when $\phi(\vebm{k})$ winds an even (odd) number of times around the equator.
	
	A $C_3T$ constraint is redundant, since $C_3T$ implies $T=(C_3T)^3$ and therefore also $C_3$ to exist separately.
	
	\subsubsection{Chern number constraint of crossings at TRIMs\label{subsq_chernnumberconstTRIMs}}
	
	Next we would like to evaluate a single band Chern number of a crossing with time-reversal symmetry, $W=T$ ($R=1$) and $T^2=\gamma$, where $\gamma=1$ for spinless and $\gamma=-1$ for spinful fermions. We split the integration-sphere around the crossing into two halves,
	\begin{align}
		\nu &= \frac{1}{\pi}\biggl(\int_{\partial S_1} A(\vebm{k}) \text{d}\vebm{k} + \int_{\partial S_2} A(\vebm{k}) \text{d}\vebm{k} \biggr) \mod 2 \\
		&= \frac{1}{\pi}\biggl(\int_{\partial S_1} A(\vebm{k}) \text{d}\vebm{k} + \int_{\partial S_2} A(-\vebm{k})\text{d}\vebm{k} \nonumber\\
		&- \int_{\partial S_2} i \alpha^*(\vebm{k}) \partial_\vebm{k} \alpha(\vebm{k}) \text{d}\vebm{k}\biggr)\mod 2 \\
		&= -\frac{1}{\pi}\int_{\partial S_2} i \alpha^*(\vebm{k}) \partial_\vebm{k} \alpha(\vebm{k}) \text{d}\vebm{k}\mod 2 ,
	\end{align}
	where $\partial S_1$ and $\partial S_2$ are paths at the halves edge running on opposite sides. Using $\alpha(\vebm{k})=e^{-i\phi(\vebm{k})}$ we have
	\begin{align}
		\nu &= -\frac{1}{\pi}\int_{\partial S_2} \partial_\vebm{k} \phi(\vebm{k}) \text{d}\vebm{k}\mod 2\\
		&= \frac{1}{\pi} (\phi(\vebm{S}) - \phi(\vebm{N})) \mod 2\label{AppEq_nu_timerev1}
	\end{align}
	with $\vebm{N}$ and $\vebm{S}$ being the north and southpole. To evaluate this expression, consider Eq.\,(\ref{AppEq_Tsymmetry_apply}). We can reinsert itself with a replacement $\vebm{k}\to-\vebm{k}$ to get
	\begin{align}
		\alpha(\vebm{k})\ket{u(-\vebm{k})}&=T\alpha^*(-\vebm{k})T\ket{u(-\vebm{k})}\\
		&=\alpha(-\vebm{k})\gamma\ket{u(-\vebm{k})}
	\end{align}
	so
	\begin{align}
		\gamma&=\alpha(\vebm{k})\alpha^*(-\vebm{k})\\
		&=e^{-i(\phi(\vebm{k}) - \phi(-\vebm{k}))} \label{AppEq_phase_difference_timerev}
	\end{align}
	which can be applied to Eq.\,(\ref{AppEq_nu_timerev1}) to arrive at $\nu = 0 \mod 2$ for the spinless case ($\gamma=1$) and $\nu = 1 \mod 2$ for the spinful one ($\gamma=-1$). So any crossing at TRIMs, which include also multifold ones, without further degeneracies away from the crossing, must have even (odd) Chern numbers without (with) SOC. We see that this constraint is explicitly fulfilled in all models found in this paper, for example in section \ref{subsq_multifoldcrossings} and in all low-energy Weyl point Hamiltonian at TRIMs in \cite{tsirkin2017compositeWeyl}.
	
	\subsection{Global constraint on band topology}\label{Sec_GlobalBandTopology}
	
	For chiral band crossings global constraints on the band topology 
	arise due to conditions on the sum of the topological charges of nodal points.
	One such global constraint is the fermion doubling theorem by Nielsen and Ninomiya, which states that for each band the sum of all chiralities has to vanish \cite{NIELSEN1983389}.
	Here, we prove a global constraint on the rotation eigenvalues, which ultimately follows from the periodicity of the BZ, i.e., the compactness of the BZ.
	To do so, we employ symmetry representations along the full rotation axis, which can be obtained by taking powers of the symmetry \cite{FURUSAKI2017SciBulletin}.
	For concreteness we consider a screw rotation symmetry $C_n(x,y,\tfrac{m}{n})$, which describes an $n$-fold rotation around the $z$ axis followed by a translation with the vector $(x, y, \tfrac{m}{n})$. 
	Taking the $n$-th power of the screw rotation we obtain
	\begin{align} \label{Eq_powerScrew}
		\big[C_n(x,y,\tfrac{m}{n})\big]^n 
		= \mathrm{e}^{i \pi s} \, T(0,0,m) 
		= \mathrm{e}^{i \pi s + i m k_z },
	\end{align}
	where $m, n \in \mathbb{Z}$, $|m|<|n|$, and  $s= 0$ ($s= 1$) for spinless (spinful) systems.  
	In the second step the translation by a full lattice vector $T(0,0,m)$ is replaced by the usual one-dimensional representation of the translation group. 
	Notably, the above and all following steps apply analogously to glide mirror operations, which would correspond to an operation with $n=2$ and either $m=0$ or $m=1$ for mirror and glide mirror symmetry, respectively.
	The symmetry eigenvalues of the $C_n$ screw rotation is found as the complex root of Eq.~(\ref{Eq_powerScrew}) yielding
	\begin{align}\label{Eq_ScrewEigVal}
		\lambda_{C_n} = \exp(i \frac{2\pi p + s\pi +  m k_z}{n} ),
	\end{align}
	where $p \in \{0, 1, \ldots, n - 1 \}$ distinguishes the $n$ different complex roots. 
	On the rotation axes invariant under the rotation $C_n(x,y,\tfrac{m}{n})$, we label the bands using $\lambda_{C_n}$ or rather, equivalently, we consider the complex phase $\varphi(k_z) = \arg \lambda_{C_n}$.
	To label a specific band $b$ that is identified by sorting the eigenvalues of the Hamiltonian according to their energy, we have to consider that Eq.~(\ref{Eq_ScrewEigVal}) does not yet include band crossings. 
	The phase $\varphi_b(k_z)$ for a specific band is given by
	\begin{align} \label{Eq_accumulatedPhase}
		\varphi_b(k_z) =  \frac{2\pi p + s\pi + m k_z }{n} + \sum_{k_c \leq k_z} \Delta\varphi_{b,c},
	\end{align}
	which must include all phase jumps $ \Delta\varphi_{b,c}$ at $k_c$ corresponding to all crossings $c$ up to $k_z$, which  may be, for example, with the bands $b-1$ or $b+1$. 
	The essential step to identify the global constraints on $\varphi_b(k_z)$, $\Delta\varphi_{b,c}$, and in extension also on all chiral crossings is the periodicity of the Brillouin zone. 
	Thus, we compare the phase $\varphi_b(k^0_z)$ at some position $k^0_z$ with the phase $\varphi_b(k^0_z + 2 \pi)$ after traversing the Brillouin and accumulating phase jumps $ \Delta\varphi_{b,c} $ at $k_c$ as well as a contribution from $ \tfrac{m k_z }{n}$,
	\begin{align} 
		\varphi_b(k^0_z + 2 \pi) - \varphi_b(k^0_z)  &= 0 \mod 2 \pi 
		\\
		\sum_{c_b} \Delta\varphi_{b,c} + 2\pi \frac{m}{n} &= 0  \mod 2 \pi. 
		\label{Eq_EigvalChanges}
	\end{align}
	The phase jumps $\Delta\varphi_{b,c}$ in Eq.~(\ref{Eq_EigvalChanges}) are not independent for different bands $b$.
	For every phase jump there should be the reverse exchange of eigenvalues in a higher or lower band. 
	Suppose we consider a system, where all band crossings are twofold, then one may iteratively substitute Eq.~(\ref{Eq_EigvalChanges}) for band $b-1$  into the equation for band $b$.
	The induction process leads to Eq.~(\ref{Eq_PhaseConstraint})
	\begin{align}
		b = 1 && \sum_{c_1} \Delta\varphi_{1,c_1} &= - 2 \pi \frac{m}{n}  \mod 2 \pi
		\nonumber\\
		b = 2 && \sum_{c_2} \Delta\varphi_{2,c_2} - \sum_{c_1} \Delta\varphi_{1,c_1} &= - 2 \pi \frac{m}{n}  \mod 2 \pi
		\nonumber\\
		&&  \Leftrightarrow \sum_{c_2} \Delta\varphi_{2,c_2} &= - 2 \cdot 2\pi \frac{m}{n}  \mod 2 \pi
		\nonumber\\
		\vdots \quad && &\vdots
		\nonumber\\
		\text{any } b &&  \sum_{c_b} \Delta\varphi_{b,c_b} &= - b \cdot 2\pi \frac{m}{n}  \mod 2 \pi.
		\label{Eq_EigvalChanges_nonSymmorphic}
	\end{align}
	This result contains the notion of filling-enforced semimetals, namely, if $b \tfrac{m}{n} \not\in \mathbb{Z}$, then there must be at least one symmetry-enforced band crossing \cite{watanabe2016gaplessness}. 
	Once the filling, i.e., the considered band $b$, is a multiple of $\tfrac{n}{m}$, band crossings do not need to exist. 
	
	\section{Applications and extensions}
	\label{sec_four}
	
	To demonstrate the power of the local and global constraints, we present a number of applications
	and discuss some extensions. 
	
	\subsection{Applications and extensions \\ of the local constraint}
	
	In the following we use the local constraint, Eq.~(\ref{Eq_ChargeFromRotEigVal}),
	to prove the existence of enforced double Weyl points
	away from TRIMs (Sec.~\ref{SubSec_doubleWeyl}).
	We generalize the local constraint to multiple rotation symmetries in Sec.~\ref{SubSec_MultipleScrew}, which enables us to infer conditions for the Chern numbers for all types of (higher-fold) chiral crossings.
	Finally, we use the local constraint to show that nodal lines
	with nonzero Chern numbers cannot be stabilized by chiral space-group symmetries (Sec.~\ref{SubSubSec_ChiralNL}).
	
	\subsubsection{Chiral crossings between identical symmetry eigenvalues}\label{SubSec_doubleWeyl}
	
	In this section we use the local constraint, Eq.~(\ref{Eq_ChargeFromRotEigVal}), to explain the existence of unusual enforced double Weyl points
	away from TRIMs~\cite{Chen2012MultiWeyl,tsirkin2017compositeWeyl}.
	First, we clarify why these Weyl points pose an open question in the understanding of chiral crossings.
	According to conventional wisdom, a stable band degeneracy can only occur  if at least one of the three following conditions is  fulfilled:
	(i) The two bands forming the crossing belong to different symmetry representations, which prevents the introduction of gap opening terms, (ii) there is a higher-dimensional representation of the little group, or (iii) there exists an anti-unitary symmetry that leaves the degeneracy point invariant, leading to   Kramers degeneracy. 
	However, in space groups 80, 98, and 210 there exist
	band crossings away from TRIMs between bands with 
	identical representations of dimension one~\cite{Hirschmann2021Tetragonals}. 
	So at first glance, all of the above three conditions for a crossing seem violated. 
	Yet, the combination of time-reversal and fourfold rotation symmetry generates, due to Kramers theorem, point-like degeneracies at high-symmetry points of certain non-primitive Brillouin zones that are not TRIMs~\cite{Hirschmann2021Tetragonals}.
	Interestingly, with SOC these crossings are known to be double Weyl points
	with Chern number $\pm 2$~\cite{Yu2021Encyclopedia}, but could until now not be understood in terms of symmetry eigenvalues \cite{Chen2012MultiWeyl,tsirkin2017compositeWeyl}.
	
\begin{figure}
		\centering
		\includegraphics[width = 0.5 \textwidth]{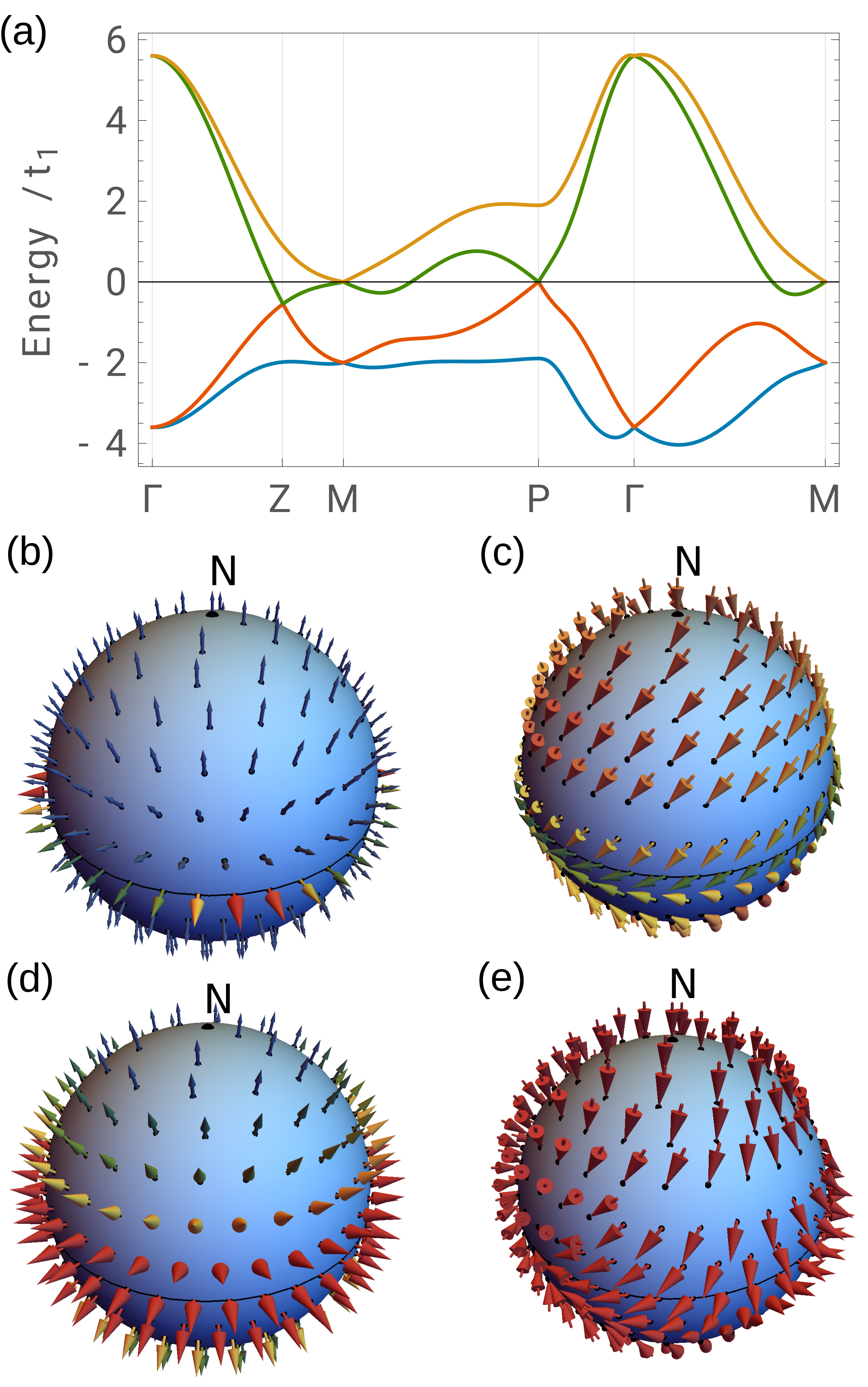}
		\caption{\label{fig_SG80_BandsAndTextures}
			\emph{Band structure, Berry curvature and spin texture for SG~80.}
			(a) Tight-binding model with SOC of SG~80 as defined in Eq.~\eqref{Eq_SG80fullHamiltonian}. The body-centered tetragonal Brillouin zone is shown in the inset of Fig.~\ref{fig_TaO2BandsAndSurfaceStates}(b).
			The double Weyl point at P is characterized by its Berry curvature and spin textures as shown in (b) and (c), respectively, which are given for the lower band of the crossing on a sphere enclosing it.
			For comparison, (d) and (e) show the Berry curvature and spin textures of a regular double Weyl point. 
			For all vector plots the north pole of the sphere is labelled by ``N".
		}
	\end{figure}

	For concreteness, let us now focus on the body-centered tetragonal SG~80 ($I4_1$),
	whose P point can host 
	two-fold degeneracies both with and without SOC
	[e.g., see Figs.~\ref{fig_SG80_BandsAndTextures}(a) and \ref{fig_SG80_spinless}]. As we will see, this band crossing can be understood by noting that the combined symmetry $T C_{4z}$, comprising time-reversal $T$ and   fourfold screw rotation $C_{4z}$, leaves the P point  invariant. Other than that, the only
	unitary symmetry that leaves P invariant is the rotation $C_{2z}$ whose symmetry eigenvalues can be used to label the bands.
	We now need to distinguish the case 
	with and without SOC, which differ slightly for SG~80. 
	Without SOC different eigenvalues  are paired by the anti-unitary operation $T C_{4z}$. 
	In our notation this corresponds to $\Delta \varphi = \pm \pi$ for the Weyl point at P which implies by Eq.~(\ref{Eq_ChargeFromRotEigVal}) a Chern number of $\nu_{P,\text{SG}80} = 1 \mod 2$.
	With SOC the representation is doubled compared to before and splits into two one-dimensional and one two-dimensional representation at P, because the Kramers theorem only applies to the latter representation, see Ref.~\cite{Hirschmann2021Tetragonals} for details.
	Since for the two-dimensional representation one eigenvalue of $C_{2z}$ is paired to itself, one finds $\Delta \varphi = 0$ implying $\nu_{P,\text{SG}80,\text{SOC}} = 0 \mod 2$.
	Taking into account that the crossing at P has been $\nu_{P,\text{SG}80} = 1$ without spin, it follows from the conservation of topological charge that $\nu_{P,\text{SG}80,\text{SOC}} = \pm 2$. 
	We have thus reached an explanation for the double Weyl point at P in terms of symmetry eigenvalues. 
	
	The discussed double Weyl point at P in SG~80 has a different origin and symmetry than any other twofold double Weyl point, which  occur either on fourfold or sixfold rotation axes or at TRIMs in the presence of spinless time-reversal symmetry \cite{Huang2016doubleWeyl, tsirkin2017compositeWeyl, Chen2012MultiWeyl}. 
	Hence, we expect that also the spin texture \cite{tan2022unified, chen2022discovery, feng2016spin, markovic2019weyl, xu2016spin, gatti2020radial, he2021kramers,hagiwara2022link} and Berry curvature
	are  distinct from the conventional double Weyl points.
	To demonstrate this, we compute the Berry curvature and spin texture of the double Weyl point in SG~80.
	For this purpose, we derive in Appendix~\ref{App_SG80_Models} a tight-binding model including SOC for SG~80.
	Figure~\ref{fig_SG80_BandsAndTextures}(a) shows the band structure of this model defined by Eq.~\eqref{Eq_SG80fullHamiltonian}. 
	As expected we find a double Weyl point of charge $\nu_{\text{P}} =2$ at each P point, which is compensated by a pair of double Weyl points on the fourfold rotation axis $\Gamma$-Z-M
	with $\nu_{\Gamma\text{-Z-M}} = -2$.
	Figures~\ref{fig_SG80_BandsAndTextures}(b) and~\ref{fig_SG80_BandsAndTextures}(c) show
	the Berry curvature and spin texture around the P point,
	respectively. 
	To contrast this with conventional double Weyl points
	we plot in Figs.~\ref{fig_SG80_BandsAndTextures}(d)
	and~\ref{fig_SG80_BandsAndTextures}(e) the Berry curvature and 
	spin texture of a conventional double Weyl point
	defined by $H(\vb{k}) = k_z \sigma_z + ( \sigma_+ (k_x - i k_y)^2 + H.C.)$, where $\sigma_+ = 1/2(\sigma_x + i \sigma_y)$ with the Pauli matrices $\sigma_j$ \cite{Zhang2020quadWeyl, chen2022discovery}.
	While the details of these textures are parameter-dependent, their symmetry properties are generic
	and dictated by the local little groups.
	In general the spin texture at P in SG~80 is anisotropic and symmetric under the \emph{anti-unitary} $C_4 T$ symmetry
	[see regions of similar color shading   in Fig.~\ref{fig_SG80_BandsAndTextures}(c)].
	In contrast, the texture of a conventional double Weyl point
	is symmetric under an \emph{unitary} (e.g., fourfold) rotaiton symmetry, see Fig.~\ref{fig_SG80_BandsAndTextures}(d,e). 
	Another difference is that the spin texture around the equator of Fig.~\ref{fig_SG80_BandsAndTextures}(c) has a unit winding, whereas the one of Fig.~\ref{fig_SG80_BandsAndTextures}(e) has   a winding of two.
	These differences in spin texture could be measured 
	experimentally, using, e.g., spin- and angle-resolved photoemission spectroscopy \cite{gatti2020radial, tan2022unified, hagiwara2022link}.

	Using a database search (see Sec.~\ref{sec_six}) we have 
	identified  NbO$_2$ and TaO$_2$ as candidate materials in SG 80 realizing the double Weyl points away from TRIMs.
	The band structure and surface states of
	these compounds are presented in Sec.~\ref{Subsec_NbO2TaO2}.
	Notably, we find that for surface terminations perpendicular
	to any of the crystal axes there appear four Fermi arcs.
	This is because for these terminations the P point
	is projected onto a symmetry related copy of itself with the same Chern number $\pm 2$, such that there emerge four Fermi arcs
	from the projeted P point in the surface BZ.
	
	The above arguments for SG~80 apply in a similar manner also
	to SG~98 ($I4_122$) and SG~210 ($F4_132$),
	for which the double Weyl points appaer at the P and W points, respectively.
	In addition, related arguments using the local constraint
	can be employed to understand the 
	charge $\nu=\pm2$ of the threefold crossings in SG~199 ($I2_13$) and SG~214 ($I4_132$) at the point P, see the discussion in Sec.~\ref{SubSubSec_ThreeAndSixfold}.

	\subsubsection{Chiral crossings with multiple rotation axes}
	\label{SubSec_MultipleScrew}
	
	Band crossing points symmetric under little groups that contain more than one rotation symmetry often exhibit larger topological charges than in the case of a single rotation symmetry \cite{bradlyn2016beyond, tang2017multiple, Schroeter2020MaximalChern}.
	Also in this case the local constraint, Eq.~(\ref{Eq_ChargeFromRotEigVal}), can be used to understand the observed topological charges.
	In the following, we extend the above arguments to multiple rotation axes and consider, for concreteness, a twofold quadruple Weyl point at $\Gamma$ in SG 195, for which a Chern number of $\nu = 4$ has been reported \cite{Zhang2020quadWeyl, Hiaoxian2021quadWeyl,Cui2021quadWeyl,Luo2021quadWeyl}.
	Other non-trivial examples of nodal points with multiple rotation symmetries are discussed in Secs.~\ref{sec_global_constraint_multifold_degen} and \ref{gen_clas_hamiltonians} in the context
	of multifold band crossings.

	For a single rotation axis one usually finds that $\nu_{c_b} =  \Delta \varphi_{b,c_b} \frac{n}{2 \pi}$ holds without the modulus operation, although the local constraint, Eq.~(\ref{Eq_ChargeFromRotEigVal}), restricts the possible charges only up to the order of the rotation $n$. 
	This is because higher topological charges would require fine tuning.
	To see this, consider a crossing point of charge $\nu = \nu_{c_b} + m n$, where $m$ is some non-zero integer.
	If this crossing is perturbed by some symmetry-allowed perturbation, the crossing may split into one with charge $\nu_{c_b}$ and $m$ sets of each $n$ Weyl points. 
	In fact, generally exactly this happens, because placing $m n$ Weyl points on the rotation axis is a fine-tuned situation.
	In other words, to achieve higher topological charges $\nu$, more lower orders in the low-energy expansion need to be set to zero, 
	which would require fine tuning.
	
\begin{figure}
		\centering
		\includegraphics[width = 0.5 \textwidth]{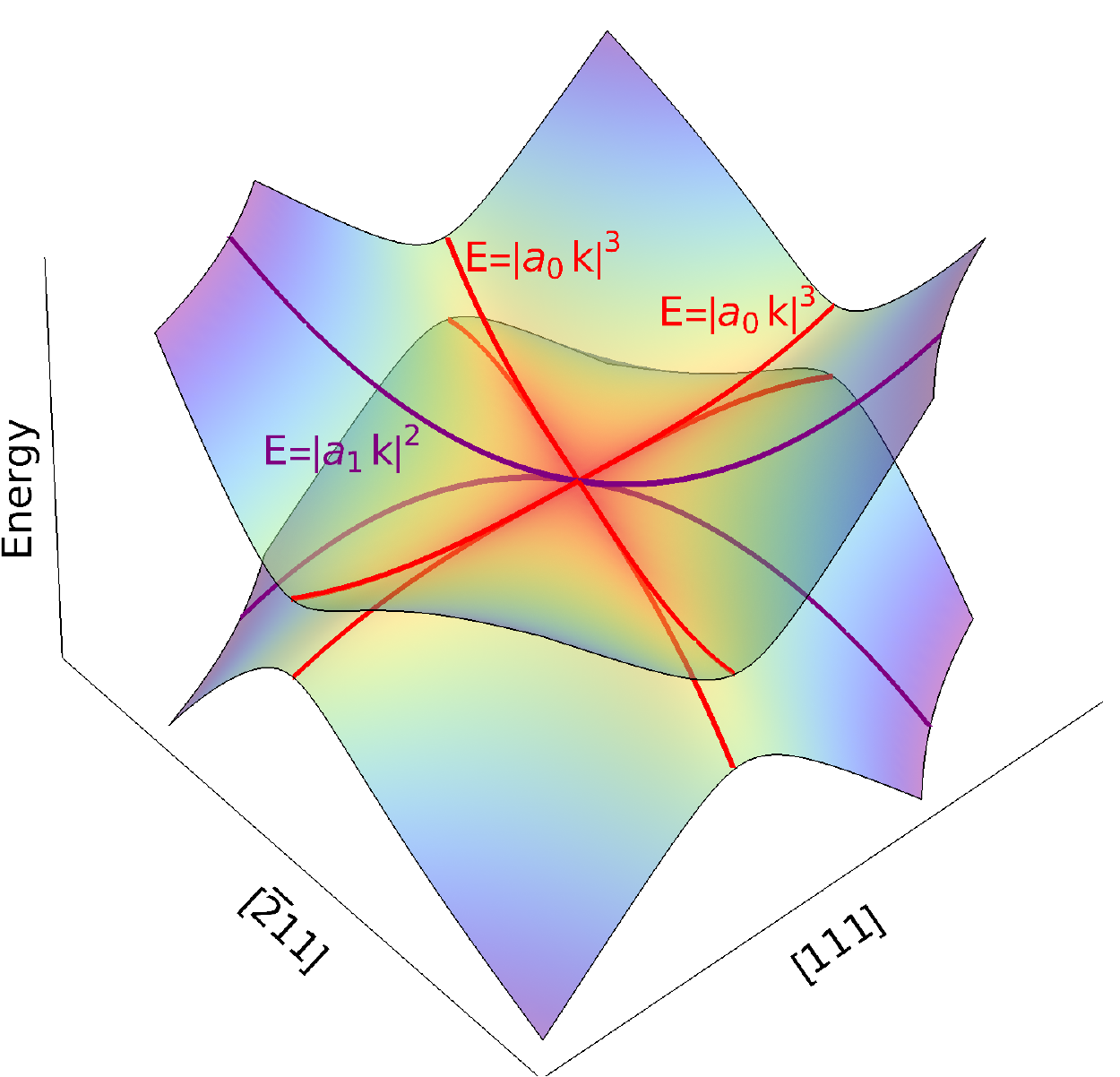}
		\caption{\label{fig_lowE_PG23_charg4}
			\emph{Quasi-symmetry-enforced Chern number in twofold quadruple Weyl points. }
			Band structure around a twofold band crossing symmetric under point group 23 and time reversal, as given by Eq.~(\ref{Eq_App_PG23}) with $d_1 = 0$. 
			The splitting of the bands along the threefold (twofold) rotation axes in red (purple) is cubic (quadratic). 
			A fourfold quasi-symmetry enforces a charge of $\nu = 4$.
		}
	\end{figure}

	In the presence of multiple rotation symmetries, however,
	there are more symmetry constraints that can lead to higher topological charges, such that the smallest possible value
	given by the local constraint~\eqref{Eq_ChargeFromRotEigVal}
	is not realized. 
	To demonstrate this, let us consider SG~195 ($P23$) with time-reversal symmetry, where a twofold quadruple Weyl point is enforced to occur at the TRIM $\Gamma$ (and also at R), if, for example, the $^1E^2E$ representation is placed on the Wyckoff position 1a \cite{elcoro2017double}.
	The corresponding little group at $\Gamma$ consist of
	the point group 23 together with time-reversal,
	which does contain a twofold and a threefold rotation, but no  fourfold rotation, different from
	the conventional twofold quadruple Weyl points
	of Refs.~\cite{Zhang2020quadWeyl, Hiaoxian2021quadWeyl,Cui2021quadWeyl,Luo2021quadWeyl}.
	From the representation of this little group, one finds that there is no exchange of the twofold rotation eigenvalues, while the threefold rotation eigenvalue switch.
	Thus, the local constraint~\eqref{Eq_ChargeFromRotEigVal} on the charge $\nu_\Gamma$ is
	\begin{align}\label{Eq_multRotSym}
		\nu_\Gamma = 0 \mod 2 \quad \text{and}\quad
		\nu_\Gamma = 1 \mod 3.
	\end{align}
	Thus, both $\nu_\Gamma = 4$ and $\nu_\Gamma = -2$ are in agreement with Eqs.~\eqref{Eq_multRotSym}. To resolve this ambiguity we construct a low-energy model $H_{\text{T}}(\vb{k})$ around $\Gamma$, which
	is symmetric under the point group~32 and time-reversal, see Appendix~\ref{App_SG195Gamma}.
	The energy bands of this model exhibit quadratic and cubic dispersions along different directions away from the crossing point, see Fig.~\ref{fig_lowE_PG23_charg4}.
	The topological phase diagram of this low-energy model contains
	only one phase with $\nu_\Gamma =  4$, in agreement with Eq.~(\ref{Eq_multRotSym}). 
	However, the lowest possible topological charge
	of $\nu_\Gamma = - 2$, cf.~Eq.~(\ref{Eq_multRotSym}),
	is not realized, in contrast to
	the conventional twofold quadruple Weyl points 
	with fourfold rotation symmetry~\cite{Zhang2020quadWeyl, Hiaoxian2021quadWeyl,Cui2021quadWeyl,Luo2021quadWeyl}.
	This raises the question, why can the charge $\nu_\Gamma = - 2$ not be realized in our low-energy model, even though 
	it would be consistent with the local constraint?  
	
	There are two ways to answer this question. 
	First, a closer look at the low-energy model presented in Appendix~\ref{App_SG195Gamma} reveals that there is a fourfold \emph{quasi-symmetry}, i.e., a symmetry of the low-energy model that is broken by terms of higher order in $k$. 
	Namely, the low-energy Hamiltonian $H_{\text{T}}(\vb{k})$, Eq.~\eqref{Eq_App_PG23}, is left invariant by $U(C_4^z)^\dagger H_{\text{T}}(\vb{k}) U(C_4^z) = H_{\text{T}}(R^{-1}\vb{k})$,
	where $U(C_4^z)$ is the representation of a fourfold rotation symmetry and
	$R$ is the corresponding transformation in real space, 
	see Eq.~\eqref{Eq_App_DefQuasiFourfold}.
	$U(C_4^z)$ is a quasi-symmetry, as it is a symmetry
	only of the lowest-order terms, but not of the full Hamiltonian.
	Yet, since the Chern number is determined exclusively by the lowest orders in $k$, at which the point crossing is well-defined, this quasi-symmetry forces the charge to be $\nu_\Gamma = + 4$ for the crossing by adding the local constraint $\nu_{\Gamma} = 0 \mod 4$.
	
	Second, the Chern number $\nu_{\Gamma} = 0 \mod 4$ can
	be understood by considering 
	how the symmetries act on the 
	Berry curvature integration.
	For this purpose we need to consider the role
	of the time-reversal symmetry together with the twofold rotation,
	to be specific the combination $C_2 T$. 
	The Berry curvature flux through the northern $S_W(N)$ and southern $S_W(S)$ half of the spherical wedge $S_W$ when considering $C_2$ are identical due to $C_2 T$ and thus
	\begin{align}
		\nu_{\Gamma}
		&=
		2 \tfrac{2}{2\pi} \int_{S_W(N)} \dd {\bf n} \cdot {\bf \Omega}({\bf k})\\
		&= \frac{2}{\pi} \int_{S_{{\bf N}{\bf P_1}}+S_\text{equator}+S_{{\bf P_2}{\bf N}}}{\bf A}\dd {\bf s}&& \text{mod } 4
	\end{align}
	where we split the path $S_W(N)$ into  $S_{{\bf N}{\bf P_1}}+S_\text{equator}+S_{{\bf P_2}{\bf N}}$ and $P_1$ and $P_2$ being endpoints of $S_\text{equator}$. The Berry connection integration on paths $S_{{\bf N}{\bf P_1}}$ and $S_{{\bf P_2}{\bf N}}$ are related by symmetry and can be evaluated in a similar way as in Eq.\,\ref{Eq_berry_connection_integration}, 
	\begin{align}
		\int_{S_{{\bf N}{\bf P_1}}+S_{{\bf P_2}{\bf N}}}{\bf A}\dd {\bf s} &= \phi_b({\bf P_1}) - \phi_b({\bf N}).
	\end{align}
	Regarding the $S_\text{equator}$ integration, consider Eq.\,(\ref{Eq_symmetrytransformA}) applied on this path
	\begin{align}
		\int_{S_\text{equator}}{\bf A}(C_2{\bf k})\dd {\bf s} &= \int_{S_\text{equator}}C_2({\bf A}({\bf k}) - \nabla\phi_b({\bf k}))\dd {\bf s}\nonumber\\
		\implies \int_{S_\text{equator}}{\bf A}(-{\bf k})\dd {\bf s} &= -\int_{S_\text{equator}}{\bf A}({\bf k}) - \nabla\phi_b({\bf k})\dd {\bf s}\nonumber\\
		\implies \int_{S_\text{equator}}{\bf A}({\bf k})\dd {\bf s} &= -\int_{S_\text{equator}}{\bf A}({\bf k}) - \nabla\phi_b({\bf k})\dd {\bf s}\nonumber\\
		\implies \int_{S_\text{equator}}{\bf A}({\bf k})\dd {\bf s} &= \frac{1}{2} \int_{S_\text{equator}} \nabla\phi_b({\bf k}) \dd {\bf s}\\
		&=\frac{1}{2}(\phi_b({\bf P_2})-\phi_b({\bf P_1}))
	\end{align}
	where all equations are valid up to mod $2\pi$ and with ${\bf A}(C_2{\bf k})={\bf A}(-{\bf k})$ and $(C_2{\bf v})\dd s=-{\bf v}\dd s$ with any vector ${\bf v}$ on the equator. We also used Eq.\,(\ref{AppEq_timerev_A}) relating ${\bf A}({\bf k})$ and ${\bf A}(-{\bf k})$ via time-reversal symmetry
	\begin{align}
		&\int_{S_\text{equator}}{\bf A}(-{\bf k})\dd {\bf s} \mod 2\pi = \nonumber\\ 
		&\int_{S_\text{equator}}{\bf A}({\bf k})\dd {\bf s}
		- i \int_{S_\text{equator}}\alpha^*(\vebm{k}) \partial_\vebm{k} \alpha(\vebm{k})\dd {\bf s} \mod 2\pi = \nonumber\\
		&\int_{S_\text{equator}}{\bf A}({\bf k})\dd {\bf s}
		+ \phi_T(\vebm{P_1}) - \phi_T(\vebm{P_2})\mod 2\pi = \nonumber\\
		&\int_{S_\text{equator}}{\bf A}({\bf k})\dd {\bf s} \mod 2\pi,
	\end{align}
	where the time-reversal symmetry sewing matrix has the form $\alpha(\vebm{k})=e^{-i\phi_T(\vebm{k})}$ and $\phi_T(\vebm{P_1})=\phi_T(-\vebm{P_1})=\phi_T(\vebm{P_2})$ in the spinless case (see Eq.\,(\ref{AppEq_phase_difference_timerev})).
	In total, we get
	\begin{align}
		\nu_\Gamma 
		&= \frac{2}{\pi} \left(-\phi_b({\bf N}) + \frac{1}{2}(\phi_b({\bf P_1})+\phi_b({\bf P_2}))\right)&& \text{mod } 4.
	\end{align}
	Explicitly calculating $\phi_b({\bf P_1})+\phi_b({\bf P_2})$ in the irrep $\Gamma_2\Gamma_3$ by plugging in the Bloch wavefunctions of the low-energy Hamiltonian into Eq.(\ref{Eq_SewingMatrix}) yields $0$.
	So
	\begin{align}
		\nu_\Gamma &= \frac{2}{\pi} \varphi({\bf N}) && \text{mod } 4.
	\end{align}
	The irrep $\Gamma_2\Gamma_3$ in $^1E^2E$ of $C_2$ is $\sigma_0$, so $\varphi({\bf N})=0$, which means we get
	\begin{align}\label{Eq_RefinedLocalConstraint_PG23}
		\nu_\Gamma &= 0 \,\text{mod } 4.
	\end{align}
	A material implementation of this WP with charge 4 was found in BaIrP, see \cite{Zhang2020quadWeyl}. There it is shown that, upon introducing SOC, this crossing evolves into 12 WPs and a 4-fold crossing at $\Gamma$ with $C=\nu_{1,2}=\pm4$, which is in section \ref{4foldcrossingssection} revealed to be the $\nu_n=\{\pm3,\pm1,\mp1,\mp3\}$ phase of the model described there.

	\subsubsection{Chiral nodal lines}
	\label{SubSubSec_ChiralNL}

	Nodal lines protected by crystalline symmetries are commonly discussed in the context of mirror symmetries, which leave a plane in the Brillouin zone invariant where they provide two distinct representations. 
	The presence of two distinct representations is sufficient to obtain accidental nodal lines.
	Furthermore, there can be symmetry-enforced line crossings, for example, if another symmetry operation anticommutes with the mirror symmetry, one finds nodal lines pinned to high-symmetry paths. 
	Alternatively, if the original mirror symmetry is nonsymmorphic this is already enough to conclude in analogy to Eq.~(\ref{Eq_PhaseConstraint}) that there must be an odd number of nodal lines crossing every other gap, which are movable in the sense that their position is parameter-dependent \cite{Young2015enforcedCrossings, Takahashi2017lines}. 
	Other cases of nodal lines include higher-fold nodal lines or almost movable nodal lines, which are only pinned to a finite number of high-symmetry points \cite{xie2021kramers, Hirschmann2021Tetragonals}.  
	
	For all of these nodal lines the Chern number vanishes because of the mirror symmetry, when calculated on a surface that fully encloses the nodal line.
	It comes to no surprise that despite the extensive research on various types of nodal lines, no example of a stable \emph{chiral} nodal line, i.e., a nodal line with Chern number, has been discussed so far. 
	Nevertheless, there are some reports of such nodal lines without mirror symmetry in the literature, which are either of unclear symmetry protection \cite{chiu2018emergence} or as in the case of the nodal lines in hexagonal AgF$_3$ \cite{gonzalez2020topological} have found to be actually weakly gapped \cite{gonzalez2021chiralonscrew}.
	Whether a chiral nodal line can exist is not only of interest due to its unique topology, but also important for the study of enforced topological nodal planes. 
	To rigorously deduce the existence of the latter, one needs to assume that a chiral nodal line does not exist. 
	In this case a non-zero sum of Weyl point chiralities within the Brillouin zone, implies a charged nodal plane, see Sec.~\ref{sec_enforced_nodal_planes}.
	
	In this section we aim to answer, whether chiral nodal lines can be stabilized by crystalline symmetries, and we will extensively apply the rotation symmetry constraint of Eq.~(\ref{Eq_ChargeFromRotEigVal}). 
	Doing so we consider points in reciprocal space lying away from any (glide) mirror planes. 
	To approach the first goal, let us assume that we have obtained a nodal line at a generic position in the Brillouin zone with a chirality $\nu_{\text{line}} < n$, where $n$ is the order of the highest rotation symmetry. 
	Suppose in this gedankenexperiment that we introduce all symmetry-allowed perturbations to the system to gap out the chiral nodal line. 
	Since the nodal line is assumed to be chiral, its topological charge has to persist in the form of Weyl points. 
	But as long as the original rotation symmetry is preserved, the condition $\nu_{\text{line}} < n$ implies that a nodal line cannot be gapped, because the number of resulting Weyl points at generic positions would be equal to $\nu_{\text{line}}$ and thus incompatible with the required multiplicity $n$. 
	Unlike nodal lines protected by a $\mathbb{Z}_2$ invariant, shrinking the nodal line to a point would not remove it, but leave a Weyl point with the same topological charge behind.
	
	Yet, despite this argument to stabilize a chiral nodal line, we will discuss in the following that the relation between rotation symmetry eigenvalues and the chirality, see Eq.~(\ref{Eq_ChargeFromRotEigVal}), strongly limit the possibility to find any nodal band feature fulfilling $\nu_{\text{line}} < n$.
	
	First, suppose there is a nodal line encircling an n-fold rotation axis. 
	Then, we can enclose the whole line by a sphere analogously to Fig.~\ref{fig_IntegrationPathAndWinding}(a), which implies by the arguments given in Sec.~\ref{Sec_ChernAndEigVal} that the chirality of all band crossings enclosed by the sphere is related to the change of rotation eigenvalues on the north and south pole of the rotation axis $\Delta \varphi_{b}$. 
	Several cases must be distinguished. 
	If there is no additional point crossing on the rotation axes, then $\Delta \varphi_{b} = 0$ leading to $\nu_{\text{line}} = 0 \mod n$, implying that the nodal line is trivial or at least unstable.
	If there are indeed additional point crossings on the rotation axis, then $\Delta \varphi_{b} \neq 0$ and one may choose the sphere to enclose only the point crossings, which implies that these crossings by themselves are responsible for the charge of $\Delta \varphi_{b} \tfrac{n}{2 \pi} \mod n $, which would be observed on the original sphere.
	In both cases the chiral nodal line is unstable. 
	
	To circumvent the objections, one may consider more intricate configurations of nodal lines. 
	If one examines a nodal line that is sufficiently extended such that it cannot be enclosed by a sphere, it is generally still possible to find a surface to enclose the line and a section of the rotation axis. 
	The proof of Eq.~(\ref{Eq_ChargeFromRotEigVal}) can then be repeated for this new surface, after the subdivision of the integration surface for the Berry curvature, the edges must be related by symmetry, see also Ref.~\cite{Fang2012ChernFromSymmetry}, where the integration surface intersects more than one rotation axis.
	Ultimately, one finds an expression depending on the changes of eigenvalues of the different rotation axes, but the symmetry representation only changes when traversing the integration contour if a crossing on a rotation axes has been enclosed. 
	Thus, either the nodal line itself has crossed a rotation axes and is responsible for the exchange of symmetry eigenvalues, such that the nodal line can be gapped out except at a set of corresponding point crossings on the rotation axes, or there will be no exchange of symmetry eigenvalues and hence at most a trivial charge. 
	In Appendix~\ref{App_chiralNL} we discuss the case of antiunitary symmetries of higher multiplicity and show that they do not circumvent the result obtained above from Eq.~(\ref{Eq_ChargeFromRotEigVal}).
	In summary, we find that all configurations of chiral nodal lines discussed here do not fit to the original proposal of a topological charge $\nu_{\text{line}} < n$, hence we find that no \emph{crystalline} symmetry is able to protect a chiral nodal line.
	Note, a chiral nodal line may still be found by considering systems with internal symmetries. 
	
\begin{figure}
		\centering
		\includegraphics[width = 0.5 \textwidth]{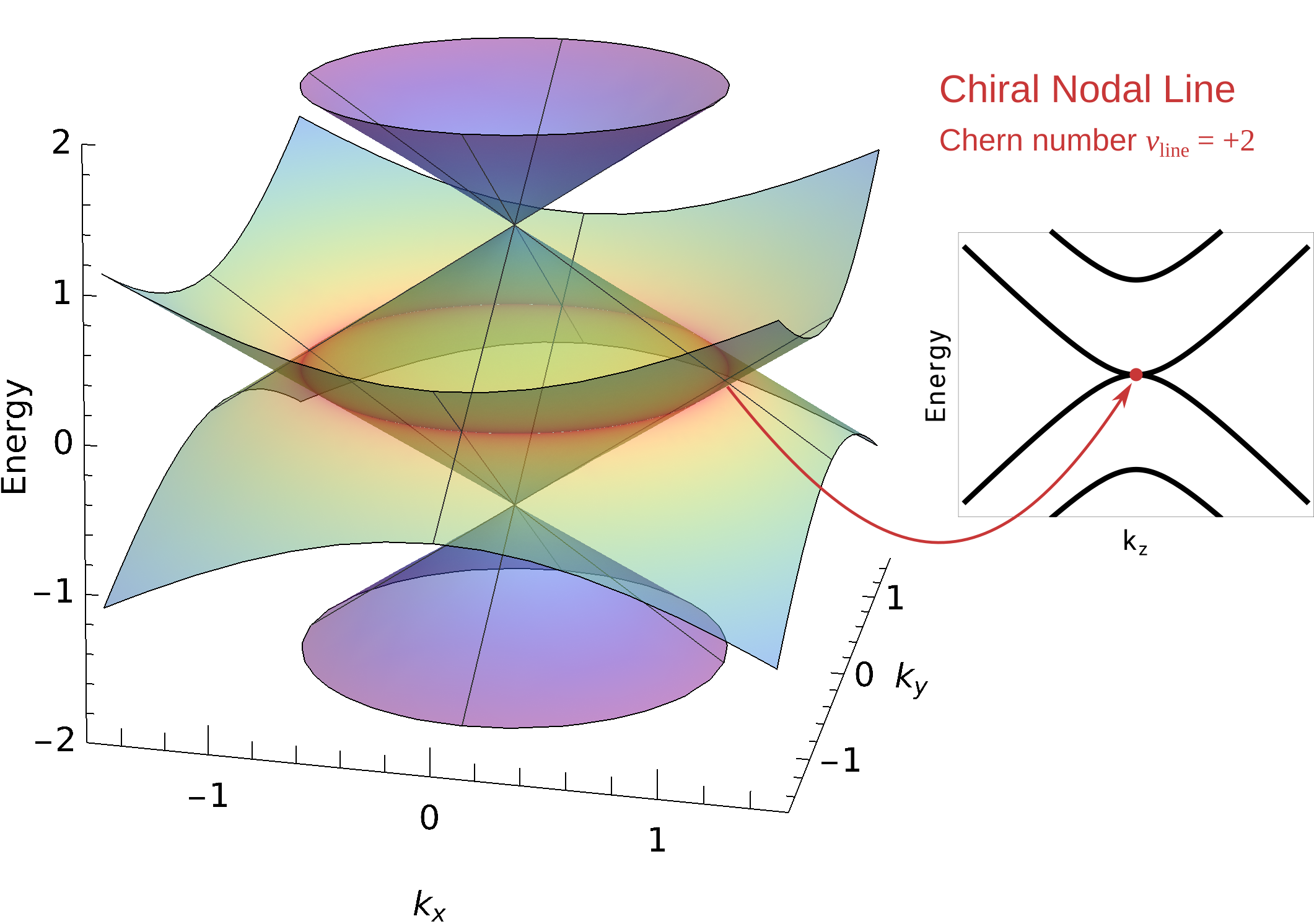}
		\caption{\label{fig_chiralNodalLine}
			\emph{Chiral nodal line.} 
			Band structure for the model chiral nodal line (red highlighting) described by Eq.~(\ref{Eq_chiralNLmodel}). 
			The dispersion of the nodal line is linear (quadratic) in radial ($k_z$) direction. 
		}
	\end{figure}

	To conclude this section we  propose a low-energy model of a chiral nodal line to illustrate how our above symmetry argument can be circumvented.
	In this construction we place two Weyl points, $W({\bf k},\epsilon)$, of different energy at the origin and couple them by the matrix $A({\bf k})$ in a way that preserves an internal symmetry $U_{\text{line}}$. 
	We define
	\begin{align}\label{Eq_chiralNLmodel}
		W({\bf k},\epsilon) &= {\bf k} \cdot {\bf \sigma} + \epsilon \sigma_0,
		\\
		A({\bf k}) &= k_z \sigma_1, \qquad \text{and}
		\\
		H({\bf k}) &= \mqty(W({\bf k},\epsilon) & A({\bf k})  \\ A({\bf k})^\dagger & W({\bf k},-\epsilon)),
	\end{align}
	where we set the energy offset to $\epsilon = 1$, ${\bf \sigma}$ is the vector of Pauli matrices, and $\sigma_0$ denotes the two-dimensional unit matrix. 
	The bands of the Weyl points intersect in a nodal sphere \cite{Tuerker2018nodalsurfaces} and are gapped by $A({\bf k})$  except at $k_z = 0$, see Fig.~\ref{fig_chiralNodalLine} for the resulting band structure. 
	This model exhibits a chiral nodal line with a Chern number of $\nu_{\text{line}} = 2$, which is inherited from the interplay of two $\nu = +1$ Weyl points. 
	It has to be noted that the charge of such a nodal line does also dependent on the hybridization away from $k_z = 0$, e.g., for $A({\bf k}) = k_z \sigma_3$ the nodal line is not charged. 
	Interestingly, there is a fourfold rotation symmetry $U(C_4) = \sigma_0 \otimes \sigma_3$, which is broken by $A({\bf k})$ for $k_z \neq 0$.
	Yet, perturbations that preserve the $U(C_4) $ symmetry gap out the nodal line, because the nodal line is not pinned to the $k_z = 0$ plane and loses its symmetry protection once moved away despite $\nu_{\text{line}} < n$.
	Nevertheless, there is an orbital symmetry in our model, namely, 
	\begin{align}
		U_{\text{line}} = \frac{1}{\sqrt{5}}\mqty(
		2 & 0 & 1 & 0 \\
		0 & 2 & 0 & 1 \\
		1 & 0 & -2 & 0 \\
		0 & 1 & 0 & -2 
		),
	\end{align}
	which fulfills $[H({\bf k}), U_{\text{line}}] = 0$. 
	The matrix is a unitary operation with the eigenvalues $\{-1,-1,+1,+1\}$ that exchange at the chiral nodal line. 
	Thus, any perturbation that respects the symmetry $U_{\text{line}}$ may deform the nodal line, but cannot introduce a gap.
	Such a chiral nodal line could be realizable for example in optical metamaterials or other synthetic systems.
	
	\subsection{Applications and extensions \\ of the global constraint}
	
	The global constraint contains the information on the possible numbers of crossings on rotation axes or mirror planes.
	This is for example a guide to the search for semimetals with few point or line crossings \cite{Hirschmann2021Tetragonals}.
	In this section we combine both constraints and the Nielsen-Ninomiya theorem \cite{NIELSEN1983389}.
	First, we will discuss a \emph{paramagnetic} space group with an enforced topological nodal plane duo.
	Secondly, we illustrate the constraints with a real band structure including accidental Weyl points, nodal planes, and multi-fold crossings.

	\subsubsection{Symmetry-enforced topological nodal planes}
	\label{sec_enforced_nodal_planes}
	
	In the following section we apply local and global constraints to the theory of symmetry-enforced topological nodal planes. 
	After a brief summary of the basic arguments that lead to (topological) nodal planes, we consider the nontrivial case of SG~94 $P4_22_12$.
	This space group is the only known case with two symmetry-enforced topological nodal planes in a paramagnetic space group, i.e., in a grey group including time reversal as a symmetry element. 
	
	We consider nodal planes as two-fold degeneracies on the surface of the Brillouin zone. 
	Such degenerate planes can be symmetry-enforced by the combined symmetry comprising time-reversal $T$ and a two-fold screw rotation $\tilde{C}_2$ \cite{wu2018nodal, chang2018topological, yu2019circumventing, wilde2021symmetry}.
	In short, the anti-unitary symmetry $T \tilde{C}_2$ fulfills Kramers theorem at every point on a plane in the Brillouin zone.
	Regions that host nodal planes are described by $k_i = \pi$ in units of the corresponding inverse lattice constant and have to be at the surface of the Brillouin zone.
	This gives rise to a natural distinction based on the number nodal planes (one, two, or three) or equivalently distinct symmetries $T \tilde{C}_2$ with eligible planes in the Brillouin zone. 
	We refer two the case of two (three) nodal planes as nodal plane duo (trio) to highlight that these nodal planes form a single connected object that can only be assigned a single Chern number. 
	
	The whole gapless structure of nodal planes may exhibit a non-zero Chern number on a surface that encloses the plane, if mirror and inversion symmetries are absent.
	For nodal planes trios, i.e., nodal planes at $k_i = \pi$ with $i\in\{x,y,z\}$, a single Kramers Weyl point at the TRIM $\Gamma$ can only be compensated by an opposite charge on the nodal planes, where one needs to consider the case of spinful time-reversal symmetry  \cite{chang2018topological, yu2019circumventing}.
	For nodal plane duos, a similar argument would result in two Kramers Weyl points that might cancel, hence it is a priori unclear, whether nodal plane duos may be nontrivial. 
	
	A topological nodal plane duo can, for example, occur due to the global constraint in a time-reversal broken state. 
	The simplest case is realized in ferromagnetic MnSi with the magnetic space group 19.27 $P2_1'2_1'2_1$.
	While the planes $k_x =\pi$ and $k_y = \pi$ exhibit a nodal plane duo, there remains only the two-fold rotation axis through $\Gamma$ that is not part of nodal planes. 
	On this axes the global constraint takes the form 
	\begin{align}
		\sum_{c_b} \Delta \varphi_{b,c_b} 
		= 
		- b \cdot \pi \mod 2 \pi,
	\end{align}
	which results in an odd number of crossings for bands with odd $b$. 
	Since each crossing exhibits a charge of $\nu = \pm 1$, cf. Eq.~(\ref{Eq_ChargeFromRotEigVal}), there is an odd overall charge within the Brillouin zone that cannot be compensated by generic crossings of even multiplicity.
	Thus, the nodal plane duo is topological with a charge of $\nu_{\text{NP}} = 1 \mod 2$, see \cite{wilde2021symmetry}.
	
	Finally, we consider the nodal plane duo enforced by SG~94 $P4_22_12$ for a spinful description with time-reversal symmetry.
	Again, the global constraint gives a non-zero sum of phase jumps for odd $b$ along the fourfold rotation axis Z-$\Gamma$-Z.
	Here, one needs the local constraint, because it is insufficient to count the number of Weyl points, which may occur as single and double Weyl points. 
	One finds
	\begin{align}
		\nu_{\text{Z-}\Gamma\text{-Z}}
		\equiv
		\sum_{c_b} \nu_{c_b}
		= 
		\sum_{c_b} \frac{n}{2\pi} \Delta \varphi_{b,c_b}  
		= 
		-2 b
		\mod 4,
	\end{align}
	where the global and local constraints, Eqs.~(\ref{Eq_ChargeFromRotEigVal}) and (\ref{Eq_PhaseConstraint}), have been substituted into the sum of all crossings on the fourfold rotation axis, and $m = 2$, $n=4$.
	Thus, for odd $b$ the Chern number of the nodal plane duo $\nu_{\text{NP}} = -\nu_{\text{Z-}\Gamma\text{-Z}} \neq 0$ independently of the details of the system.
	
\begin{figure}
		\centering
		\includegraphics[width = 0.49 \textwidth]{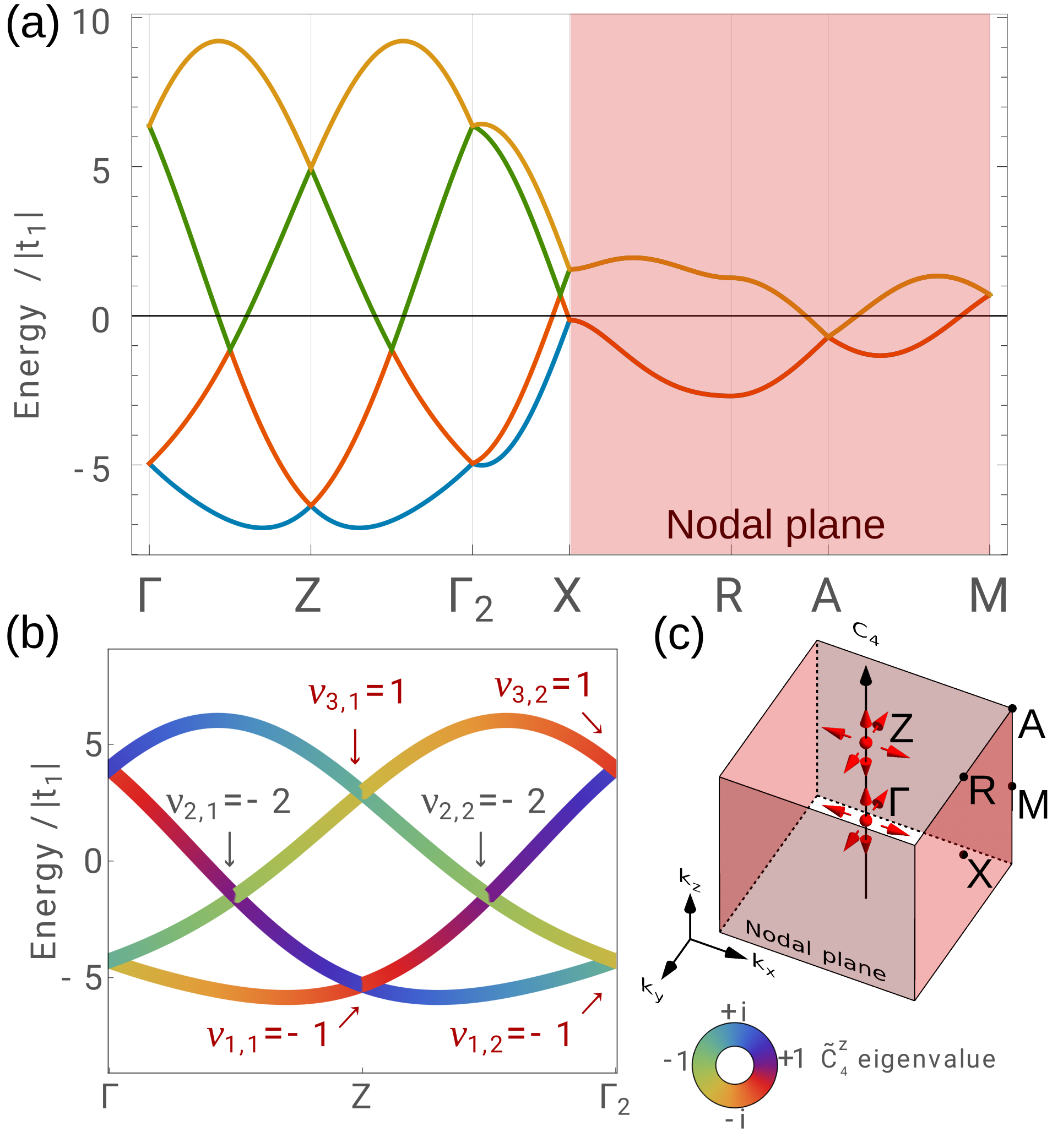}
		\caption{\label{fig_SG94_NPDuo}
			\emph{Tight-binding model of SG~94.}
			(a) Band structure of the model defined in Eq.~\eqref{Eq_SG94_TBmodel}.
			(b) Symmetry eigenvalues of the fourfold screw rotation $\tilde{C}_4$ along the full rotation axis $\Gamma-Z-\Gamma_2$. 
			The chiralities $\nu$ are compensated by an opposite charge contained within the nodal plane shaded in red in (a) and in the Brillouin zone (c).
			In (c) the arrows show exemplarily the Berry curvature associated to the point crossings at $\Gamma$ and Z.
		}
	\end{figure}

	To illustrate these results we have devised a Hamiltonian of SG~94, see Appendix~\ref{App_SG94_NP_Duo}.
	This model has minimal set of four connected bands with symmetry-enforced hourglass band structures along $\Gamma$-Z and $\Gamma$-X, see Fig~\ref{fig_SG94_NPDuo}(a), and two nodal planes covering the surfaces defined by $k_x = \pi$ or $k_y = \pi$, Fig~\ref{fig_SG94_NPDuo}(c).
	The chiralities of Weyl pointson the $\Gamma$-Z-$\Gamma_2$ follow our local constraint Eq.~\eqref{Eq_ChargeFromRotEigVal}, see Fig~\ref{fig_SG94_NPDuo}(b).
	As predicted the nodal planes are topologically charged.
	For example, for the lowest band the chiralities $\nu_{1,1} = \nu_{1,2} = -1$ at $\Gamma$ and Z, respectively, add up and are also not compensated by charges at generic positions. 
	Thus, the lower nodal planes carries the opposite Chern number $\nu_{1,\text{NP}} = +2$.
	This concludes our discussion of SG~94, which is the only known space group that enforces a pair of enforced topological nodal planes without magnetism.
	
	\subsubsection{Global constraints for multi-fold degeneracies}
	\label{sec_global_constraint_multifold_degen}
	
	While in the previous example the absence of a multi-fold crossing simplified the exposition, we are going to consider in the following the opposite case, where several multi-fold crossings occur and local and global constraints may not directly substituted into each other. 
	The salient difference is that at multi-fold crossings the exchange of bands described by $\Delta \varphi$ may occur with lower (or higher bands) that are not necessarily adjacent to the considered band, i.e., not only bands $b-1$ (or $b+1$).
	
	To illustrate the constraints with multi-fold crossings in a real band structure we discuss the cubic compound BaAsPt (SG~198 $P2_13$) exhibiting an unusual multi-fold crossing point including a $\nu =5$ band. 
	The material will be closer examined in Sec.~\ref{Sec_BaAsPt}.
	
	Here, our goal is to give some intuition on how the global constraint is fulfilled, while respecting the Nielsen-Ninomyia theorem.
	For the latter one may pick in principle any subset of bands to determine the relevant chiralities of crossings by the non-Abelian generalization of the Chern number for this set of bands.
	Here, we consider for simplicity only the band $b'$ that bounds the red shaded area in Fig.~\ref{fig:BaAsPt_bandstruct_wannier} from below.
	
	Since we encounter multi-band crossings, e.g., at $\Gamma$, M, and R, we have to use the general form of the global constraint introduced in Eq.~(\ref{Eq_EigvalChanges}).
	It implies that each two-fold rotation axis should exhibit a total phase $ \sum_{c_{b'}} \Delta \varphi_{b',c_{b'}} = \pi \mod 2\pi$, whereas the symmorphic three-fold rotations require a phase change of $ \sum_{c_{b'}} \Delta \varphi_{b',c_{b'}} = 0 \mod 2 \pi$.
	Note that in Eq.~(\ref{Eq_EigvalChanges}) the summations include crossings to lower and to higher bands on the considered rotation axis.
	
	One half of the threefold rotation axis $\Gamma$-R, cf.points Fig.~\ref{fig:BaAsPt_bandstruct_wannier}, exhibits three Weyl points to higher and three to lower bands together with multi-fold crossings at $\Gamma$ and R (one of the latter is in close proximity to $\Gamma$).
	Since between each two crossing on $\Gamma$-R there is a crossing in the next lower gap, all crossings to the higher (lower) band have an identical phase jump $\Delta \varphi$ and the numerical calculation determines $\Delta \varphi = +2\pi/3 $ ($-2\pi/3$).
	When taking the position relative to band $b'$ into account all five Weyl points thus contribute $\nu = +1$.
	As a side remark, the lower crossing of charge $\nu = -1$ in the inset appears together with three generic Weyl points with $\nu = +1$ in its close proximity.
	Due to their close proximity, we have labeled the crossing on the $\Gamma$-R axis with the total charge of the four crossings on at next to the rotation axis, i.e., $\nu = +2$ for the band that bounds the red-shaded region from below. 
	While these generic crossings are not symmmetry-enforced similar arrangements of crossings around a threefold rotation axis have been predicted before in an analysis of CoSi, which has the same SG~198 \cite{huber2022network}.   
	For the full threefold axis $\Gamma$-R there are 12 phase jumps adding up to a phase shift of  $\Delta \varphi = +24\pi/3 = 0  \mod 2 \pi$.
	The multi-fold crossings at $\Gamma$ and R exhibit $\Delta \varphi = \pm 2 \pi/3$ such that in total the phase equals $0 \mod 2 \pi$ on each threefold rotation axis.
	In total the band $b'$ comprises thus 48 Weyl points of charge $\nu = +1$ on generic points of its threefold rotation axes.
	
	On the two-fold rotation axis along $\Gamma$-X there is one crossing contributing $\Delta \varphi = \pi$.
	Since $\Gamma$ and X are time reversal invariant and the twofold rotation eigenvalues are complex at $\Gamma$ and real at X, it is clear prior to any calculation that the phase changes at $\Gamma$ but not at X.
	Thus, a full twofold rotation axis X-$\Gamma$-X exhibits an odd number of phases as expected.
	Overall there are 6 Weyl points of $\nu = +1$ on the twofold axes through $\Gamma$.
	The two-fold bands on the nodal planes along the R-M line exhibit two distinct representations that are characterized by two-fold rotation eigenvalues like on $\Gamma$-X, thus also here the global constraint applies.
	On R-M there is a crossing to a lower band as well as a pinned crossing at M, both exhibit an exhange of bands, whereas none occurs at R. 
	Along R-M-R there is an odd number of crossings fulfilling the global constraint
	Although we encounter chiral crossings on R-M, these do not contribute an Abelian Chern number to the band $b'$, because a surface enclosing them is gapless due to the presence of the nodal planes.
	But for the Nielsen-Ninomiya theorem applied to band $b'$, one has to consider the contribution to the nodal plane of $\nu = 3$.
	
	In summary, the chiral charges on the band that bounds the red-shaded region from below are $6 \nu_{b',\Gamma\text{-X}} = 6$, $8 \nu_{b',\Gamma\text{-R}} = 48$, as well as $\nu_{b',\Gamma} = -5$, $\nu_{b',\text{R}} = -4$ and $\nu_{b',\text{NP},b'} = +3$, which adds up to 48.
	Note, that despite the relevance of the sixfold crossing at R to the band $b'$, one needs to use the non-Abelian Chern number calculation to determine the charge contributed to the red-shaded gap, see Ref.~\cite{huber2022network} for the details of such a calculation.
	By using the Nielsen-Ninomiya theorem for band $b'$ we can infer that there are at least two set of Weyl points at generic positions. 
	Indeed, by a closer inspection of the band structure we find that there are additional Weyl points close to the  $\Gamma$-R axis axes.
	As mentioned before there are 24 Weyl points of charge  $\nu = -1$ in the vicinity R as well as another set of Weyl points with also  $\nu = -1$ close to $\Gamma$.
	
	While we had to consider the charge of the nodal plane explicitly, in absence of nodal planes it is possible to infer the existence of Weyl points at generic positions based on symmetries alone, e.g., in a spinful representation of SG~19 or the magnetic SG~19.27 for the movable fourfold double Weyl points as noticed for a tight-binding model in Ref.~\cite{chang2018topological, wilde2021symmetry}. 
	It is thus possible to use the local and global constraints together with the Nielsen-Ninomiya theorem to deduce the existence of Weyl points at generic positions within the Brillouin zone.

	\section{Generation and classification of low-energy Hamiltonians for the multifold crossing case\label{gen_clas_hamiltonians}}
	
	As we have already seen in a previous section \ref{SubSec_MultipleScrew}, combinations of different symmetries, including time-reversal, can lead to surprising results. Up until now we considered only Weyl points. So the next question is how the non-abelian constraints affect multifold crossings in this regard. Here we do not only want to restrict ourselves on just the evaluation of constraints, but to explicitly calculate Chern numbers in all topological phases of all multi-fold crossings, as the solution to constraints derived for the non-abelian case (see section \ref{Sec_ChernAndEigValNonAbelian}) are not unique and larger Chern numbers than the minimal ones fulfilling the given constraints can, due to the higher symmetry, be no longer excluded. We can see these cases directly when such a topological classification is carried out explicitly.
	
	This complete topological classification of all multifold crossings in all space groups follows a three phase approach. First all irreducible representations (irreps) with dimensions higher than 2 were found at all high symmetry points using the Bilbao Crystallographic Server\,\cite{aroyo2006bilbao}. Since we included time-reversal symmetry in all of our analysis, the search can be restricted to double space groups with broken inversion symmetry, since only there topological charges are allowed to be nonzero in presence of time-reversal symmetry.
	Then, low energy Hamiltonians were generated for all irreps found in the last step, such that these Hamiltonians respect all symmetries at the given high symmetry points. Finally, the whole parameter space of these Hamiltonian are topologically classified.
	
	We note that there is an alternative approach for generating low-energy Hamiltonians than the one shown in this section based on\,\cite{tang2021exhaustive}, where all possible Hamiltonian terms are tabulated. We used the method described here, since we found it more convenient to lookup a small number of symmetry generators and their representations instead of all possible Hamiltonian terms. See also\,\cite{luttinger1956quantum,greschphd,jiang2021k} for more alternative algorithms.
	
	\subsection{Automatic generation of low-energy Hamiltonians from irreps}
	
	A general low energy Hamiltonian up to second order in wave-vector $k$ has the following form
	\begin{align}
		H_{nm}(\vebm{k}) &= \sum_{h p_1} \alpha_{h,1} H^1_{h p_1 nm} k_{p_1} \nonumber
		\\
		& \quad + \sum_{h p_1 p_2} \alpha_{h,2} H^2_{h p_1 p_2 nm} k_{p_1} k_{p_2} + \mathcal{O}(k^3). \label{Ham}
	\end{align}
	$n$ and $m$ enumerate the orbital degrees of freedom. $\alpha_{h,q}$ are the free parameters of $H$ at order $q$. $H^q_{h p_1...p_q nm}$ are the linearly independent terms in $H$. The goal of the following algorithm is to compute these terms.
	
	The starting point of the automatic generation are symmetry generators and the corresponding irrep at a given high symmetry point. With these generators we build up the whole little group $G$ at his high symmetry point and the representation $D(g)$ of those symmetries $g\in G$. The only constraint of a low energy Hamiltonian at this point must be that it is symmetric 
	\begin{align}
		\forall g \in G : H(g \vebm{k}) &= D(g) H(\vebm{k}) D(g^{-1}).
	\end{align}
	We can symmetrize the Hamiltonian in Eq.~(\ref{Ham}) via
	\begin{align}
		\tilde {H}^1_{h v_1ij} &= \frac{1}{|G|} \sum_{g\in G} g^{-1}_{vp_1} D(g)_{in} H^1_{h p_1nm} D(g^{-1})_{mj} \label{s3:eqsymmetrized}\\
		\tilde {H}^2_{h v_1 v_2ij} &= \frac{1}{|G|} \sum_{g\in G} g^{-1}_{v_1p_1} g^{-1}_{v_2p_2} D(g)_{in} H^2_{h p_1 p_2nm} D(g^{-1})_{mj} \label{s3:eqsymmetrized2}\\
		...\nonumber
	\end{align}
	where $g_{ab}$ is the real space representation of $g$ and Einstein notation was used. Then $\tilde {H}^q_{h p_1...p_q nm}$ are the new terms of a symmetric $H$. Note that $D(g)$ can be anti-unitary, which is the case when $g$ is for example the time-reversal symmetry. In this case, $D(g)=U\mathcal{K}$ with $U$ unitary and $\mathcal{K}$ being the complex conjugation operator. The latter one can be eliminated by commuting it through all term in Eq.\,\ref{s3:eqsymmetrized} and\,\ref{s3:eqsymmetrized2} until we can use $\mathcal{K}^2=1$. 
	
	The algorithm starts by generating a set of random complex $H^q_{h p_1...p_q nm}$ terms, with $h \in \{1, ..., N\}$ and $N$ being the total amount of randomly generated terms. These are then symmetrized via \ref{s3:eqsymmetrized} to produce $N$ symmetrized $\tilde {H}^q_{h p_1...p_q nm}$ terms. Only the linearly independent terms are kept, which is done using a Gram-Schmidt orthogonalization, during which the terms are treated as vectors by flattening them to a single index. This also reduces the number of terms $\tilde{N}\leq N$ to the maximal set of symmetric and linearly independent terms. The number of free parameters of this Hamiltonian at order $q$ is also $\tilde{N}$. 
	
	For better handling of these terms, we would like to normalize the real or imaginary part of as many of their entries to 1, since they are still filled with random numerical values of arbitrary magnitude. We can not normalize all entries to 1, since not all are linearly independent. This normalization is done by first gathering all nonzero columns in $\text{Re}(\tilde {H}^q_{h p_1...p_q nm})$ and $\text{Im}(\tilde {H}^q_{h p_1...p_q nm})$ in a new matrix $M_{h r}$ with size $(\tilde{N}, P)$, with $P$ the number of nonzero columns. Rows that are linearly dependent on other rows are removed in $M$, such that $M$ is quadratic and invertible. The final terms of $H$ are then computed with
	\begin{align}
		H'^q_{h' p_1...p_q nm} = \sum_{h} M^{-1}_{h' h} \tilde{H}^q_{h p_1...p_q nm}.
	\end{align}
	Due to the inversion of $M$, the real or imaginary part of all nonzero entries in $\tilde{H}$ which are chosen to build up the $M$ matrix are normalized to 1 in only one of the terms while they are set to 0 in all other. Entries that are not part of the final $M$ are either a fraction or a fraction consisting of square roots. The last step of the algorithm is to convert the numerical values of $H'$ into analytical expressions by comparing the entries to the values of those analytical expressions and also to project $H'$ to Pauli or Gellmann matrices. To test if this conversion worked, the symmetry of the resulting Hamiltonian is checked.
	
	\subsection{Classification of all multifold crossings at high-symmetry points\label{subsq_multifoldcrossings}}
	
	Using the algorithm described in the previous section, a Hamiltonian for each entry in the compiled list of all irreps with dimension $>2$ was generated. Since we only want to study the topological charge of the crossing at the high symmetry point in question, it is sufficient, with only one exception as we will see later, to generate only the terms up to linear order in $k$, since higher orders could only produce additional crossings away from the high symmetry point and do not alter the topological charge of the multifold crossing. Some of the generated Hamiltonian are equivalent or equivalent up to a transformation, so these cases can be grouped and classified together. The transformations either have no effect on or flip the topological charge.
	
	The determination of every bands topological phase diagrams of the Hamiltonians all follow the same idea of first finding all points in parameter space where the current band in question of the given Hamiltonian becomes gapless. These are the only points where topological phase transitions can happen, i.e. the topological charge of the multifold crossing can change. These points make up subspaces in parameter space which separate different topological phases, which were found by considering the characteristic polynomial of $H$ and comparing it to a characteristic polynomial describing a Hamiltonian in a gapless phases. 
	
	So after finding these subspaces it is possible to determine the topological charges of every phase by evaluating it numerically deep in a given phase. This way one can color in the whole phase by the determined topological charge. Since no other topological phases are possible, we can enumerate all possible topological charges for all multifold crossings.
	
	During this topological classification, the Chern number of single bands is sometimes undefined. This happens due to band degeneracies, for example nodal planes, which by symmetry persist to all orders in $\vebm{k}$. In most of these cases, one can still define a non-abelian Chern number, see Eq.~(\ref{Eq_nonabelianOmega}). In the case of 4-fold crossings on nodal planes, we compute non-abelian Chern numbers $\nu_{b,b+1}$, where bands $b$ and $b+1$ are part of the nodal plane.

	\subsubsection{4-fold crossings \label{4foldcrossingssection}}
	The main results for all 4-fold crossings are summarized in tables \ref{s3table1} and \ref{s3table2}. The topological charge of the lowest band $\nu_1$ is undefined in most irreps, since there the lowest two bands can be shown to be always twofold degenerate at some k-points away from $0$ at all orders in $\vebm{k}$ due to symmetry constraints. Where this is not the case, an unusually high Chern number of $\pm5$ can be observed.
	
\begin{figure}
		\centering
		\includegraphics[width = 0.49\textwidth]{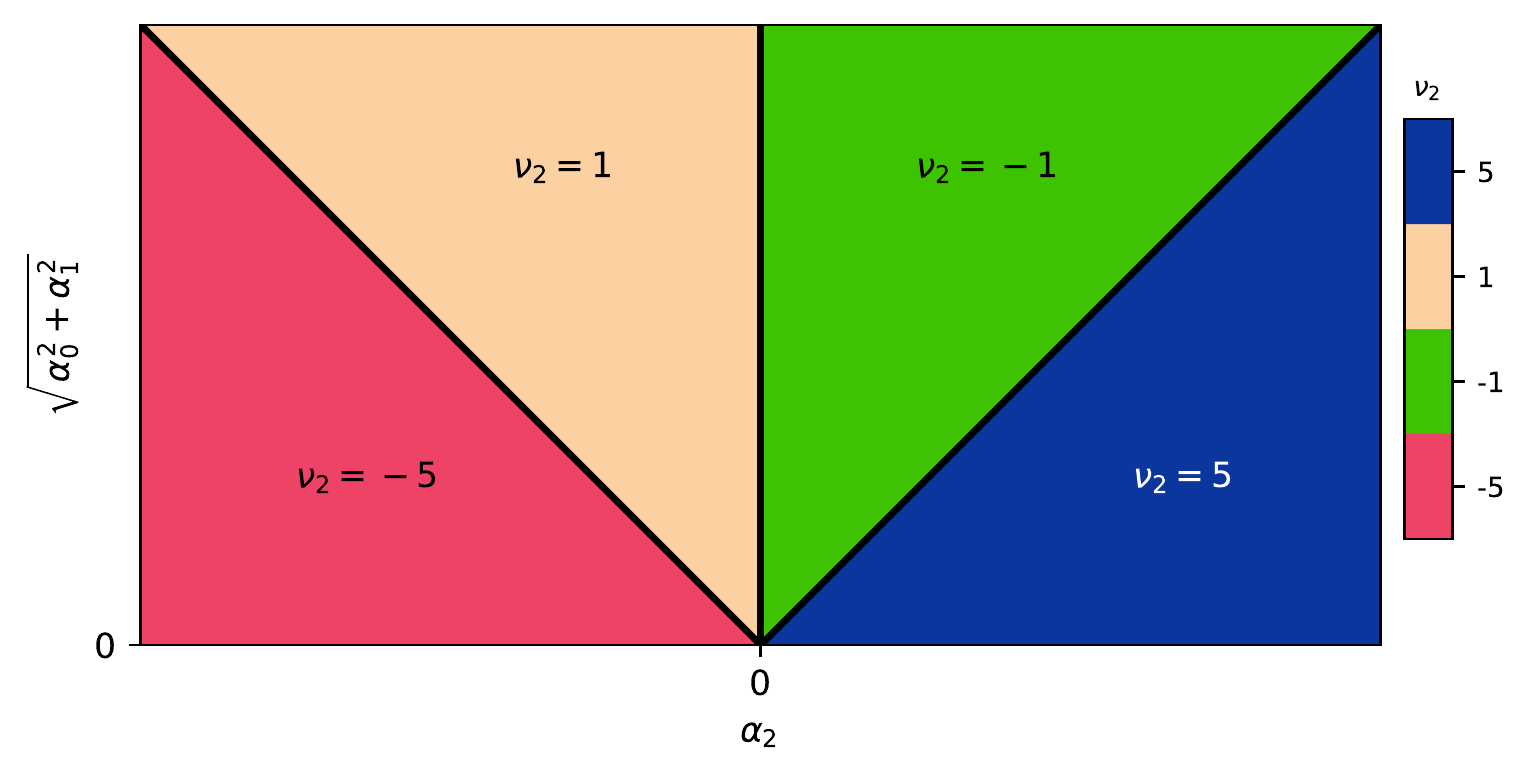}
		\caption{Topological phase diagram of the 4-fold crossing described by Hamiltonian \ref{model1} with three parameters $\alpha_0, \alpha_1$ and $\alpha_2$. Depicted is the Chern number of band 2 $\nu_2$. An unusually high absolute Chern number of $5$ can be found in regions where $\alpha_2>\sqrt{\alpha_0^2+\alpha_1^2}$ or $\alpha_2<-\sqrt{\alpha_0^2+\alpha_1^2}$.}
		\label{fig:s3phasediagramC5}
	\end{figure}

	As this result is quite unexpected, we explicitly show the topological phase diagram and its derivation of one of the two Hamiltonians, the model for the $\bar{\Gamma}_6\bar{\Gamma}_7$ irrep, that describe these cases. This irrep can be found in SG 195-199. The little group contains $C_2^{x,y,z}$, $C_3^{(\pm1,\pm1,\pm1)}$ and time-reversal symmetry. In the following, all used representations are equivalent to the ones on the Bilbao Crystallographic Server\,\cite{aroyo2006bilbao}. The Hamiltonian generated by the algorithm described in the previous section is
	\begin{align}
		H= &\alpha_0\biggl[
		2k_x\sigma_x\tau_z + 
		k_y\left(-\sqrt{3}\sigma_x\tau_0 - \sigma_y\tau_0\right) + \nonumber\\&
		k_z\left(\sigma_x\tau_x + \sqrt{3}\sigma_y\tau_x\right)
		\biggr] +\nonumber\\&
		\alpha_1\biggl[
		-2k_x\sigma_y\tau_z + 
		k_y\left(-\sigma_x\tau_0 + \sqrt{3}\sigma_y\tau_0\right) + \nonumber\\&
		k_z\left(\sqrt{3}\sigma_x\tau_x - \sigma_y\tau_x\right)
		\biggr]+\nonumber\\&
		2\alpha_2\biggl[
		k_x\sigma_z\tau_x + 
		k_y\sigma_0\tau_y + 
		k_z\sigma_z\tau_z
		\biggr] \label{model1}
	\end{align}
	with $\tau$ and $\sigma$ being Pauli matrices. It is possible to show (see Appendix~\ref{appendixSG198Gamma6Gamma7}) that the Hamiltonian is only gapless for $\alpha_2=\pm\sqrt{\alpha_0^2+\alpha_1^2}$, $\alpha_2=0$ or $\sqrt{\alpha_0^2+\alpha_1^2}=0$ at points away from $k=(0,0,0)$. We can assign the spaces in between gapless planes in parameter space $\alpha_n$ with precomputed Chern numbers to arrive at the topological phase diagram of the $\bar{\Gamma}_6\bar{\Gamma}_7$ irrep model. See figure \ref{fig:s3phasediagramC5} for the phase diagram for band 2. We find that band 1 has two phases, for $\alpha_2<0$ the Chern number is $\nu_1=-3$, for $\alpha_2>0$ it is $\nu_1=3$. For bands 3 and 4 use $\nu_3=-\nu_2$ and $\nu_4=-\nu_1$.
	
	In the appendix of\,\cite{Zhang2020quadWeyl}, a symmetry equivalent Hamiltonian to Eq.~(\ref{model1}) has been derived, although there the whole topological phase diagram has not been mapped out. Previously\,\cite{tang2017multiple}, this 4-fold crossing has been described by a Rarita-Schwinger-Weyl spin-3/2 Hamiltonian\,\cite{rarita1941theory,PhysRevB.94.195205,PhysRevB.93.045113} $H(\vebm{k})=\alpha\vebm{S}\cdot\vebm{k}$, which only supports the $\nu_n=\{\pm 3, \pm1, \mp1, \mp3\}$ phase, or a distinction was made\,\cite{chang2017unconventional} but only this phase was considered. Here we see that this description is incomplete. To our knowledge, the $\nu_n=\{\pm 3, \mp5, \pm5, \mp3\}$ topological phase has not been observed yet.
	
	We find no 3-fold symmetry eigenvalue phase jumps for the lowest/highest band. A phase jump of $2\pi/3$ for $\alpha_2<0$ and of $4\pi/3$ for $\alpha_2>0$ was observed for band 2. For all bands, a phase jump of $\pi$ was found for both 2-fold symmetries, which constraints all Chern numbers to be odd $\nu_n=1\,\text{mod }2$. Further, the 3-fold symmetry constraints the Chern number of the lowest and highest band to $\nu_{1/4}=0\,\text{mod }3$, which is consistent with $\nu_{1/4}=\pm3$. Then, for $\alpha_2<0$ we have the constraint $\nu_{2/3}=1\,\text{mod }3$ and for $\alpha_2>0$ we have $\nu_{2/3}=2\,\text{mod }3$, which are fulfilled in all phases in figure \ref{fig:s3phasediagramC5}.
	
	The transition between $\nu_2=\pm1$ and $\nu_2=\mp5$ phases is facilitated (see Appendix~\ref{appendixSG198Gamma6Gamma7}) by a gap closing of the middle two bands. This suggests, that the mechanism behind this Chern number switch is an absorption/emission of 6 Weyl points on the $C_2$ invariant axes into/out of the multifold crossing. This implies an exchange of $C_2$ symmetry eigenvalues of bands $2$ and $3$ between these two phases, which we confirmed by a direct calculation.
	
	The phase transition $\nu_n$ to $-\nu_n$ at $\alpha_2=0$ takes place by a simultaneous gap closing on $C_2$ invariant lines of the outer band pairs $(1,2)$ and $(3,4)$ as well as on $C_3$ invariant lines of band pair $(2,3)$. Since corresponding symmetry eigenvalues switch on both invariant lines, 6 WPs on the outer band pairs and 8 WPs on the middle bandpair fuse with or emerge from the 4-fold crossing. The WPs on $C_2$ invariant lines with total charge $\pm6$ switch the sign of the lowest/highest Chern number $\nu_{1/4}=\mp3\to\pm3$. For the middle two bands a combined total Chern number of $\mp6\pm8=\pm2$ switches the sign of the middle Chern numbers $\nu_{2/3}=\mp1\to\pm1$. 
	
	\begin{figure}
		\centering
		\includegraphics[width = 0.49\textwidth]{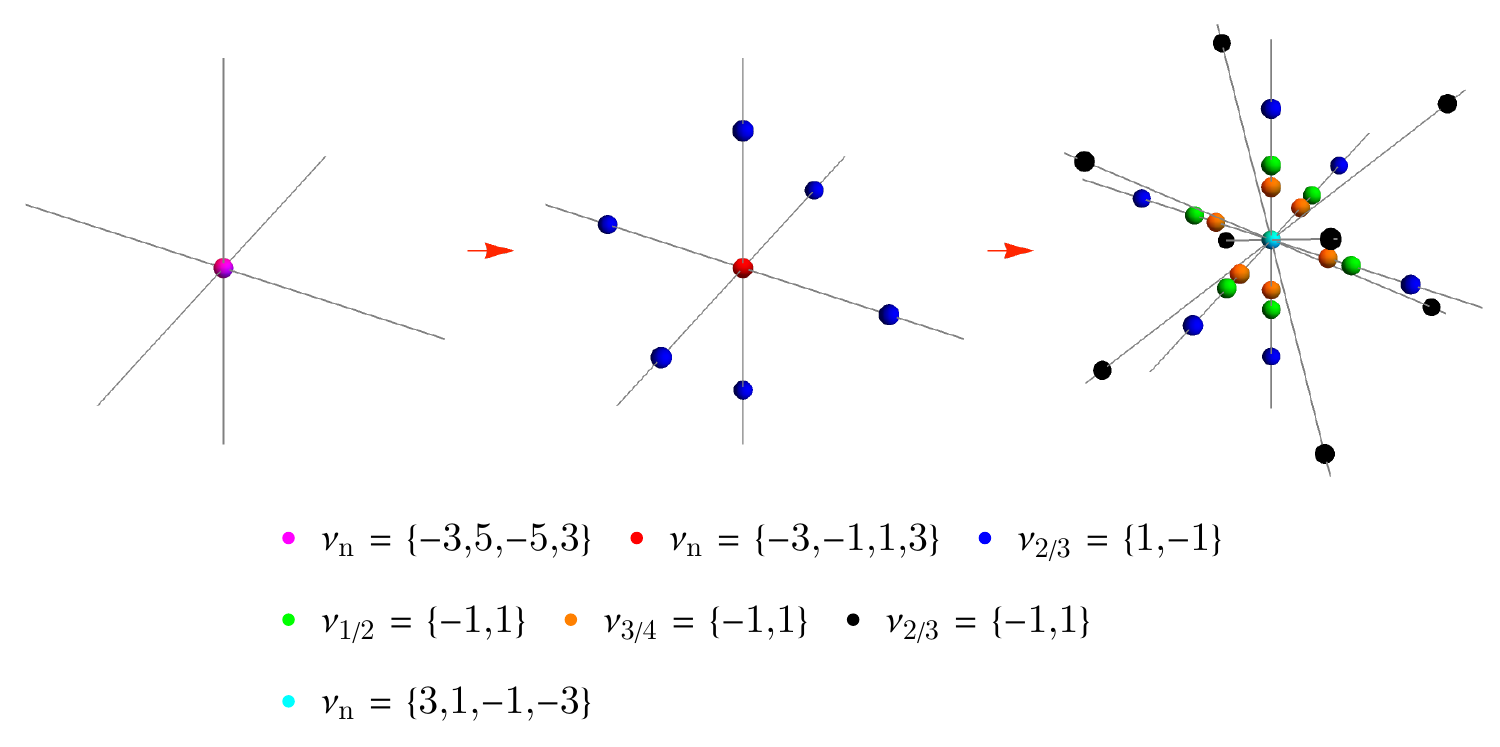}
		\caption{Starting from a singe 4-fold crossing described by the $\bar{\Gamma}_6\bar{\Gamma}_7$ irrep model (see Eq.\,\ref{model1}) in the $\nu_2=5$ phase, a phase transition to $\nu_2=1$ over $\nu_2=-1$ produces a total of $26$ WPs across the 3 bandpairs. WP and multifold point charges are color-coded.}
		\label{fig:s3phasetransitionWPproduction}
	\end{figure}
	
	This means, consecutive topological phase transitions over $\nu_2=5\rightarrow-1\rightarrow1$ produce a total of $26$ WPs distributed across the 3 bandpairs,  $6$ in the lower and upper bandpair respectively and $6+8=14$ in the middle bandpair, provided there are no other crossings at the start in the $\nu_2=5$ phase, since these could also be merged into the multi-fold point to carry out the phase transition. This process is visualised in figure\,\ref{fig:s3phasetransitionWPproduction}.
	
	\begin{figure}
		{\centering Band 1}
		\begin{tabular}{cc}
			$\nu_2=-1$ & $\nu_2=5$ \\
			\begin{minipage}{.25\textwidth}\includegraphics[width = \linewidth]{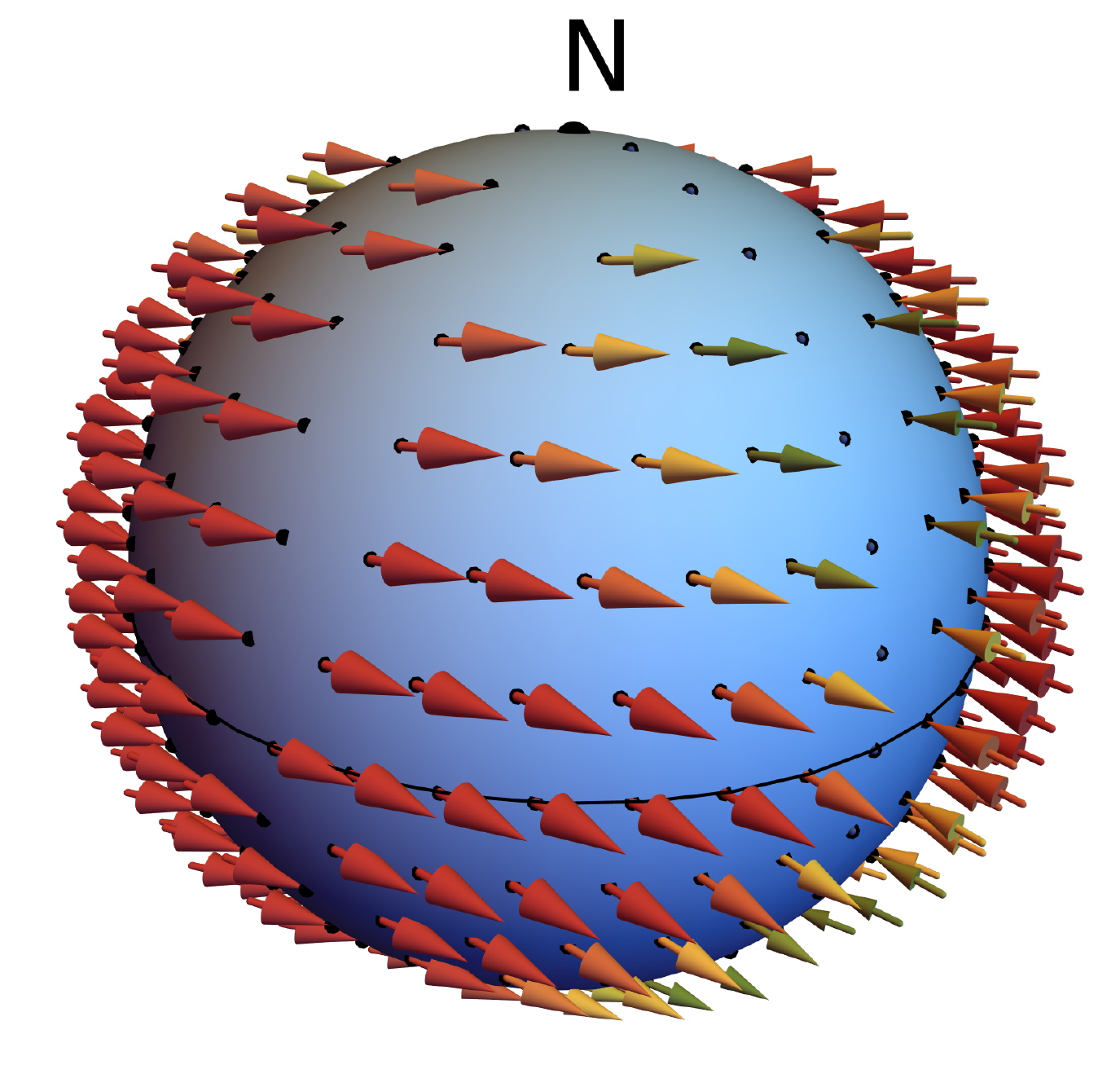}\end{minipage} 
			& \begin{minipage}{.25\textwidth}\includegraphics[width = \linewidth]{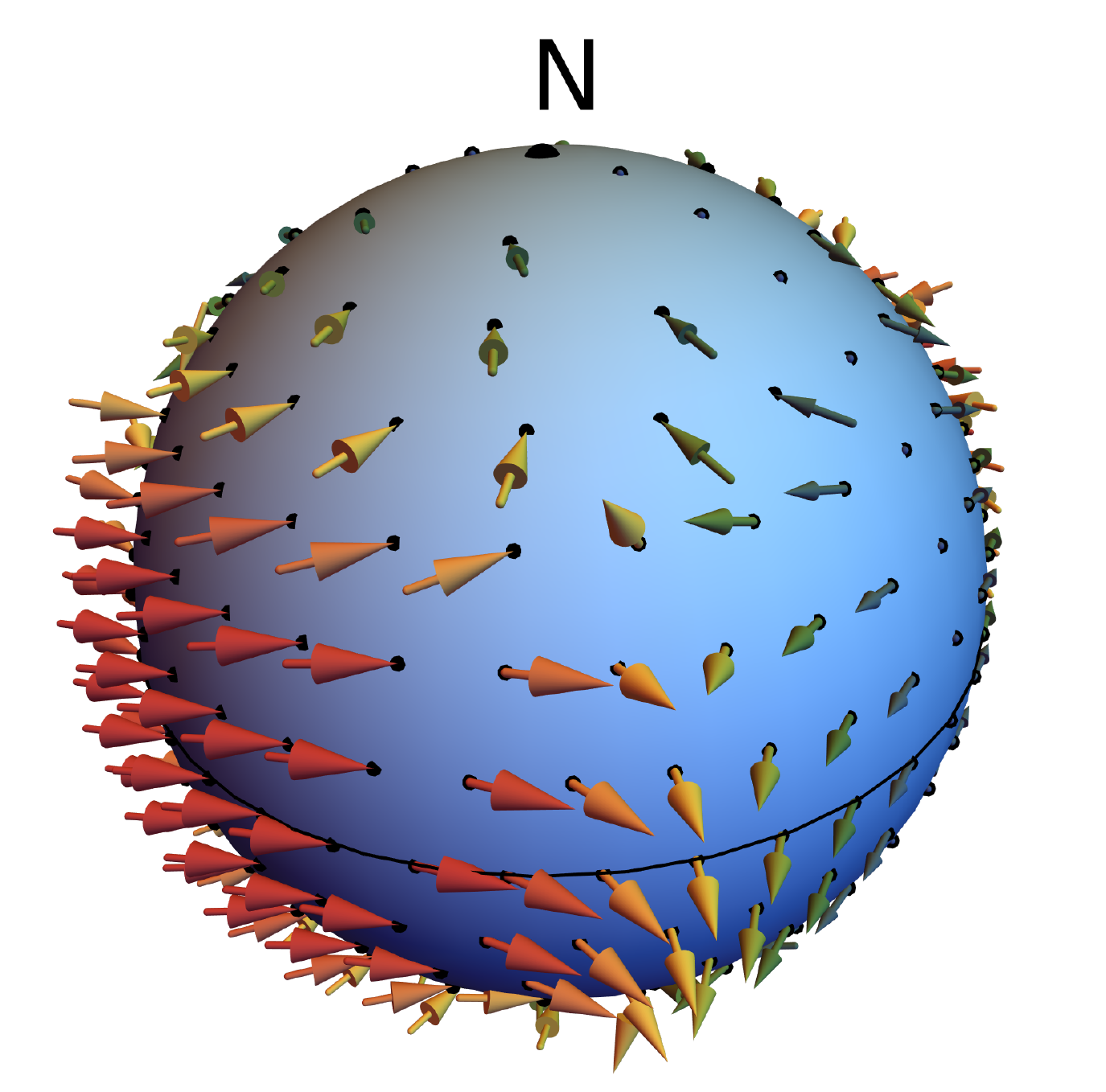}\end{minipage}
		\end{tabular}
		{\centering Band 2}
		\begin{tabular}{cc}
			\begin{minipage}{.25\textwidth}\includegraphics[width = \linewidth]{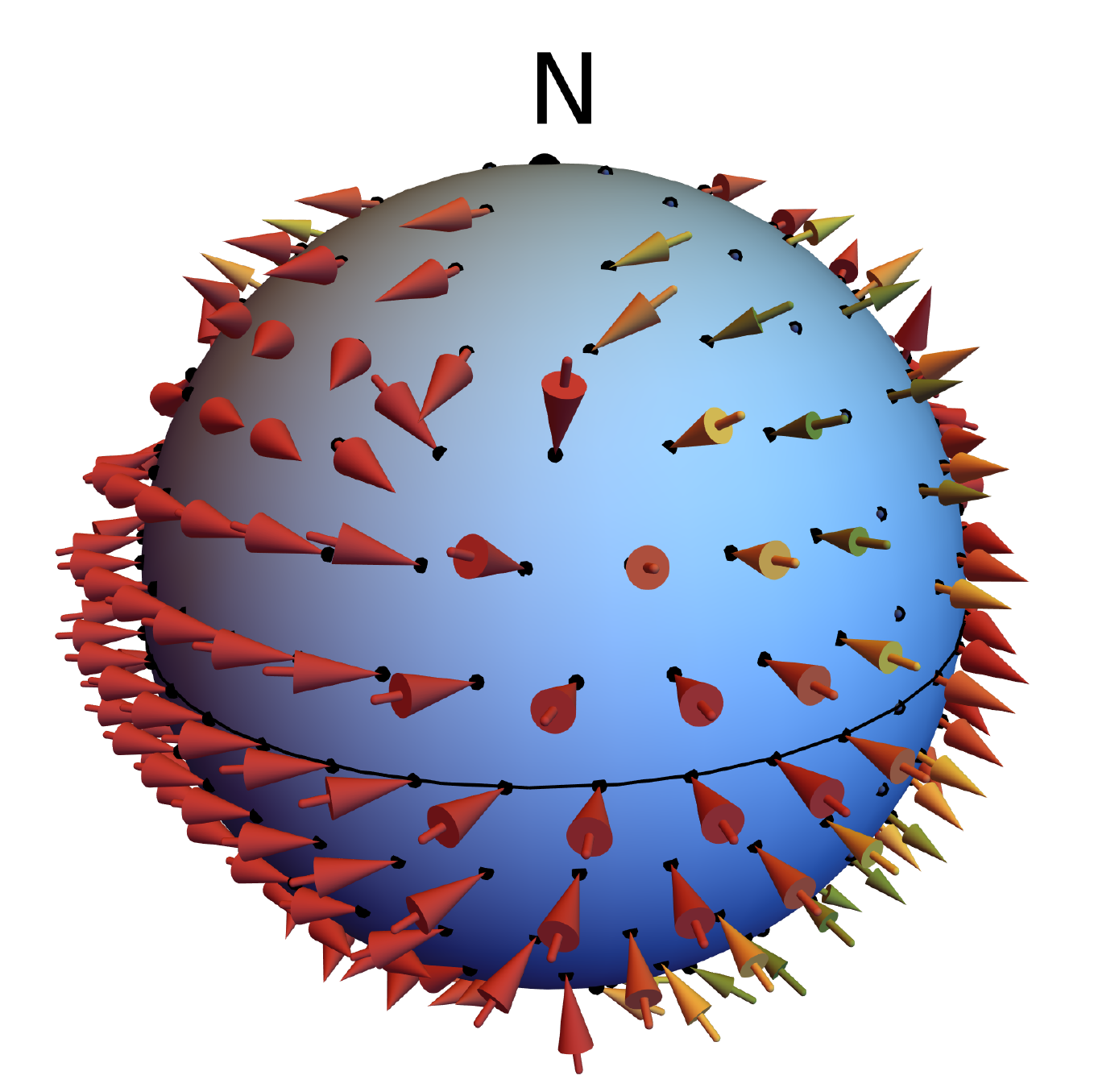}\end{minipage} 
			& \begin{minipage}{.25\textwidth}\includegraphics[width = \linewidth]{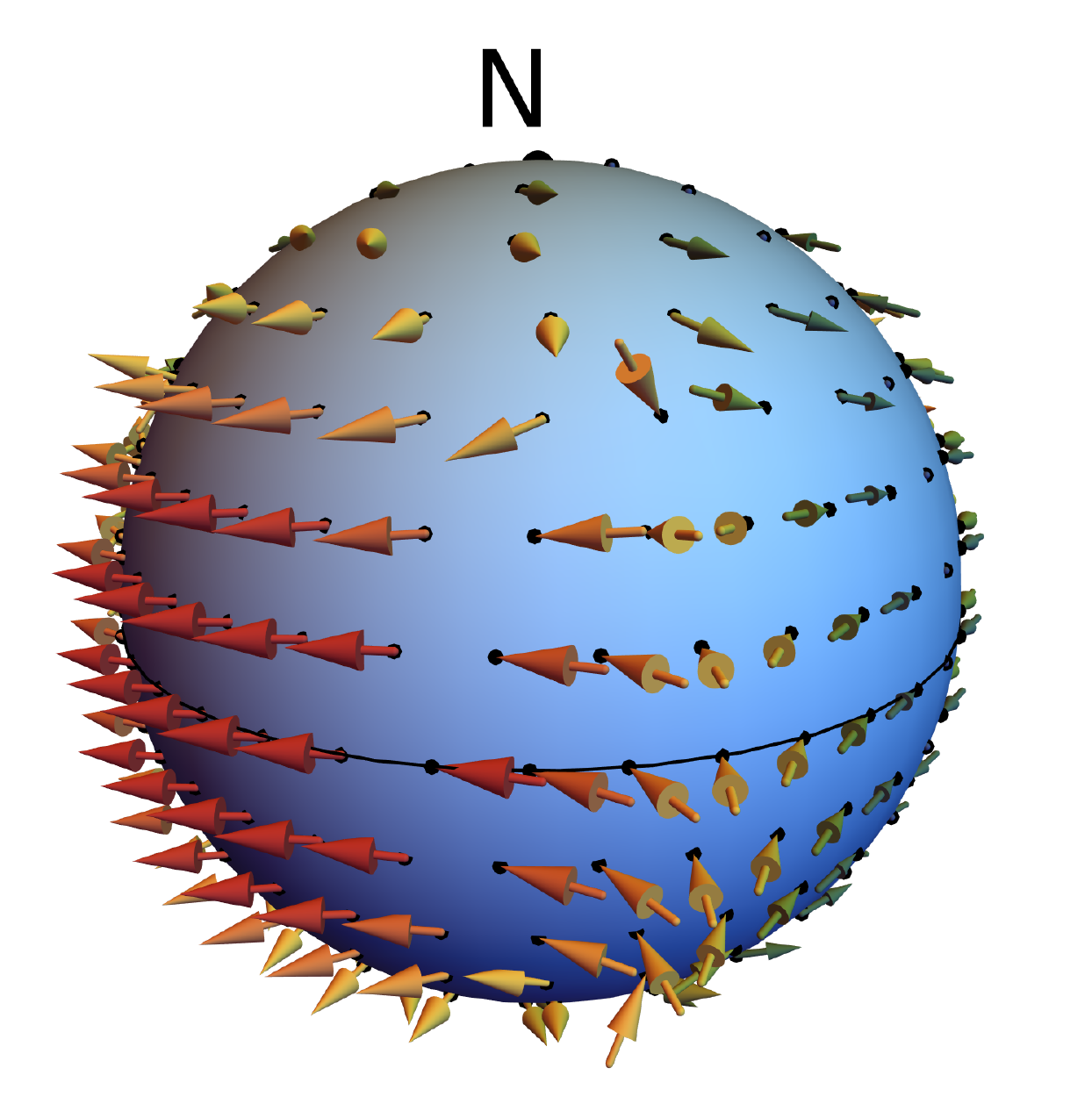}\end{minipage}
		\end{tabular}
		\caption{
			Spin texture from the expectation value of $\sigma_a\tau_0$ for the 4-fold crossing described by Eq.\,(\ref{model1}) for bands 1 and 2 in the $\nu_2=-1$ and $\nu_2=5$ phase. The arrow size and color depend on the magnitude of the spin, red being maximal.
			\label{berrycurvature4foldtable}}
	\end{figure}
	
	In figure \ref{berrycurvature4foldtable} the spin texture of this 4-fold crossing in the $\nu_2=-1$ and $\nu_2=5$ phase are compared for bands 1 and 2. For this, the spin expectation value $\langle \vebm{\sigma}\tau_0 \rangle$ is used, since the time-reversal irrep $-i\sigma_y\tau_0$ suggests that $\vebm{\sigma}$ is the spin degree of freedom of this crossing. The parameters chosen for the $\nu_2=-1$ are $\alpha_0=0.1$, $\alpha_1=0$ and $\alpha_1=1$, while for the $\nu_2=5$ phase the parameters are $\alpha_0=1$, $\alpha_1=0$ and $\alpha_1=0.1$. We note that the spin texture is not symmetric under the little group at this crossing, since SOC mixes the spin and orbital degrees of freedom in the irrep $\bar{\Gamma}_6\bar{\Gamma}_7$. The spin texture differences especially in the first band might be measurable in a spin-resolved ARPES experiment\,\cite{gatti2020radial,hagiwara2022link}. In this way, these two topological phases can be distinguished.
	
	\subsubsection{3 and 6-fold crossings} 
	\label{SubSubSec_ThreeAndSixfold}
	
	The Hamiltonians for 3 and 6-fold crossings found by the procedure described above reproduce the ones listed in\,\cite{bradlyn2016beyond}. We also find the Hamiltonians for all 3-fold crossings
	\begin{align}
		H=\begin{pmatrix}
			0 & k_z(\alpha-i\beta) & -k_y(\alpha+i\beta)\\
			& 0 & k_x(\alpha-i\beta)\\
			\ldots & & 0\\
		\end{pmatrix} \label{s3eq7}
	\end{align}
	to be equivalent up to transformations. Since these transformations and the explicit dependence of the topological charge were left out of\,\cite{bradlyn2016beyond}, we included them here in the tables \ref{s3table3} and \ref{s3table4} and in the Appendix~\ref{appendix3foldmodels}.
	
	The 6-fold crossing Hamiltonian of SG 198 $\bar{R}_7\bar{R}_7$ irrep is equal to
	\begin{align}
		H=&\alpha_0\biggl[
		k_x\sigma_z\lambda_7 -
		k_y\sigma_z\lambda_6 +
		k_z\sigma_z\lambda_3
		\biggr] +\nonumber\\&
		\alpha_1\biggl[
		k_x\sigma_0\lambda_5 + 
		k_y\sigma_0\lambda_2 + 
		k_z\sigma_0\lambda_1
		\biggr] +\nonumber\\&
		\alpha_2\biggl[
		k_x\sigma_x\lambda_7 -
		k_y\sigma_x\lambda_6 +
		k_z\sigma_x\lambda_3
		\biggr]  +\nonumber\\&
		\alpha_3\biggl[
		k_x\sigma_y\lambda_7 -
		k_y\sigma_y\lambda_6 +
		k_z\sigma_y\lambda_3
		\biggr]\label{6foldmodel}
	\end{align}
	with $\sigma_n$ and $\lambda_n$ being Pauli and Gellmann matrices (see the Appendix~\ref{appendix:gellmanndefinition} for a definition). This Hamiltonian is equivalent up to a unitary transformation to the one found in \cite{bradlyn2016beyond}, with $e^{i\phi}=\alpha_0+i\alpha_1$ and $b=\alpha_2+i\alpha_3$. There it was also shown that you can arrive at the Hamiltonian for the SG 212 and 213 $\bar{R}_7\bar{R}_8$ irrep by setting $\phi=\frac{\pi}{2}$. Due to nodal planes crossing these points, Chern numbers for odd fillings can not be defined. The non-abelian Chern number for the middle two bands $\nu_{34}=0$ remains trivial, while the Chern number for the remaining bands are $\nu_{12}=-\nu_{56}=\pm4$. The exact phase diagram and its derivation can be found in the Appendix~\ref{appendix6foldmodels}.
	
	\begin{table*}
		\begin{tabular}{ |l|l|l|l|l|l|l|}
			\hline SG & Irrep & $\nu_1$& $\nu_2$ & $\nu_{12}$& Model & Transformation  \\\hline
			19 & $R_1R_1$ & - & - & $\pm2$ & \ref{SG19R1R1NOSOC}& -\\\hline
			92 & $A_1A_2$ & - &  - & $\pm2$ & \ref{SG90A6A7SOC}&$k_x\to k_y$,$k_y\to -k_x$,$\alpha_2=0$,$\alpha_3=0$\\\hline
			96 & $A_1A_2$ & - &  - & $\pm2$ & \ref{SG90A6A7SOC}&$k_y\to -k_y$,$\alpha_2=0$,$\alpha_3=0$\\\hline
			198 & $R_1R_3$/$R_2R_2$ & - &  - & $\pm2$ & \ref{SG198R1R3NOSOC} & -\\\hline
			212/213 & $R_1R_2$ & - &  - & $\pm2$ & \ref{SG198R1R3NOSOC} & -\\\hline
			212/213 & $R_3$ & - &  - & $\pm2$ & \ref{SG212R3NOSOC} & -\\\hline
		\end{tabular}
		\caption{All possible topological charges of all 4-fold crossings in the spinless case (without SOC). Their Hamiltonians are either listed under the Model column or are obtained after a possible transformation applied on the given Hamiltonian. \label{s3table1}}
	\end{table*}
	
	\begin{table*}
		\begin{tabular}{ |l|l|l|l|l|l|l|}
			\hline SG & Irrep & $\nu_1$& $\nu_2$ & $\nu_{12}$& Model & Transformation  \\\hline
			18 & $\bar{S}_5\bar{S}_5$/$\bar{R}_5\bar{R}_5$ & - & - & $\pm2$ &\ref{SG198MSOC} & - \\\hline
			19 & $\bar{S}_5\bar{S}_5$ & - & - & $\pm2$ &\ref{SG198MSOC} & - \\\hline
			19 & $\bar{T}_5\bar{T}_5$ & - & - & $\pm2$ &\ref{SG198MSOC} & $k_x\to k_y, k_y\to k_z, k_z\to k_x$ \\\hline
			19 & $\bar{U}_5\bar{U}_5$ & - & - & $\pm2$ &\ref{SG198MSOC} & $k_y\to k_z, k_z\to k_y$ \\\hline
			90 & $\bar{A}_6\bar{A}_7$/$\bar{M}_6\bar{M}_7$ & - & - & $\pm2$ & \ref{SG90A6A7SOC} & - \\\hline
			92/94/96 & $\bar{M}_6\bar{M}_7$ & - & - & $\pm2$ & \ref{SG90A6A7SOC} & - \\\hline
			92/96 & $\bar{R}_5\bar{R}_5$ & - & - & $\pm2$ & \ref{SG198MSOC} & $k_x\to k_y, k_y\to k_z, k_z\to k_x$ \\\hline
			92/96 & $\bar{A}_7\bar{A}_7$ & - & - & $\pm4$ & \ref{SG92A7A7SOC} & - \\\hline
			94 & $\bar{A}_6\bar{A}_7$ & - & - & $\pm2$ & \ref{SG90A6A7SOC} & $k_x\leftrightarrow k_y$ \\\hline
			195/196/197/198/199 & $\bar{\Gamma}_6\bar{\Gamma}_7$ & $\pm3$ & $\pm1$,$\mp5$ & $\pm4$,$\pm2$ & \ref{4foldcrossingssection} & - \\\hline
			195 & $\bar{R}_6\bar{R}_7$ & $\pm3$ & $\pm1$,$\mp5$ & $\pm4$,$\pm2$ & \ref{4foldcrossingssection} & - \\\hline
			197 & $\bar{H}_6\bar{H}_7$ & $\pm3$ & $\pm1$,$\mp5$ & $\pm4$,$\pm2$ & \ref{4foldcrossingssection} & - \\\hline
			198 & $\bar{M}_5\bar{M}_5$ & - & - & $\pm2$ & \ref{SG198MSOC} & - \\\hline
			199 & $\bar{H}_6\bar{H}_7$ & $\pm3$ & $\pm1$,$\mp5$ & $\pm4$,$\pm2$ & \ref{4foldcrossingssection} & $U=\sigma_z\tau_x$ \\\hline
			207/208/209/210 & $\bar{\Gamma}_8$ & $\pm3$ & $\pm1$,$\mp5$ & $\pm4$,$\pm2$ & \ref{SG207GammaSOCSEC} & - \\
			/211/212/213/214& & &&&& \\\hline
			207/208 & $\bar{R}_8$ & $\pm3$ & $\pm1$,$\mp5$ & $\pm4$,$\pm2$ & \ref{SG207GammaSOCSEC} & - \\\hline
			211 & $\bar{H}_8$ & $\pm3$ & $\pm1$,$\mp5$ & $\pm4$,$\pm2$ & \ref{SG207GammaSOCSEC} & - \\\hline
			212 & $\bar{M}_6\bar{M}_7$ & - & - & $\pm2$ & \ref{212M6M7} & - \\\hline
			213 & $\bar{M}_6\bar{M}_7$ & - & - & $\pm2$ & \ref{212M6M7} & $k_y\to -k_y$ \\\hline
			214 & $\bar{H}_8$ & $\pm3$ & $\pm1$,$\mp5$ & $\pm4$,$\pm2$ & \ref{SG207GammaSOCSEC} & $U=\sigma_0\tau_x$,$\alpha_1\to-\alpha_1$ \\\hline
		\end{tabular}
		\caption{All possible topological charges of 4-fold crossings in all spinful SGs and their Hamiltonians. \label{s3table2}}
	\end{table*}
	
	\begin{table}
		\begin{tabular}{| c | c | c |}
			\hline SG & Irrep & Transformation\\\hline
			195...199 & $\Gamma_4$ & $\beta\to-\beta$,$k_x\to k_z$,$k_y\to -k_x$,$k_z\to k_y$\\\hline
			195 & $R_4$ & $\beta\to-\beta$,$k_x\to k_z$,$k_y\to -k_x$,$k_z\to k_y$\\\hline
			197 & $H_4$, $P_4$ & $\beta\to-\beta$,$k_x\to k_z$,$k_y\to -k_x$,$k_z\to k_y$\\\hline
			199 & $H_4$ & $\beta\to-\beta$,$k_y\to -k_y$\\\hline
			207...214 & $\Gamma_4$, $\Gamma_5$ & $\alpha=0$,$\beta\to-\beta$,$k_x\to k_z$,$k_y\to -k_x$,$k_z\to k_y$\\\hline
			207,208 & $R_4$, $R_5$ & $\alpha=0$,$\beta\to-\beta$,$k_x\to k_z$,$k_y\to -k_x$,$k_z\to k_y$\\\hline
			211,214 & $H_4$, $H_5$ & $\alpha=0$,$\beta\to-\beta$,$k_x\to k_z$,$k_y\to -k_x$,$k_z\to k_y$\\\hline
		\end{tabular}
		\caption{All 3-fold crossings without SOC and corresponding transformations, which generate their Hamiltonian from Eq.~(\ref{s3eq7}). \label{s3table3}}
	\end{table}
	\begin{table}
		\begin{tabular}{| c | c | c |}
			\hline SG & Irrep & Transformation\\\hline
			199,214 & $\bar{P}_7$ & -\\\hline
		\end{tabular}
		\caption{All spinful 3-fold crossings. The Hamiltonian is described in Eq.~(\ref{s3eq7}). \label{s3table4}}
	\end{table}

	\section{Materials}
	\label{sec_six}
	
	Here we discuss two material examples. Details on the calculations can be found in the Appendix \ref{App_details_calculation}.
	
	\subsection{BaAsPt and related compounds (SG 198)}\label{Sec_BaAsPt}

	\begin{figure}[t]
		\centering
		\includegraphics[width = 0.5\textwidth]{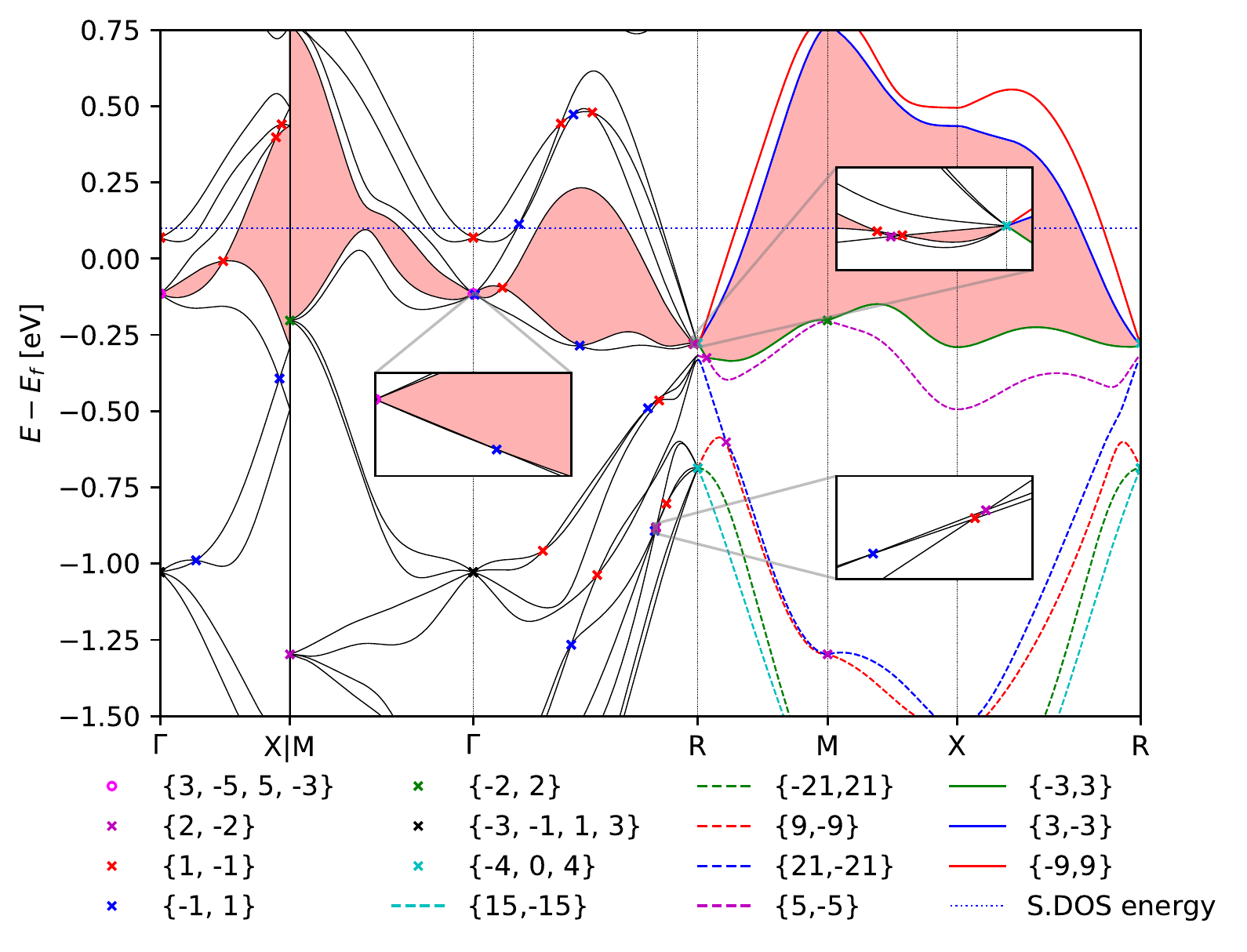}
		\caption{Bandstructure of BaAsPt and all band crossings on high symmetry lines. The bandgap between the bands with a Chern number of $\pm5$ at the 4-fold $\Gamma$ crossings is indicated by the pink region. This topological bandgap is responsible for the Fermi arcs in figure \ref{fig:s4BaAsPt_surface_band7_0_1eV}. The blue dashed line is the energy slice taken there. We also observe NPs with Chern numbers up to $21$ at energies below $E_f$.}
		\label{fig:BaAsPt_bandstruct_wannier}
	\end{figure}
	\begin{figure}[t]
		\centering
		\includegraphics[width = 0.4\textwidth]{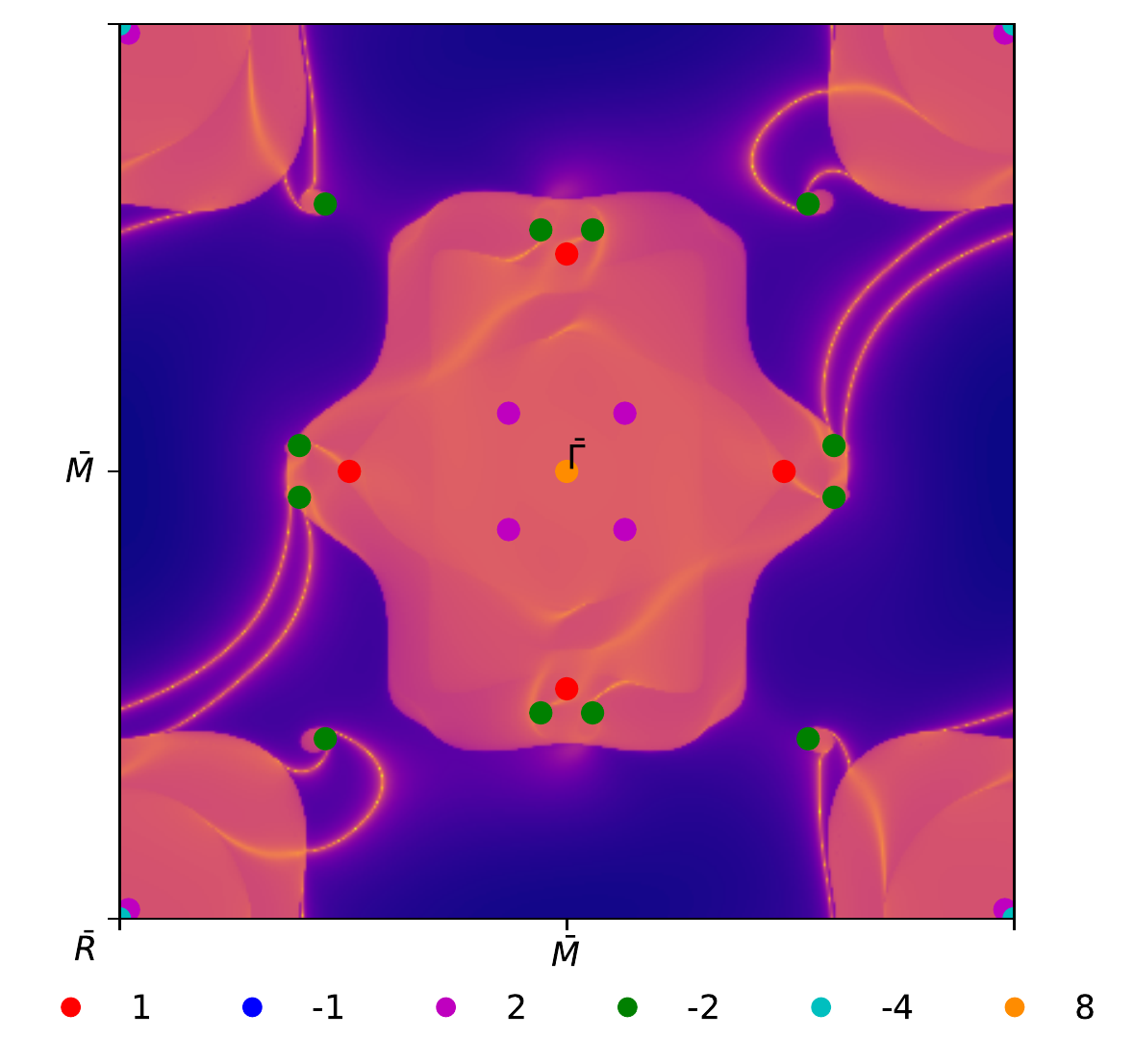}
		\caption{Surface density of states (DOS) of BaAsPt at $0.1\,$eV. Also shown are the projected topological charges, resulting from a bandfilling in between the bandpair with charge $\nu=\pm5$ at $\Gamma$ from the bulk. This is the bandfilling fulfilled in the parts of the surface DOS without bulk states and where the Fermi arcs reside. With this filling, the charge of the $\Gamma$ point is $3-5=-2$, which is compensated by two WPs on the $\Gamma X$-line. Further, there are 8 WPs on the $\Gamma R$-line very close to the $\Gamma$ point which we included into the total charge of $8$ of the projected $\bar{\Gamma}$ point. The total charge of bulk states around $\bar{\Gamma}$ is $4$, so there are 4 Fermi arcs connecting these states to the $\bar{R}$ point with topological charge $\nu=-4$. Note that the $\nu=2$ charges near $\bar{R}$ are slightly away from $\bar{R}$, so there are 4 copies at every corner. These charges are connected with 8 Fermi arcs to bulk states with $\nu=-2$ points between $\bar{\Gamma}$ and $\bar{R}$.}
		\label{fig:s4BaAsPt_surface_band7_0_1eV}
	\end{figure}
	
	A material search for a 4-fold crossing with a Chern number of $5$ sufficiently close to the Fermi energy was done in space groups 195-199 and 207-214. First materials from the materials project\,\cite{Jain2013} are screened for 4-fold crossings near the Fermi energy. The Chern number of this point was directly computed\,\cite{huber2022network} using density-functional theory, in particular Quantum Espresso\,\cite{giannozzi2009quantum}. This search was stopped at the first material found, which was BaAsPt in SG 198. There a 4-fold point with $\nu=\pm5$ was found at $\Gamma$ at $-100\,$meV, see figure\,\ref{fig:BaAsPt_bandstruct_wannier}. We note that BaAsPt belongs to a whole class of materials in SG 198, referred to as LaIrSi-type materials in\,\cite{Zhang2020quadWeyl}, consisting of three elements with similar bandstructures, as seen on the materials project\,\cite{Jain2013}, and likely similar orbital characteristics, such that $\nu=\pm5$ might also be found in those, though at different distances to $E_F$, due to variations of the Fermi energy in these compounds.
	
	A full topological classification\,\cite{huber2022network} of the 6 bands crossing $E_F$ has been carried out. We enumerate these bands from 1 to 6 in descending order in energy. The charge of the nodal planes, which occur in SG 198 at the BZ boundary, is shown in figure\,\ref{fig:BaAsPt_bandstruct_wannier} in solid colored lines. The figure also shows all crossings on high symmetry lines. WPs at generic positions have been found for band pair $(4,5)$ (the band pair with $\nu=\pm5$ at $\Gamma$ and whose bandgap is shaded red) at $\vebm{k}=2\pi(-0.0289,-0.2699, -0.2988)^T$ and all symmetry related points with $\nu_{5}=-1$. Another generic WP was found at $\vebm{k}=2\pi(0, -0.2085, -0.291)^T$ with $\nu_4=-1$ for band pair $(3,4)$. The fermion doubling theorem is fulfilled when counting in the topological charges of the WPs, multifold crossings and NPs found by the full topological classification of the 6 bands crossing $E_F$.
	
	Further, a large topological band gap shaded in red separating the two bands with $\nu=\pm5$ can be seen. A surface DOS calculation at $100\,$meV shows a large number of Fermi arcs, see figure\,\ref{fig:s4BaAsPt_surface_band7_0_1eV}, despite a screening of the topological charge from the 4-fold point, which due to the filling of this topological band gap is $-3+5=-2$, by charges on $\Gamma-X$, which sum up to $2$. 4 copies of these WPs appear on the projection $\bar{\Gamma}-\bar{M}$. Very close to the $\Gamma$ point on the $\Gamma-R$ line there are 8 WPs, which we included into the charge of $\bar{\Gamma}$. The total charge of the bulk bands surrounding $\bar{\Gamma}$ of 4 give rise to 4 Fermi arcs emerging from the bulk states at $\bar{\Gamma}$ and running to the $\bar{R}$ point with charge $-4$. The remaining Fermi arcs are entirely explained by projected topological crossings of the band pair $(4,5)$, namely a small pocket between $\bar{\Gamma}$ and $\bar{R}$ with containing a charge of $-2$ connecting via 2 Fermi arcs to bulk bands with a charge of $2$ near $\bar{R}$. In total, we are counting 12 Fermi arcs.

	\subsection{NbO$_2$ and TaO$_2$ (SG 80)}\label{Subsec_NbO2TaO2}
	
	\begin{figure}
		\centering
		\includegraphics[width=\linewidth]{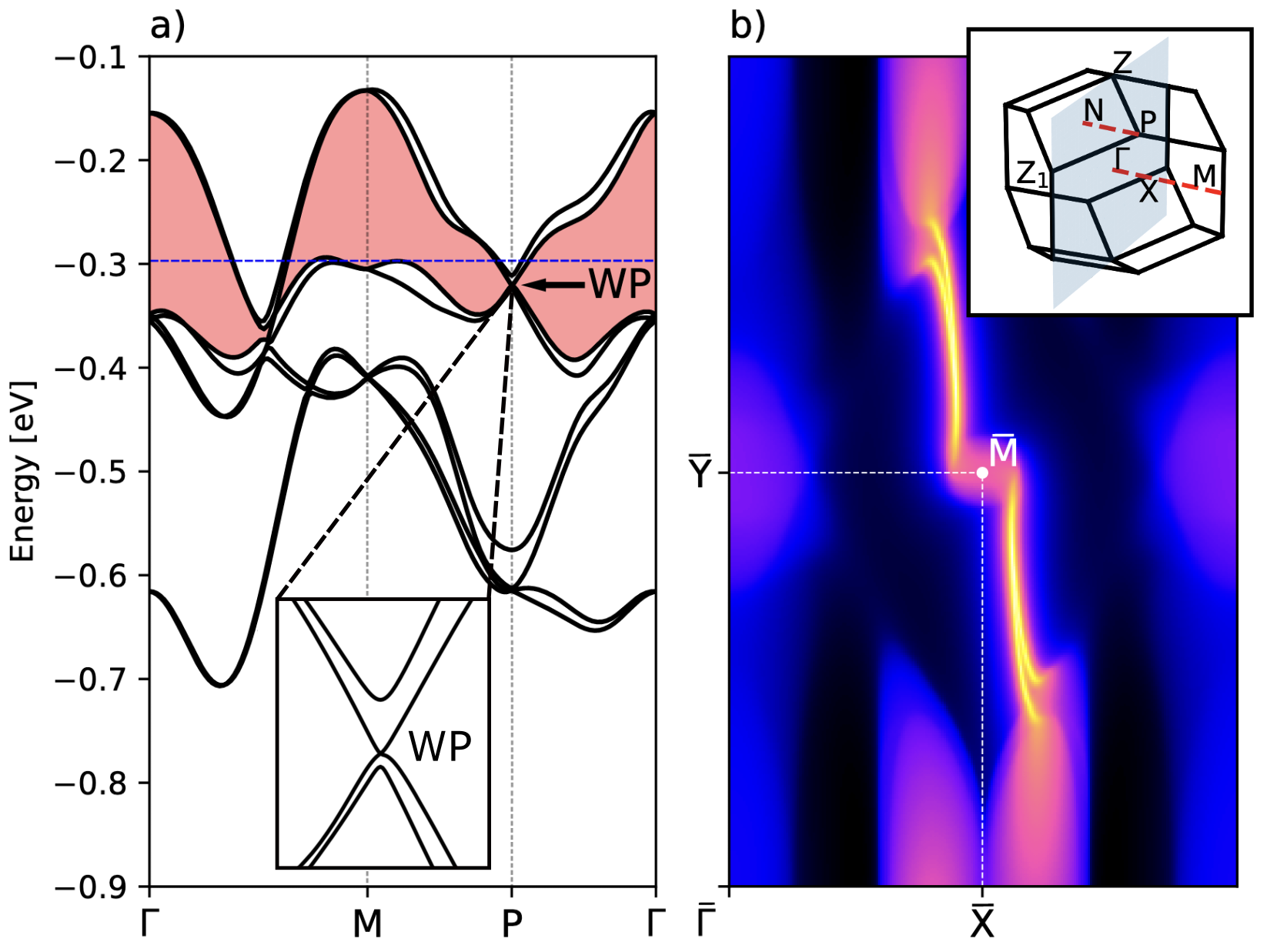}
		\caption{a) DFT bulk band structure of the distorted TaO$_2$ compound with the topological band gap colored in red and an arrow pointing to the Weyl point (WP). The inset is showing a zoom of the double Weyl point. b) Surface spectral density at $-0.297$ eV (blue dashed line in a)) with a termination projecting $M$ onto $\Gamma$ ($\bar{\Gamma}$) and $P$ onto $N$ ($\bar{M}$) as shown in the inset. Two Fermi arcs connect the projection of the Weyl point at $P$ with the bulk bands. }
		\label{fig_TaO2BandsAndSurfaceStates}
	\end{figure}
	
	Niobium dioxide was first synthesized in 1941 and was shown to crystallize in a rutile structure with tetragonal space group symmetry 136\cite{brauer1941oxyde}. Further research revealed the existence of a distorted lower-symmetry phase $\beta$-NbO$_2$\cite{magneli1955note}. During the structural transition, pairs of niobium atoms dimerize along the c-axis, and although the nature of the transition is believed to be of Peierls type, the specifics have been the subject of an extensive amount of research throughout the years\cite{o2014electronic, seta1982calorimetric, shapiro1974neutron, rao1973semiconductor, jacob2010thermodynamic}. Slightly sub-stochiometric single crystals of $\beta$-NbO$_2$ can be synthesized in oxygen-deficient environments, and its crystal structure has space group symmetry 80\cite{schweizer1982darstellung}. Much later, $\beta$-NbO$_2$ was proposed as a potential realization of a topological chiral crystal with Kramers-Weyl fermions in its bulk and the corresponding boundary modes on its surface\cite{chang2018topological}. 
	
	Since the topological band gap in $\beta$-NbO$_2$ is small and the crossing of interest is overshadowed by spectral weight of other bands in its vicinity, we propose two alterations to the compound to improve its usefulness as a topological semimetal.
	First, to increase the effect of spin-orbit coupling, we consider the hypothetical compound $\beta$-TaO$_2$, which is expected to have the same crystal structure since tantalum and niobium have very similar ionic radii and electron configurations\cite{schulz2017niobium}. Second, we enhance the distortion mode that connects the rutile and the reported lower-symmetry phase of NbO$_2$. To do this, we compare the crystal structures of the parent and the distorted compound, calculate the irreducible representations of the distortions and identify the linear combination of modes that connects the two configurations using the ISODISTORT tool\cite{isodistort2}. The computed distortion is then exaggerated by a factor of 1.5, retaining the space group symmetry of $\beta$-NbO$_2$. 
	Potential routes to synthesize the proposed crystal include growing it at higher temperatures or in a more oxygen-deficient environment\cite{schweizer1982darstellung}.
	
	The band structure and the surface states of $\beta$-TaO$_2$ are shown in Fig.~\ref{fig_TaO2BandsAndSurfaceStates}.
	In the vicinity of the Fermi energy there are two time-reversal-related double Weyl points protected by fourfold rotation symmetry, one of which is seen on the line $\Gamma$-M, as well as two double Weyl points pinned to the points labeled P.
	Our calculation shows that Weyl points on $\Gamma$-M with charge $\nu = +2$ compensate the ones at P with $\nu = -2$. 
	To our knowledge this is the first example, where double Weyl points are enforced away from a TRIM but pinned to a lower symmetry point. These doubly charged degeneracies on a two-fold rotation axis contradict previous suggestions that double Weyl points require four- or six-fold rotational symmetry\cite{Chen2012MultiWeyl}, and can only be understood from our argument relating symmetry eigenvalues. 
	
	\section{Conclusion}
	\label{sec_conclusions}
	
	In this paper we have derived two fundamental laws of chiral band crossings: A local constraint relating the Chern number to phase jumps of rotation eigenvalues (Sec.~\ref{sub_sec_local_constraint}), and a global constraint that restricts the number of chiral crossings on rotation axes (Sec.~\ref{sub_sec_global_constraint}). 
	To demonstrate the strength of these laws, we have applied 
	them to determine the existence of enforce double Weyl points, nodal planes, and other band topologies (Sec.~\ref{sec_four}). 
	Complementing these arguments by an exhaustive 
	classification of low-energy models, we have determined
	the generic topological phase diagrams of all multifold crossings (Sec.~\ref{gen_clas_hamiltonians}). Our analysis reveals, among others, that there are fourfold crossing 
	points with Chern number 5 (Sec.~\ref{4foldcrossingssection}).
	To illustrate some of the derived topological band features,
	we have discussed two material examples (Sec.~\ref{sec_six}):
	BaAsPt in SG 198 with fourfold crossing of Chern number 5 and NbO$_2$/TaO$_2$ in SG 80 with  double Weyl points.
	
	There are several directions for future work. 
	First, the local and global constraints can be applied in a straightforward  manner to magnetic space groups. For example,
	the local constraint can be used to infer the existence of double Weyl points away from TRIMs in magnetic space groups, similar to Sec.~\ref{SubSec_doubleWeyl}. 
	Second, our fundamental laws can be employed to
	study (multifold) nodal points and nodal planes of
	bosonic band structures, e.g., phonon or magnon bands.
	Third, our results have implications for topological response functions that are influenced by the Berry curvature,
	e.g., anomalous Hall currents, photogalvanic effects, and magnetooptic Kerr effects. Working out signatures of the 
	discussed band topologies (e.g., the nodal planes or the fourfold crossings with $\nu=5$) in these response functions would be an interesting task
	for future study.

	\section*{Acknowledgements}
	We are thankful to Douglas Fabini, and Johannes Mitscherling for enlightening discussions.
	We acknowledges the support by the Max Planck-UBC-UTokyo Center for Quantum Materials.
	\appendix
	\renewcommand{\appendixname}{APPENDIX}
	
	\section{Proof of Eq.~(\ref{Eq_symmetrytransformA})}\label{app_symmetrytransfromA}
	
	Since we chose a different sign in the definition of the Berry connection (see Eq.\,(\ref{eq_nonabelianberryconnection})) as compared to \cite{Fang2012ChernFromSymmetry}, we here present the derivation of Eq.\,(\ref{Eq_symmetrytransformA}) (Eq.\,(83) in \cite{Fang2012ChernFromSymmetry}).
	\begin{align}
		&{\bf A}(R\cdot{\bf k})=i\braket{n(R\cdot{\bf k})}{R\cdot\grad_{{\bf k}}|m(R\cdot{\bf k})}\nonumber\\
		&=i R \sum_{c,d} \mathcal{B}_{C_n}({\bf k})_{nc} \braket{c({\bf k})}{\hat{R}^{-1}\grad_{{\bf k}}\mathcal{B}_{C_n}^*({\bf k})_{md}\hat{R}\biggl|d({\bf k})}\nonumber\\
		&=i R \sum_{c,d} \mathcal{B}_{C_n}({\bf k})_{nc} \braket{c({\bf k})}{\grad_{{\bf k}}\mathcal{B}_{C_n}^*({\bf k})_{md}\biggl|d({\bf k})}\nonumber\\
		&=R(\mathcal{B}_{C_n}{\bf A}({\bf k})\mathcal{B}_{C_n}^{-1}+i\mathcal{B}_{C_n}\grad_{{\bf k}}\mathcal{B}_{C_n}^{-1}).
	\end{align}
	
	\section{Chiral nodal lines from magnetic symmetries}
	\label{App_chiralNL}
	
	In Sec.~\ref{SubSubSec_ChiralNL} we discussed the possibility of a nodal line characterized by a non-zero Chern number. 
	In the following we generalize this discussion to symmetries, which comprise both time reversal $T$ and an $n$-fold rotation $C_n$ around the z direction. 
	The arguments excluding the possibility of chiral nodal lines for $C_4T$ and $C_6T$ follow from the constraints in Eqs.~(\ref{Eq_constrantiunitaryC4T}) and (\ref{Eq_constrantiunitaryC6T}) in the same way as in the main text.
	
	If there is a nonzero sewing matrix phase difference $\Delta\phi_{c_b}$, the relations imply a point-like band crossing on the axis.
	Since this implies that the only possible chiral charges of the line are equal to the multiplicity of Weyl points, such lines would be unstable.
	To see this, consider a case where $\Delta\phi_{c_b}$ is nonzero and does not change when the size of the sphere surrounding the $C_nT$ invariant point shrinks to zero. $\Delta\phi_{c_b}\neq0$ implies with the local constraints $\nu_{c_b}\neq0$. If $\nu_{c_b}$ is nonzero, only a point-like crossing can carry the charge implied by the arbitrary small integration sphere.
	
	Alternatively, if $\Delta\phi_{c_b}$ changes discontinuously when the sphere shrinks to zero, the crossings carrying the charge difference implied by the constraints, Eqs.~(\ref{Eq_constrantiunitaryC4T}) or (\ref{Eq_constrantiunitaryC6T}), must lie on the rotation axis and can not be attributed to a chiral nodal line. To see this, one must deform the integration sphere into an spheroid while keeping the intersection points on the axis constant, such that $\Delta\phi_{c_b}$ remains unchanged. Thereby, the equatorial radius of the spheroid can be reduced to zero to exclude any finite size nodal line from the enclosed region, such that the topological charges may only lie on the rotation axis in form of WPs located where $\Delta\phi_{c_b}$ changes.
	
	The case of $C_2T$ differs, as the constraint derived from it does not include a term in the form of $\Delta\phi_{c_b}$, see Eq.\,(\ref{Eq_constrantiunitaryC2T}). Instead it involves the winding of $\phi$ around the rotation axis on a $C_2T$ invariant path. 
	A nonzero winding, which results in a $\nu_{c_b}=1\mod2$ constraint, does not imply a charged nodal line of charge 1, since this nodal line is able to gap out into just a single WP on the $C_2T$ invariant plane.
	
	\section{Tight-binding model for SG 80} \label{App_SG80_Models}
	
	To illustrate the band topology induced by SG~80, we give a minimal tight-binding model for the  spinless and spinful case, as discussed in Sec.~\ref{SubSec_doubleWeyl}.
	We consider a generic model for the 2a Wyckoff position and take all symmetry-allowed terms up to 2nd nearest neighbors into account.  
	We use the phase convention of Bloch functions for the tight-binding orbitals \cite{vanderbilt2018berry} and the primitive vectors as basis for $\vb{k}$ \cite{setyawan2010high}.
	Our model takes the form
	\begin{align}
		(H_{\text{SG80},\uparrow})_{11} &=
		2 t'_1 \big( \cos(k_1 + k_2 + k_3) + \cos(k_3) \big) 
		\nonumber\\
		&+ 2 t'_2 \big( \cos(k_1) + \cos(k_2) \big),
		\\[0.2cm]
		(H_{\text{SG80},\uparrow})_{22} &=
		2 t'_1 \big(\cos(k_1) + \cos(k_2) \big) 
		\nonumber\\
		&+ 2 t'_2 \big(\cos(k_1 + k_2 + k_3) +\cos(k_3) \big),
		\\[0.2cm]
		(H_{\text{SG80},\uparrow})_{12} &= (H_{\uparrow})_{21}^* \\
		&\hspace{-1.5cm}= 
		t_1 \left(1 + \mathrm{e}^{i k_1} + \mathrm{e}^{i (k_1 + k_3)}  +  \mathrm{e}^{i (k_1 + k_2 + k_3)} \right) 
		\nonumber\\
		&\hspace{-1.5cm}+ t_2 \left(
		\mathrm{e}^{i (k_1 + k_2)}  + \mathrm{e}^{-i k_2}  +  \mathrm{e}^{i(2 k_1 + k_2 + k_3)} + \mathrm{e}^{i k_3} \right), 
	\end{align}
	where the parameters $t_1, t_2, t'_1, t'_2 \in \mathbb{R}$. 
	With spin-orbit coupling the full Hamiltonian becomes
	\begin{align}
		H_{\text{SG80}}(\vb{k})
		= 
		\mqty(
		H_{\text{SG80},\uparrow}^{\text{SOC}} & H_{\text{SG80},\uparrow\downarrow}^{\text{SOC}} \\
		(H_{\text{SG80},\uparrow\downarrow}^{\text{SOC}})^\dagger & H_{\text{SG80},\downarrow}^{\text{SOC}} ), 
		\label{Eq_SG80fullHamiltonian}
	\end{align}
	where $H_{\text{SG80},\uparrow}^{\text{SOC}}$ and $H_{\text{SG80},\downarrow}^{\text{SOC}}$ is obtained by $H_{\text{SG80},\uparrow}$ by adding hopping terms that differ for the two spin directions. 
	We introduce this by
	\begin{align}
		H_{\text{SG80},\uparrow}^{\text{SOC}}
		&= 
		H_{\text{SG80},\uparrow} 
		+ \Delta H_{\text{SG80},\uparrow},
		\\[0.2cm]
		(\Delta H_{\text{SG80},\uparrow})_{11} 
		&=
		\big(
		\tilde{t}'_1 ( \mathrm{e}^{i(- k_1 - k_2 - k_3)}  + \mathrm{e}^{i k_3}) 
		\nonumber\\
		&+ \tilde{t}'_2 ( \mathrm{e}^{-i k_1}  + \mathrm{e}^{-i k_2} )
		\big)
		+ c.c.,
		\\[0.2cm]
		(\Delta H_{\text{SG80},\uparrow})_{22} 
		&=
		\big(\tilde{t}'_1 (\mathrm{e}^{-i k_1}  + \mathrm{e}^{-i k_2} ) 
		\nonumber\\
		&+ 
		\tilde{t}'_2 ( \mathrm{e}^{i(- k_1 - k_2 - k_3)} + \mathrm{e}^{i k_3} ) \big) 
		+ c.c.,
		\\[0.2cm]
		(\Delta H_{\text{SG80},\uparrow})_{12} 
		&= (\Delta H_{\text{SG80},\uparrow})_{21}^*
		\nonumber\\
		&=
		\tilde{t}_1 ( \mathrm{e}^{i k_1}  + \mathrm{e}^{i (k_1 + k_2 + k_3)} ) 
		\nonumber\\
		&+ \tilde{t}_2 ( \mathrm{e}^{i k_1 + i k_2} + \mathrm{e}^{i ( 2 k_1 + k_2 + k_3)} ) 
		\nonumber\\
		&+ \tilde{t}_1^* ( 1 + \mathrm{e}^{i ( k_1 + k_3 )} ) 
		\nonumber\\
		&+ \tilde{t}_2^* ( \mathrm{e}^{-i k_2}  + \mathrm{e}^{i k_3} ),
	\end{align}
	where $\tilde{t}'_1, \tilde{t}'_2 \in \mathbb{C} $.
	The corresponding matrix elements in $H_{\text{SG80},\downarrow}^{\text{SOC}}$ can be obtained from $H_{\text{SG80},\uparrow}^{\text{SOC}}$ by replacing in the diagonal entries $\vb{k} \rightarrow - \vb{k}$ and in the off-diagonal entries $\tilde{t}_1 (\tilde{t}_2) \rightarrow \tilde{t}_1^* (\tilde{t}_2^*)$ for all occurrences.
	The coupling between spins takes the form 
	\begin{align}
		&(H_{\text{SG80},\uparrow\downarrow})_{11}
		\nonumber\\
		&=
		2i l'_1 \left(\sin(k_1 + k_2 + k_3) + \sin(k_3) \right)
		\nonumber\\
		&+ 2i l'_2 \left( \sin(k_1) - \sin(k_2) \right),
		\\[0.2cm]
		&(H_{\text{SG80},\uparrow\downarrow})_{22}
		\nonumber\\
		&=
		2 l'_1 
		\left( \sin(k_1) -  \sin(k_2)  \right)
		\nonumber\\
		&+ 2 l'_2 
		\left( - \sin(k_1 + k_2 + k_3) - \sin(k_3)  \right),
		\\[0.2cm]   
		&(H_{\text{SG80},\uparrow\downarrow})_{12}
		\nonumber\\ 
		&=
		l_1 \left( 1 + i \mathrm{e}^{i k_1} - \mathrm{e}^{i (k_1 + k_3)} - 
		i \mathrm{e}^{i (k_1 + k_2 + k_3)} \right)
		\nonumber\\
		&+ l_2 \left( - i \mathrm{e}^{i( k_1 + k_2)}  + \mathrm{e}^{-i k_2} + i \mathrm{e}^{i(2 k_1 + k_2 + k_3)} - \mathrm{e}^{i k_3} \right),
		\\[0.0cm]  
		&(H_{\text{SG80},\uparrow\downarrow})_{21}
		\nonumber\\  
		&=
		-(H_{\text{SG80},\uparrow\downarrow}(\vb{k} \rightarrow - \vb{k}))_{12}
	\end{align}
	where the parameters $l_1, l_2,  l'_1, l'_2 \in \mathbb{C}$. 
	
	For the remaining discussion we use the parameters defined in units of $t_1$ as
	$
	t_2 = 0.2,\,
	t'_1 = 0.3,\, t'_2 = -0.2
	$
	and if spin-orbit coupling is included we add
	$\tilde{t}_1= -0.1 + 0.2 i,\, \tilde{t}_2 = 0.05 - 0.05 i,\, 
	l_1 = 0.4 i,\, l_2 = 0.05 - 0.15 i,\,
	\tilde{t}'_1 = 0.1,\, \tilde{t}'_2 = 0.05,\, 
	l'_1 = l'_2 = 0.1 i$.
	A possible minimal set of parameters that realizes the same band topology is given by $\tilde{t}_1 = 0.2 + 0.4 i,\, l_1 =0.8 i,\, \tilde{t}'_1 = 0.4$, where all other parameters (except $t_1 = 1$) are set to 0.
	
	\begin{figure}
		\centering
		\includegraphics[width = 0.45 \textwidth]{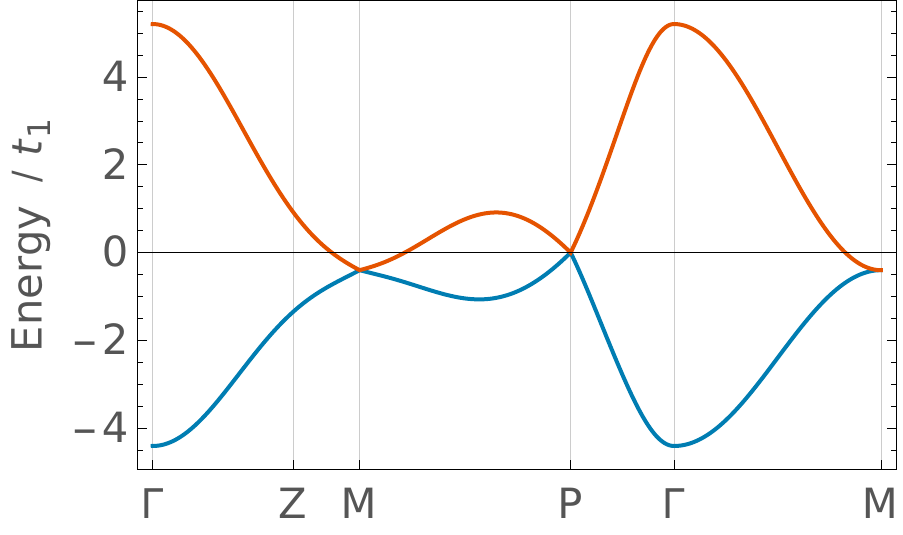}
		\caption{\label{fig_SG80_spinless}
			Band structure for the model defined in Eq.~\eqref{Eq_SG80fullHamiltonian} respecting SG~80 \emph{without SOC}. The Brillouin zone is shown in the in-set of Fig.~\ref{fig_TaO2BandsAndSurfaceStates}(b).
		}
	\end{figure}
	
	In the limit of vanishing spin-orbit coupling our model exhibits only three Weyl points for each spin sector, one at each of the two distinct points P with $\nu = +1$ and a double Weyl point at the TRIM M with $\nu = -2$, see Fig.~\ref{fig_SG80_spinless}. 
	Without the spin degeneracy the unity charge at the point P is different from the possible charges of twofold crossings at TRIMs, which are always $\nu = \pm 2$ in spinless systems \cite{tsirkin2017compositeWeyl}.
	The double Weyl point at M shows on the $\Gamma$-M path, i.e., where $k_z = 0$ is constant, the typical quadratic dispersion expected perpendicular to the rotation axis $\Gamma$-Z-M \cite{Zhang2020quadWeyl}.
	Note, our model for SG~80 without SOC is a counter example to the commonly conceived notion that there have to be at least four Weyl points in the presence of time-reversal symmetry \cite{2016_fourWPwithTRS, belopolski2017_fourWPwithTRS}.
	Here, the number of Weyl points in our model is lower than four without additional band crossings \cite{Hirschmann2021Tetragonals}.
	
	Once spin-orbit coupling is taken into account the bands at P (M) split into a double Weyl point and two non-degenerate bands (two double Weyl points on the $\Gamma$-Z-M path), a detailed description is given in Sec.~\ref{SubSec_doubleWeyl}.
	
	For the numerical determination of Chern numbers we use the Wilson loop approach on a discretized Brillouin zone as described in Ref.~\cite{fukui2005chern}.
	
	\section{Local constraints in the presence of quasi-symmetries}
	\label{App_SG195Gamma}
	
	In the following we derive a low-energy model for a twofold crossing that appears among the spinless representations of the point group T 23, describing the rotation symmetries of a tetrahedron. 
	The resulting model applies to SGs 195-199 at the TRIMs $\Gamma$ and R, L or H within the primitive, face-centered, or body-centered cubic unit cells in the nomenclature of the Bilbao crystallographic server \cite{Aroyo2014}. 
	All possible terms up to third order that are compatible with the twofold and threefold rotation of the cubic point groups symmetry are given in the model $H_{\text{T}}({\bf k})$ as
	\begin{align}
		H_{\text{T}}({\bf k}) &= \left( d_0 + d_1 (k_x^2 + k_y^2 + k_z^2) \right)\sigma_0 \nonumber\\
		&+ a_0 k_x k_y k_z \sigma_z \nonumber\\ 
		&+ \mqty( 0 & a_1 (k_x^2 + k_y^2 \mathrm{e}^{- i 2 \pi/3}  + k_z^2 \mathrm{e}^{- i 4 \pi/3} ) \\ H.C. & 0),
		\label{Eq_App_PG23}
	\end{align}
	where the Pauli matrices $\sigma_i$ have been used and the first term is diagonal that shifts and bends both bands with $d_0, d_1 \in \mathbb{R}$. 
	The chirality $\nu$ of this twofold crossing is $\nu = 4 \text{sgn}(a_0)$ and $a_1 \in \mathbb{C}$.
	In the given basis, the twofold and threefold rotations and time-reversal are represented by 
	\begin{align}
		&U(C^x_2) = \sigma_0, 
		\qquad
		U(C^{111}_3) = \mqty( \mathrm{e}^{i 2\pi/3} & 0\\ 0 & \mathrm{e}^{- i 2\pi/3}), 
		\\
		&\text{and } 
		\qquad T = \sigma_1 K, 
	\end{align}
	respectively.
	
	As discussed in the main text, $\nu = \pm 4$ is a peculiar value.
	For example for $a_0 > 0$, the local constraints, Eq.~(\ref{Eq_ChargeFromRotEigVal}), derived from the twofold and threefold rotations imply $\nu = 0 \mod 2 = 1 \mod 3 $, which is consistent with $\nu = 4$ but would imply the simpler possibility of $\nu = -2$. 
	This can not only be understood by a refined local constraint, Eq.~(\ref{Eq_RefinedLocalConstraint_PG23}), but also by the emergence of a yet recognized quasi-symmetry that mimics a fourfold rotation. 
	The existence of a fourfold symmetry, can be motivated pictorially by noticing that the symmetries $C^z_2$ and $C^z_2 T$ together reduce the integration surface of the Chern number into a quarter of the full sphere, as discussed in Sec.~\ref{SubSec_MultipleScrew}.
	To gap out the bands away from ${\bf k} = 0$ and to obtain a well-defined chirality, it is sufficient to take all the symmetry-allowed terms up to cubic order in $k_i$, as we did in $H_{\text{T}}({\bf k})$.
	Interestingly, the fourfold rotation symmetry of the model is only broken at the fourth order in $k_i$, which is irrelevant for the value of the Chern number. 
	
	The model, Eq.~(\ref{Eq_App_PG23}), exhibits a parameter-dependent fourfold rotation symmetry $U(C^z_4)$ defined as
	\begin{align}\label{Eq_App_DefQuasiFourfold}
		U(C^z_4)
		=
		\mqty(0 & \mathrm{e}^{i(\arg(a_2) + 2 \pi/3 )} \\ \mathrm{e}^{-i(\arg(a_2) + 2 \pi/3 )} & 0), 
	\end{align}
	fulfilling 
	\begin{align}
		U(C^z_4)^\dagger 
		H_{\text{T}}(k_x,k_y,k_z)
		U(C^z_4)
		= 
		H_{\text{T}}(k_y,-k_y,k_z).
	\end{align}
	The symmetry eigenvalues of this operation are independent of $a_1$ and equal to $\lambda(C^z_4) \in \{ +1 , -1 \}$. 
	We made a choice choice of complex phase in Eq.~(\ref{Eq_App_DefQuasiFourfold}) that also does not affect the eigenvalues, but has been used to ensure that $U(C^z_4)$ commutes with time-reversal symmetry $T$. 
	The commuting property together with the reality of eigenvalues implies that at the twofold degeneracy the symmetry eigenvalues of $U(C^z_4)$ do not exchange, i.e., the phase jump is $\Delta \varphi = 0$. 
	And thus we can apply our local constraint according to Eq.~(\ref{Eq_ChargeFromRotEigVal}) to the fourfold rotation to obtain $\nu = 0 \mod 4$ resulting in $\nu = +4$ for $a_0 > 0$.
	
	Let us compare this result to the low-energy model, again for a spinless system, described by the octahedral point group O 432, where the twofold axes of tetrahedral group T are replaced by  fourfold rotations. 
	We can turn $H_{\text{T}}({\bf k})$ into a model of point group O by setting $a_1 = - 2 \pi/3$ in Eq.~(\ref{Eq_App_PG23}).
	This choice turns the quasi-symmetry $U(C^z_4)$ into a  representation of the actual fourfold rotation of the octahedral point group. 
	Naturally, for this real fourfold rotation $U(C^z_4)$ our argument on the local constraint it opposes on the chirality is unchanged and one finds $\nu = \pm 4$ as well. 
	
	In summary, we find that the existence of a fourfold quasi-symmetry in the low-energy model explains the charge $\nu = \pm 4$ of the twofold degeneracies with the tetrahedral point group T. 
	
	\section{Tight-binding model for SG~94}\label{App_SG94_NP_Duo}
	
	In the following we define the model of SG~94 including spin and with time-reversal symmetry, which is used to create Fig.~\ref{fig_SG94_NPDuo}. 
	We take the 2a Wyckoff position with the sites $(0,\, 0,\, 0),\, (1/2,\, 1/2,\, 1/2)$ and a spin-1/2 as internal degree of freedom on each site.
	To keep the model simple, we pick three independent terms that are sufficient to avoid any accidental band degeneracies from the set of hopping terms between up to next-nearest neighbors.
	Our 4$ \times $4 Hamiltonian thus takes the form
	\begin{align}
		\label{Eq_SG94_TBmodel}
		H_{\text{SG94},\uparrow}
		=
		\mqty( H_{\text{SG94},\uparrow}  & H_{\text{SG94,SOC}} \\ H_{\text{SG94,SOC}}^\dagger & H_{\text{SG94},\downarrow}  ),
	\end{align}
	where the matrix blocks are 
	\begin{align}
		(H_{\text{SG94},\uparrow})_{11} &= (H_{\text{SG94},\uparrow})_{22} =
		t_2 \mathrm{e}^{-i k_z} + t_2^* \mathrm{e}^{i k_z},
		\\
		(H_{\text{SG94},\uparrow})_{12} 
		&=
		(1 + \mathrm{e}^{i k_x}) (1 + \mathrm{e}^{i k_y}) (t_1 + t_1^* \mathrm{e}^{i k_z}),
		\\[0.3cm]
		H_{\text{SG94},\downarrow} &= H_{\text{SG94},\downarrow}(t_1 \leftrightarrow t_1^*, t_2 \leftrightarrow t_2^*)
	\end{align}
	for hopping that preserves spin and 
	\begin{align}
		&(H_{\text{SG94,SOC}})_{11}
		= (H_{\text{SG94,SOC}})_{22} = 0, 
		\\
		&(H_{\text{SG94,SOC}})_{12} =
		\nonumber\\
		&\quad
		t_{\text{SOC}}^* (-i \mathrm{e}^{i k_x} + i \mathrm{e}^{i k_y} - \mathrm{e}^{i k_z} + \mathrm{e}^{i (k_x + k_y + k_z)}) 
		\nonumber\\
		&\quad+ t_{\text{SOC}} (i - i \mathrm{e}^{i (k_x + k_y)} - \mathrm{e}^{i (k_x + k_z)} + \mathrm{e}^{i (k_y + k_z)}), 
		\\
		&(H_{\text{SG94,SOC}})_{21}
		=
		\mathrm{e}^{-i (k_x + k_y + k_z)} \big( 
		\nonumber\\
		&\quad t_{\text{SOC}}^* (-1 + \mathrm{e}^{i (k_x + k_y)} - i \mathrm{e}^{i (k_x + k_z)} + i \mathrm{e}^{i (k_y + k_z)}) 
		\nonumber\\
		&\quad+ t_{\text{SOC}} (-\mathrm{e}^{i k_x} + \mathrm{e}^{i k_y} + i \mathrm{e}^{i k_z} - i \mathrm{e}^{i (k_x + k_y + k_z)}) 
		\big)
	\end{align}
	for spin-orbit coupling terms.
	For Fig.~\ref{fig_SG94_NPDuo} we use the parameters $t_1 = 1 + i,\, t_2 = 0.5 + 0.8 i,\, t_{\text{SOC}} = -0.2 + 0.5 i $.

	\section{4-fold models\label{appendix4foldmodels}} 
	
	\subsection{SG 198 $\bar{\Gamma}_6\bar{\Gamma}_7$ model with SOC \label{appendixSG198Gamma6Gamma7}}
	
	In the following we show that the phase transitions of the low-energy Hamiltonian (Eq.\,(\ref{model1})) describing the $\bar{\Gamma}_6\bar{\Gamma}_7$ irrep are the ones shown in figure\,\ref{fig:s3phasediagramC5}. We first apply a transformation into a basis diagonal in the 3-fold rotation
	\begin{align}
		\begin{pmatrix}
			q_+\\
			q_-\\
			q_z
		\end{pmatrix}
		=\frac{1}{2\sqrt{3}}
		\begin{pmatrix}
			2 & -1+i\sqrt{3}& -1-i\sqrt{3} \\
			2 & -1-i\sqrt{3}& -1+i\sqrt{3}  \\
			-2 & -2 &  -2
		\end{pmatrix}\cdot\vebm{k}.
	\end{align}
	Due to scale invariance, we can parameterize the following way
	\begin{align}
		\alpha_0 &= \cos(G)\cos(Z),\\
		\alpha_1 &= \cos(G)\sin(Z),\\
		\alpha_2 &= \sin(G).
	\end{align}
	with $0\leq Z< 2\pi$ and $-\frac{\pi}{2}\leq G \leq\frac{\pi}{2}$. We find all points in parameter space $\alpha_n$ where the energy levels of $H$ become degenerate away from $\vebm{k}=0$ by considering the characteristic polynomial of $H$
	\begin{align}
		\chi(E) = E^4 + a(q_+,q_z)E^2+\text{det}H.
	\end{align}
	Since there are only even powers of $E$ in $\chi$, the spectrum is particle-hole symmetric. It is gapless when $\text{det}H=0$ or $a^2-4\text{det}H=0$. We find that $\chi$ and $\text{det}H$ do not depend on $Z$, so the topological phase diagram must be rotationaly invariant. Further we parameterize
	\begin{align}
		q_+ &= \exp(iK)\cos(M)\\
		q_z &= \sin(M)
	\end{align}
	due to scale-invariance without loss of generality. 
	\newline
	\newline
	\textit{$\text{det}H=0$ case:} We find 
	\begin{align}
		\text{det}H&=p_1(G,M)\cos(K)^3\nonumber\\ &+p_2(G,M)\cos(K)+p_3(G,M).
	\end{align}
	Particle-hole symmetry implies $\text{det}H\geq0$, so $\text{det}H=0$ must be a minimum. Therefore
	\begin{align}
		\frac{\text{d\,}\text{det}H}{\text{d}K} &= -\sin(K)(3p_1\cos(K)^2+p_2) = 0\\
		\implies (\sin(K)&=0 \implies K\in\{0,\pi\}\\
		\lor \cos(K)&=\pm\sqrt{\frac{-p_2}{3p_1}}=\pm\frac{1}{2}\\
		\implies K&\in\{\frac{\pi}{3},\frac{2\pi}{3}\}).
	\end{align}
	So $\{0,\pi,\frac{\pi}{3},\frac{2\pi}{3}\}$ are the only $K$ values where $H$ can get gapless via $\text{det}H=0$.
	
	With $K=0$, $\text{det}H$ has the form
	\begin{align}
		\text{det}H=f_4(M)\sin(G)^4+f_2(M)\sin(G)^2+f_0(M).
	\end{align}
	It is zero when
	\begin{align}
		\sin(G)^2=\frac{-f_2 \pm\sqrt{f_2^2-4f_4f_0}}{2f_4}=U_\pm(M).
	\end{align}
	A solution for $G$ is found, when $U_\pm$ is real and $0\leq U_\pm\leq 1$. We find this is the case for $M\in\{-\frac{\pi}{2}, \arctan\left(\frac{1}{2}\right), \frac{\pi}{2}\}$, where $U_+=0 \implies G=0$, and $M=-\frac{\pi}{4}$ with $U_+=\frac{1}{2} \implies G=\pm\frac{\pi}{4}$. $U_-=\sin(G)^2$ only reproduces the $M=-\frac{\pi}{4}$ solution.
	
	The $M\in\{-\frac{\pi}{2}, \arctan\left(\frac{1}{2}\right), \frac{\pi}{2}\}$ solutions with $K=0$ correspond (under symmetry and scale transformations) to degeneracies on the $k=(t,t,t)$ line, while $M=-\frac{\pi}{4}$ corresponds to $k=(t,0,0)$.
	
	For the case $K=\pi$ we find after a transformation with $M\rightarrow-M$ and $U_\pm \rightarrow U_\mp$ the same solutions. All other cases can be found by applying the $3$-fold symmetry found in the little group at this $k$-point. Therefore, with $\text{det}H=0$, degeneracies only happen on high symmetry lines and there are gapless points at $G\in\{0,\pm\frac{\pi}{4}\}$. We find for $G=0$ a gap closing of the bandpair $(2,3)$ on $C_3$ invariant lines and at $G=\pm\frac{\pi}{4}$ a gap closing on $C_2$ invariant lines for the same bandpair.
	\newline
	\newline
	\textit{$a^2-4\text{det}H=0$ case:} We find due to particle-hole symmetry $a=-E_1^2-E_2^2$, so $a^2-4\text{det}H\geq0$, since
	\begin{align}
		a^2&\geq 4\text{det}H\nonumber\\
		\iff (E_1^2+E_2^2)^2&\geq 4E_1^2E_2^2\nonumber\\
		\iff (E_1^2-E_2^2)^2&\geq 0.
	\end{align}
	So the point where $a^2-4\text{det}H=0$ must be a minimum of $a^2-4\text{det}H$. We find the following form of $a^2-4\text{det}H$
	\begin{align}
		a^2-4\text{det}H&=s_1(G,M)\cos(K)^3\nonumber\\ &+s_2(G,M)\cos(K)+s_3(G,M).
	\end{align}
	An optimum can be found via
	\begin{align}
		\frac{\text{d} (a^2-4\text{det}H)}{\text{d}K} &= -\sin(K)(3s_1\cos(K)^2+s_2) = 0\\
		\implies K&\in\{0,\pi\}\\
		\lor \cos(K)&=\pm\sqrt{\frac{-s_2}{3s_1}}=\pm\frac{1}{2}\\
		\implies K&\in\{\frac{\pi}{3},\frac{2\pi}{3}\}.
	\end{align}
	So once again $\{0,\pi,\frac{\pi}{3},\frac{2\pi}{3}\}$ are the only $K$ values where $H$ can get gapless, now via $a^2-4\text{det}H=0$. With $K=\frac{n\pi}{3}$ and $n\in\mathbb{N}$ we find
	\begin{align}
		a^2-4\text{det}H&=(r_1(n,M)\cos(G)^2 + r_2(n,M))\cos(G)^2
	\end{align}
	solutions to $\cos(G)^2=0$ and $\cos(G)^2=\frac{-r_2}{r_1}$ only exist for $G\in\{-\frac{\pi}{2},0,\frac{\pi}{2}\}$. For $G=0$ we find a gap closing of the outer bandpairs at $C_2$ invariant lines. For $G=\pm\frac{\pi}{2}$ the gap closing of the same bandpair occurs for any $q$, as we find that $a^2-4\text{det}H=0$ for any $M,K$.
	
	We therefore showed that the Hamiltonian is only gapless for $G\in\{-\frac{\pi}{2},-\frac{\pi}{4},0,\frac{\pi}{4},\frac{\pi}{2}\}$ at points away from $k=(0,0,0)$, which translates to $\alpha_2=\pm\sqrt{\alpha_0^2+\alpha_1^2}$, $\alpha_2=0$ or $\sqrt{\alpha_0^2+\alpha_1^2}=0$.
	
	\subsection{SG 207 $\bar{\Gamma}_8$ model with SOC\label{SG207GammaSOCSEC}}
	The generator of this irrep contains a time-reversal symmetry $T$, three 2-fold symmetries $2_{001}$, $2_{010}$, $2_{110}$ and a 3-fold symmetry $3_{111}$. The first-order Hamiltonian generated from these symmetries is
	\begin{align}
		H=&\alpha_0\biggl[
		\frac{k_x}{2}\left(\sqrt{3}\sigma_x\tau_0 + \sigma_x\tau_z + \sigma_y\tau_0 - \sqrt{3}\sigma_y\tau_z\right) + \nonumber\\&
		\frac{k_y}{2}\left(-\sqrt{3}\sigma_x\tau_0 - \sigma_x\tau_z + \sigma_y\tau_0 - \sqrt{3}\sigma_y\tau_z\right) + \nonumber\\&
		k_z\left(\sigma_x\tau_x - \sigma_x\tau_y\right)
		\biggr] +\nonumber\\&
		\frac{\alpha_1}{\sqrt{2}}\biggl[
		k_x\left(\sigma_0\tau_x + \sigma_0\tau_y + \sigma_z\tau_x - \sigma_z\tau_y\right) + \nonumber\\&
		k_y\left(\sigma_0\tau_x + \sigma_0\tau_y - \sigma_z\tau_x + \sigma_z\tau_y\right) + 
		2k_z\sigma_z\tau_z
		\biggr].\label{SG207GammaSOC}
	\end{align}
	The characteristic polynomial of this Hamiltonian is identical to the one of model \ref{model1} from the main text when $\alpha_1=0$ and $\alpha_2\to \alpha_1$. Therefore the topological phase separating points in parameter space of $\alpha_0$ and $\alpha_1$ are reproduced. Computing the Chern numbers, one finds that topological phase diagram as a whole stays the same, except now, one axis corresponds to only $\alpha_0$, instead of $\sqrt{\alpha_0^2+\alpha_1^2}$. The statements about the symmetry eigenvalue phase jumps also stay the same.
	
	This model also describes SG 207 $\bar{R}_8$, SG 208 $\bar{\Gamma}_8$, $\bar{R}_8$, SG 209 $\bar{\Gamma}_8$, SG 210 $\bar{\Gamma}_8$, SG 211 $\bar{\Gamma}_8$, $\bar{H}_8$, SG 212 $\bar{\Gamma}_8$, SG 213 $\bar{\Gamma}_8$, SG 214 $\bar{\Gamma}_8$ with SOC. 
	
	The SG 214 $\bar{H}_8$ model with SOC can be found by applying $U=\sigma_0\tau_x$ and $\alpha_1\to-\alpha_1$ to Hamiltonian \ref{SG207GammaSOC}.
	
	\subsection{SG 198 $\bar{M}_5\bar{M}_5$ model with SOC \label{SG198MSOC}}
	The generator of this irrep contains a time-reversal symmetry $T$ and two 2-fold symmetries $2_{001}$, $2_{010}$. The generated Hamiltonian is
	\begin{align}
		H=&\alpha_0
		k_z\sigma_0\tau_y
		+
		\alpha_1
		k_x\sigma_z\tau_z
		+
		\alpha_2
		k_x\sigma_x\tau_z
		+
		\alpha_3
		k_x\sigma_y\tau_z
		+\nonumber\\&
		\alpha_4
		k_y\sigma_z\tau_x
		+
		\alpha_5
		k_y\sigma_x\tau_x
		+
		\alpha_6
		k_y\sigma_y\tau_x
	\end{align}
	Its spectrum is particle-hole symmetric as the characteristic polynomial is $\chi(E)=E^4+aE^2+\text{det}H$.
	
	We can rewrite $H$ with
	\begin{align}
		\vebm{\alpha}_1 = (\alpha_1, \alpha_2, \alpha_3)^T\\
		\vebm{\alpha}_2 = (\alpha_4, \alpha_5, \alpha_6)^T
	\end{align}
	such that
	\begin{align}
		H=&\alpha_0
		k_z\tau_y\sigma_0
		+
		k_x\tau_z(\vebm{\alpha}_1\cdot\vebm{\sigma})
		+
		k_y\tau_x(\vebm{\alpha}_2\cdot\vebm{\sigma}).
	\end{align}
	Due to the two nodal planes on the $k_x=0$ and $k_y=0$ planes, the Chern number of single bands are always undefined. The Chern number for a filling of 2 can still be computed when there are no 4-fold degeneracies away from $\vebm{k}=(0,0,0)^T$. Therefore we only need to find these 4-fold degenerate points in parameter space, that is when $\text{det}H=0$.
	
	If $\alpha_0=0$, then $H$ is 4-fold degenerate on the $k_z$ line.
	If $\alpha_0\neq0$, $\vebm{\alpha}_1\neq0$ and $\vebm{\alpha}_2\neq0$, we can normalize $|\vebm{\alpha}_{1/2}|=1$ by using scale invariance of $H$. We rotate $\vebm{\sigma}$ with a unitary transformation $U^\dag\vebm{\sigma}U=R\vebm{\sigma}$ with $R$ being a rotation matrix, such that
	\begin{align}
		U^\dag(\vebm{\alpha}_1\cdot\vebm{\sigma})U&=\vebm{\alpha}_1\cdot(R\vebm{\sigma})\nonumber\\
		&=(R^{-1}\vebm{\alpha}_1)\cdot\vebm{\sigma}\nonumber\\
		&=\sigma_z
	\end{align}
	where we chose $R$ such that $R^{-1}\vebm{\alpha}_1=(1,0,0)^T$. Then
	
	\begin{align}
		H'=&U^\dag HU\nonumber\\=&\alpha_0
		k_z\tau_y\sigma_0
		+
		k_x\tau_z\sigma_z
		+
		k_y\tau_x(\tilde{\alpha}_2\cdot\vebm{\sigma}).
	\end{align}
	with $\tilde{\alpha}_2=R^{-1}\vebm{\alpha}_2$. We parameterize without loss of generality
	\begin{align}
		\tilde{\alpha}_2 &= (\cos P\cos R, \cos P \sin R, \sin P)^T\\
		\vebm{k} &= (\cos G\cos Z, \cos G \sin Z, \sin G)^T
	\end{align}
	We find
	\begin{align}
		|H'|=\text{det}H=A(P,R,Z)\cos(G)^4+1
	\end{align}
	and $\text{det}H\geq0$. The point $\text{det}H=0$ must be a minimum of $\text{det}H$. We find optima at $G\in\{-\frac{\pi}{2},0,\frac{\pi}{2}\}$. $G=\pm\frac{\pi}{2}$ can be excluded, since there $\text{det}H=1\neq0$. This leaves $G=0$. With this, we find the constraint for $A$ such that $\text{det}H=0$
	\begin{align}
		A(P,R,Z)&=4(\cos^4Z-\cos^2Z)(1-\cos^2R\cos^2P)\nonumber\\
		&=-1
	\end{align}
	We find optima of $\text{det}H$ with $\frac{\text{d}\text{det}H}{\text{d}Z}|_{G=0}=\frac{\text{d}A}{\text{d}Z}=0$ at $Z=n\frac{\pi}{4}$ with $n\in\mathbb{Z}_8$. We can exclude $Z=\frac{\pi}{2},\pi$ and $\frac{3\pi}{2}$, since there $A=0$. Inserting the remaining $Z$ values, we get
	\begin{align}
		\cos^2R\cos^2P=0.
	\end{align}
	Note that
	\begin{align}
		\vebm{\alpha}_1\cdot\vebm{\alpha}_2&=(R^{-1}\vebm{\alpha}_1)\cdot(R^{-1}\vebm{\alpha}_2)\nonumber\\
		&=(1,0,0)^T\cdot\tilde{\alpha}_2\nonumber\\
		&=\cos P \cos R.
	\end{align}
	Therefore $\vebm{\alpha}_1\cdot\vebm{\alpha}_2=0$ represents a surface in parameter space separating different topological phases. This is also true when $|\vebm{\alpha}_{1/2}|=0$. 
	
	If $|\vebm{\alpha}_1|=0$ we find the characteristic polynomial of $H$ with the same parametrization of $\vebm{k}$ as above and $|\vebm{\alpha}_2|=1$
	\begin{align}
		\chi(E)=(E^2+\cos^2G\cos^2Z-1)^2.
	\end{align}
	Note that $\chi(E)$ is independent of $\vebm{\alpha}_2$ since it can always be rotated by a unitary transformation, such that $H$ is rotationally invariant in $\vebm{\alpha}_2$. We find a 4-fold point at $E_{1,2,3,4}=0$. This is the case, when $G=0$ and $Z\in\{0,\pi\}$, so $H$ is 4-fold degenerate on the $k_x$ line.
	
	If $|\vebm{\alpha}_2|=0$ we find the characteristic polynomial of $H$ with $|\vebm{\alpha}_1|=1$
	\begin{align}
		\chi(E)=(E^2+\cos^2G\sin^2Z-1)^2.
	\end{align}
	We find a 4-fold point at $G=0$ and $Z\in\{\frac{\pi}{2},\frac{3\pi}{2}\}$, so $H$ is 4-fold degenerate on the $k_y$ line.
	
	In conclusion, we found that the topology of this model is entirely dependent on the sign of $\alpha_0$ and $\vebm{\alpha}_1\cdot\vebm{\alpha}_2$. We find the Chern number of the lowest two bands to be $\nu=2(-1)^{\theta(\alpha_0)+\theta(\vebm{\alpha}_1\cdot\vebm{\alpha}_2)}$.
	
	For both 2-fold symmetries, the constraint of Eq.~(\ref{Eq_nonabeliansewingjump}) amounts to $\nu_{12}=0\,\text{mod}\,2$ by direct calculation. This can be seen directly on the $(001)$ line, where the same symmetry eigenvalues are paired by $2_{001}$. This way, a symmetry eigenvalue jump can be defined for the two degenerate bands and is equal to $\pi$ everywhere in the phase diagram. Using Eq.~(\ref{Eq_nonabeliansymjump}) one arrives once again at $\nu_{12}=\frac{2}{2\pi}(\pi+\pi)\,\text{mod}\,2=0\,\text{mod}\,2$.
	
	This Hamiltonian also describes SG 18 $\bar{S}_5\bar{S}_5$, $\bar{R}_5\bar{R}_5$, SG 19 $\bar{S}_5\bar{S}_5$ with SOC. By applying a Chern number preserving $k_x\to k_y, k_y\to k_z, k_z\to k_x$ rotation, one arrives at the SG 19 $\bar{T}_5\bar{T}_5$, SG 92 $\bar{R}_5\bar{R}_5$ and SG 96 $\bar{R}_5\bar{R}_5$ Hamiltonian. By applying a $k_y\to k_z, k_z\to k_y$ reflection, during which Chern numbers are flipped, one arrives at the SG 19 $\bar{U}_5\bar{U}_5$ model.
	
	\subsection{SG 212 $\bar{M}_6\bar{M}_7$ model with SOC \label{212M6M7}}
	The little groups generator contains a time-reversal symmetry $T$ and three 2-fold symmetries $2_{001}$, $2_{010}$, $2_{110}$. The low-energy Hamiltonian generated from these symmetries is
	\begin{align}
		H=&\alpha_0
		k_z\sigma_z\tau_z +\nonumber\\&
		\alpha_1\biggl[
		k_x\sigma_0\tau_y -
		k_y\sigma_z\tau_x
		\biggr] +\nonumber\\&
		\alpha_2\biggl[
		k_x\sigma_x\tau_z -
		k_y\sigma_y\tau_0
		\biggr] +\nonumber\\&
		\alpha_3\biggl[
		k_x\sigma_y\tau_z +
		k_y\sigma_x\tau_0
		\biggr]
	\end{align}
	
	The lower and upper two bands of this Hamiltonian always have doubly degenerate points, such that the Chern number of the lower and upper band is undefined. This means, we only have to look at all points in parameter space where $\text{det}H=0$. This is the case when $|(\alpha_1,\alpha_2,\alpha_3)|=0$. There, $H=0$ on the $k_z=0$ plane. We also find $H=0$ for $\alpha_0=0$ on the $k_z$ line. Due to these considerations and scaling properties of $H$, we first consider the $\alpha_0=1$ case, which corresponds to $\alpha_0>0$.
	
	We also set $k_z=1$, such that all now reachable $k$ points correspond under rescaling to the upper half of the unit sphere in $k$ space. We parameterize
	\begin{align}
		\alpha_1 &= r_1\cos P\cos R\nonumber\\\nonumber\alpha_2 &= r_1\cos P \sin R\\\alpha_3&=r_1\sin P
	\end{align}
	with $r_1\geq0$. $\text{det}H$ takes on the following form
	\begin{align}
		\text{det}H&=r_1^4(k_x^4+k_y^4 + (16F(R,P)+2)k_x^2k_y^2) \nonumber\\&+ 2r_1^2(k_x^2 + k_y^2) + 1
	\end{align}
	where we find $-\frac{1}{4}\leq F(R,P)\leq0$ with $F(R,P)=\cos^2 P\cos^2 R(\cos^2 P\cos^2 R - 1)=\alpha_1^2(\alpha_1^2-r_1^2)/r_1^4$. Inserting the minimal value of $F$ in $\text{det}H$, we get
	\begin{align}
		\text{det}H=r_1^4(k_x^2-k_y^2)^2 + 2r_1^2(k_x^2 + k_y^2) + 1.
	\end{align}
	There $\text{det}H=0$ has no real solution, as $\text{det}H>0$. Therefore, for all $F\in[-\frac{1}{4},0]$, $\text{det}H>0$. Due to symmetry, this means, that also on the $k_z=-1$ plane, no degeneracies of the middle bandpair can occur. 
	
	\begin{figure}
		\centering
		\includegraphics[width=0.49\textwidth]{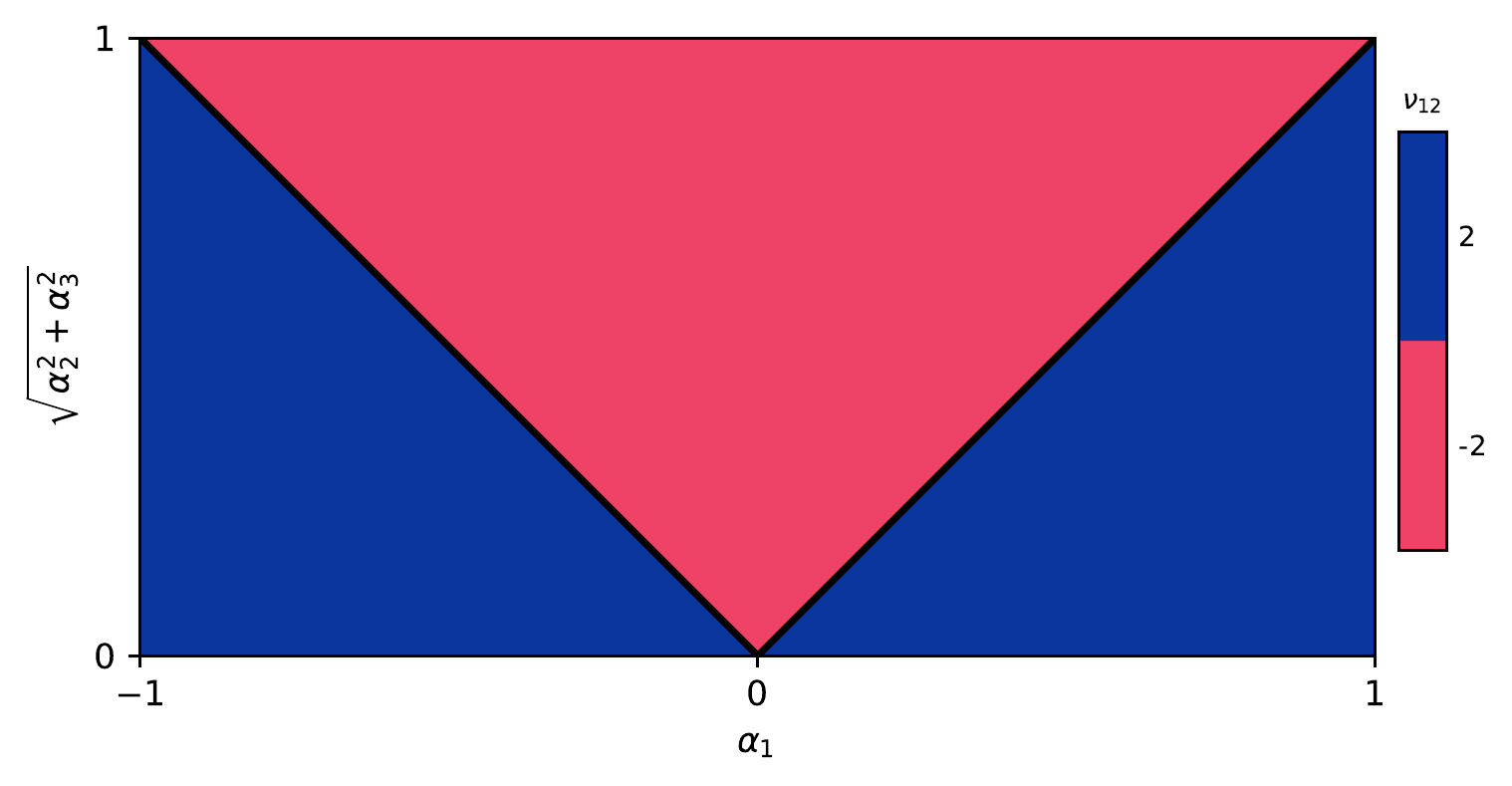}
		\caption{Topological phase diagram of the lower two bands Chern number $\nu_{12}$ of the SG 212 $\bar{M}_6\bar{M}_7$ Model with $\alpha_0>0$.\label{Fgg1}}
	\end{figure}
	The only way left, is the $k_z=0$ with $r_1>0$ and $|(k_x,k_y)|\neq 0$ case. We enforce this, by parameterizing $k_x=r_2\cos(G),k_y=r_2\sin(G)$ with $r_2>0$. The determinant with $k_z=0$ becomes
	\begin{align}
		\text{det}H= &r_1^4r_2^4(1 - 4(1-\cos(P)^2\cos(R)^2)\nonumber\\
		&\cos(P)^2\cos(R)^2\sin(2G)^2).
	\end{align}
	At $\text{det}H=0$, this becomes
	\begin{align}
		|\sin(2G)|=\frac{1}{2\sqrt{1-\left(\frac{\alpha_1}{r_1}\right)^2}|\frac{\alpha_1}{r_1}|}=K\left(\alpha_1/r_1\right)
	\end{align}
	The only solutions with $0\leq K \leq 1$ and $K\in\mathbb{R}$, are $\alpha_1=\pm\frac{r_1}{\sqrt{2}}$ with $K=1$, such that $G\in\{\frac{\pi}{4},\frac{3\pi}{4},\frac{5\pi}{4},\frac{7\pi}{4}\}$. The topological phase diagram is therefore rotationally invariant in the $(\alpha_1,\alpha_2,\alpha_3)$ parameter space around the $\alpha_1$ axis. The phase separating lines are $\alpha_1=\pm\sqrt{\alpha_2^2+\alpha_3^2}$. The topological phase diagram can be determined the same way as in the previous models, see figure \ref{Fgg1}. For $\alpha_0<0$, the Chern numbers in figure \ref{Fgg1} switch signs.
	
	By direct calculation, we find $\nu_{12}=0\,\text{mod}\,2$. The symmetry eigenvalue phase jump for $2_{001}$ is defined, since on the $(001)$ the same symmetry eigenvalues are paired, and $\pi$ for all bands. The symmetry eigenvalue jump of $2_{110}$ is also defined and $\pi$. Both lead to the $\nu_{12}=0\,\text{mod}\,2$ constraint once again. Symmetry eigenvalue jumps are undefined for $2_{010}$, as there different symmetry eigenvalues are paired.
	
	We get the SG 213 $\bar{M}_6\bar{M}_7$ Model with SOC by mapping $k_y\to -k_y$. This flips the sign of the Chern numbers in the topological phase diagram of this model.

	\subsection{SG 198 $R_1R_3$ model without SOC \label{SG198R1R3NOSOC}}
	This model is symmetric under a time-reversal symmetry $T$, two 2-fold symmetries $2_{001}$, $2_{010}$ and a 3-fold symmetry $3_{111}$.
	\begin{align}
		H=&\alpha_0\left[
		-k_x\sigma_z\tau_x + 
		k_y\sigma_0\tau_y + 
		k_z\sigma_z\tau_z.
		\right]
	\end{align}
	Due to the nodal plane, the Chern number of the lower and upper bands are undefined. The Chern number with a filling of 2 is $\nu_{12}=2(-1)^{\theta(\alpha_0)}$. Evaluating Eq.~(\ref{Eq_nonabeliansewingjump}) yields the constraints $\nu_{12}=0\,\text{mod}\,2$, $\nu_{12}=1\,\text{mod}\,3$ for $\alpha_0>0$ and $\nu_{12}=2\,\text{mod}\,3$ for $\alpha_0<0$, which are all fulfilled in this model. This model also describes SG 198 $R_2R_2$, SG 212 $R_1R_2$, SG 213 $R_1R_2$ without SOC.
	
	\subsection{SG 212 $R_3$ model without SOC \label{SG212R3NOSOC}}
	This model is symmetric under a time-reversal symmetry $T$, three 2-fold symmetries $2_{001}$, $2_{010}$,$2_{110}$ and a 3-fold symmetry $3_{111}$. The low-energy Hamiltonian is
	\begin{align}
		H=&\alpha_0\biggl[
		k_x\left(-\sigma_0\tau_x + \sigma_0\tau_y - \sigma_z\tau_x - \sigma_z\tau_y\right) + \nonumber\\&
		k_y\left(-\sigma_0\tau_x + \sigma_0\tau_y + \sigma_z\tau_x + \sigma_z\tau_y\right) + \nonumber\\&
		2k_z\sigma_z\tau_z
		\biggr]
	\end{align}
	Due to the nodal plane, the Chern number of the lower and upper bands are undefined. The Chern number with a filling of 2 is $\nu_{12}=2(-1)^{\theta(\alpha_0)}$. From Eq.~(\ref{Eq_nonabeliansewingjump}) we get the fulfilled constraints $\nu_{12}=0\,\text{mod}\,2$, $\nu_{12}=1\,\text{mod}\,3$ for $\alpha_0>0$ and $\nu_{12}=2\,\text{mod}\,3$ for $\alpha_0<0$. This model also describes SG 213
	$R_3$ without SOC.
	
	\subsection{SG 90 $\bar{A}_6\bar{A}_7$ model with SOC \label{SG90A6A7SOC}}
	This model is symmetric under a time-reversal symmetry $T$, two 2-fold symmetries $2_{001}$, $2_{010}$ and a 4-fold symmetry $4_{001}$. The Hamiltonian is
	\begin{align}
		H=&\alpha_0
		k_z\sigma_z\tau_z +\nonumber\\&
		\alpha_1\biggl[
		k_x\left(\sigma_0\tau_y + \sigma_z\tau_x\right) +
		k_y\left(\sigma_0\tau_y - \sigma_z\tau_x\right)
		\biggr] +\nonumber\\&
		\frac{\alpha_2}{\sqrt{2}}\biggl[
		k_x\left(-\sigma_x\tau_0 - \sigma_x\tau_z + \sigma_y\tau_0 - \sigma_y\tau_z\right) + \nonumber\\&
		k_y\left(-\sigma_x\tau_0 + \sigma_x\tau_z + \sigma_y\tau_0 + \sigma_y\tau_z\right)
		\biggr] +\nonumber\\&
		\frac{\alpha_3}{\sqrt{2}}\biggl[
		k_x\left(\sigma_x\tau_0 - \sigma_x\tau_z + \sigma_y\tau_0 + \sigma_y\tau_z\right) + \nonumber\\&
		k_y\left(\sigma_x\tau_0 + \sigma_x\tau_z + \sigma_y\tau_0 - \sigma_y\tau_z\right)
		\biggr]
	\end{align}
	
	We only need to look at the $\text{det}H=0$ points, due to the double degeneracy of the first two bands at some k lines. When $\alpha_0=0$, $H=0$ on the $k_z$ line, so $\alpha_0=0$ divides the topological phase diagram. Therefore we set $\alpha_0=1$, which corresponds to the $\alpha_0>0$ region. The Chern numbers we get will either remain unchanged or flip for $\alpha_0<0$. We also first look at the $k_z=1$ plane. Then we get
	\begin{align}
		\text{det}H&=A(k_x,k_y,r)\sin(P)^4+B(k_x,k_y,r)\sin(P)^2\nonumber\\&+C(k_x,k_y,r)
	\end{align}
	where we parameterized $\alpha_1=r\sin(P), \alpha_2=r\cos(P)\cos(R), \alpha_3=r\cos(P)\sin(R)$ and $r>0$.
	We find that the determinant fulfills $B^2-4AC\leq0$. So the only possible real solutions of $\text{det}H=0$ are when $B^2-4AC=0$, which is the case when $k_x=0 \lor k_y=0$. At those points $\text{det}H=0$ can not be fulfilled, since $A(0,k_y,r)=A(k_x,0,r)=B(0,k_y,r)=B(k_x,0,r)=0$ and $C(0,k_y,r)\geq1$ and $C(k_x,0,r)\geq1$. The only place left is the $k_z=0$ plane. We parameterize $k_x=\cos(G),k_y=\sin(G)$, where we used the scaling properties of $H$. We get
	\begin{align}
		\text{det}H&= r^4(-64\sin(G)^4\sin(P)^4 + 64\sin(G)^4\sin(P)^2 \nonumber\\&+ 64\sin(G)^2\sin(P)^4 - 64\sin(G)^2\sin(P)^2 + 4)
	\end{align}
	The only real solutions are $G\in\{\frac{\pi}{4},\frac{3\pi}{4},\frac{5\pi}{4},\frac{7\pi}{4}\}$ with $P=\pm\frac{\pi}{4}$. The topological phase diagram is identical to the one of the SG 212 $\bar{M}_6\bar{M}_7$ model, see figure \ref{Fgg1}, with $\alpha_0>0$. The Chern numbers flip when $\alpha_0<0$.
	
	The symmetry eigenvalue jump of $2_{001}$ is always $\pi$. The one of $4_{001}$ and $2_{010}$ can not be defined, since different eigenvalues are paired. So we need to use Eq.~(\ref{Eq_nonabeliansewingjump}) to derive constraints. We get $\nu_{12}=0\,\text{mod}\,2$ and $\nu_{12}=2\,\text{mod}\,4$, both fulfilled in all topological phases of this model.
	
	This model also describes SG 90 $\bar{M}_6\bar{M}_7$, SG 92 $\bar{M}_6\bar{M}_7$, SG 94 $\bar{M}_6\bar{M}_7$, SG 96 $\bar{M}_6\bar{M}_7$ with SOC. The reflection $k_x\leftrightarrow k_y$ can be applied, during which Chern numbers are flipped, to get the SG 94 $\bar{A}_6\bar{A}_7$ model with SOC. Apply the Chern number preserving rotation $k_x\to k_y$ and $k_y\to -k_x$ and set $\alpha_2=0$ and $\alpha_3=0$ to get the SG 92 $A_1A_2$ model without SOC, where the Chern number with a filling of 2 is just $\nu=-2(-1)^{\theta(\alpha_0)}$. The SG 96 $A_1A_2$ model is found by applying a $k_y\to -k_y$ reflection and also by setting $\alpha_2=0$ and $\alpha_3=0$. This flips the sign of the Chern number.
	
	\subsection{SG 92 $\bar{A}_7\bar{A}_7$ model with SOC \label{SG92A7A7SOC}}
	This model is symmetric under a time-reversal symmetry $T$, two 2-fold symmetries $2_{001}$, $2_{010}$ and a 4-fold symmetry $4_{001}$. The generated Hamiltonian is
	\begin{align}
		H=&\alpha_0
		k_z\sigma_z\tau_z +
		\alpha_1
		k_z\sigma_x\tau_0 +
		\alpha_2
		k_z\sigma_y\tau_0
	\end{align}
	No Chern numbers can be defined. This Hamiltonian also describes SG 96 $\bar{A}_7\bar{A}_7$ with SOC. This Hamiltonian must be expanded to $k^2$ to extract Chern numbers. This was done in \cite{wu2020higher,Hirschmann2021Tetragonals}. One arrives at $\nu=\pm4$ for a filling of 2. Other Chern numbers can not be defined due to degeneracies. See \cite{wu2020higher} for the exact topological phase diagram.
	
	\subsection{SG 19 $R_1R_1$ model without SOC \label{SG19R1R1NOSOC}}
	This model is symmetric under a time-reversal symmetry $T$ and two 2-fold symmetries $2_{001}$, $2_{010}$. The Hamiltonian is
	\begin{align}
		H=&\alpha
		k_x\sigma_z\tau_z+
		\beta
		k_y\sigma_z\tau_x+
		\gamma
		k_z\sigma_0\tau_y
	\end{align}
	$H$ is always doubly degenerate due to the nodal planes. Therefore we only need to look for $\text{det}H=0$.
	\begin{align}
		\text{det}H=(k_x^2\alpha^2+k_y^2\beta^2+k_z^2\gamma^2)^2
	\end{align}
	This leads to the condition
	\begin{align}
		k_x^2\alpha^2+k_y^2\beta^2+k_z^2\gamma^2=0
	\end{align}
	There exist $\vebm{k}$ points away from $\vebm{k}=0$ where this is the case when $\alpha=0\lor\beta=0\lor\gamma=0$. 
	We get $\nu_{12}=-2(-1)^{\theta(\alpha)+\theta(\beta)+\theta(\gamma)}$ for the lower two bands. The symmetry eigenvalue phase jump is always $\pi$ for both $2_{001}$ and $2_{010}$. Therefore we get the condition $\nu_{12}=0\,\text{mod}\,2$ from Eq.~(\ref{Eq_nonabeliansymjump}), which is fulfilled.
	
	\section{3-fold models\label{appendix3foldmodels}}
\begin{figure}[t]
		\centering
		\includegraphics[width=0.45\textwidth]{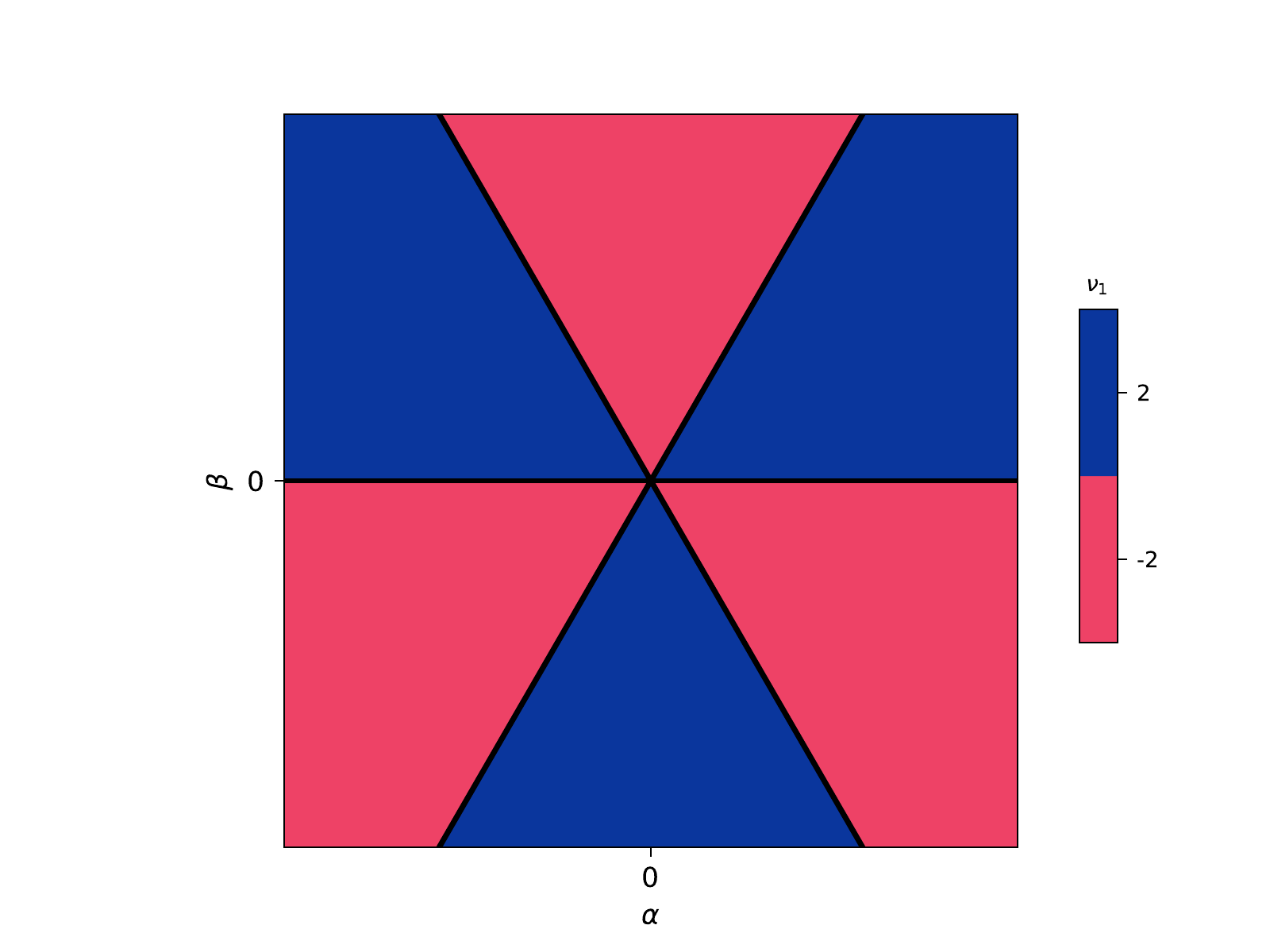}
		\caption{Topological phase diagram of band 1 of the SG 199 $\bar{P}_7$ model.\label{appendix3foldmodelstopphase}}
	\end{figure}

	Introduce $\kappa=\alpha+i\beta$. Normalize it without loss of generality $\kappa=e^{i\gamma}$.
	Then
	\begin{align}
		H=\begin{pmatrix}
			0 & k_ze^{-i\gamma} & -k_ye^{i\gamma}\\
			& 0 & k_xe^{-i\gamma}\\
			\ldots & & 0\\
		\end{pmatrix}
	\end{align}
	The characteristic polynomial is
	\begin{align}
		\chi(\lambda)&=-\lambda^3+\lambda(k_x^2+k_y^2+k_z^2)-2k_xk_yk_z\cos(3\gamma)\nonumber\\
		&=-\lambda^3+3\lambda-2k_xk_yk_z\cos(3\gamma)
	\end{align}
	where we set normalized $k_x^2+k_y^2+k_z^2=3$ using the scale invariance of $H$. 
	The characteristic polynomial with two degenerate energies is
	\begin{align}
		\chi(\lambda)&=-(\lambda-E_1)^2(\lambda-E_2)\nonumber\\
		&=-\lambda^3+\lambda^2(E_2+2E_1)-\lambda(E_1^2+2E_1E_2)+E_1^2E_2
	\end{align}
	We see that
	\begin{align}
		E_2&=-2E_1\\
		3&=-E_1^2-2E_1E_2=3E_1^2\nonumber\\
		&\implies E_1=\pm 1 \land E_2=\mp 2.
	\end{align}
	The only solutions to
	\begin{align}
		k_x^2k_y^2k_z^2\cos(3\gamma)^2=1 \land k_x^2+k_y^2+k_z^2=3
	\end{align}
	are
	\begin{align}
		k_{x/y/z}=\pm1 \land \gamma=\frac{\pi n}{3} \text{\,with\,} n\in\mathbb{Z}
	\end{align}
	so all lines in $\alpha\beta$ parameter space going from the origin and in $\frac{\pi}{3}$ angle to each other, starting at the $\beta=0$ line are the only lines separating different topological phases. We get the phase diagram for $\nu_1$ shown in figure \ref{appendix3foldmodelstopphase}, and $\nu_2=-\nu_1$. The $\nu_1=-2$ phases coincide with a symmetry eigenvalue phase jump of $\Delta\varphi_1=2\pi/3$, $\Delta\varphi_2=0$ and $\Delta\varphi_3=4\pi/3$ for the 3-fold rotation. The first bands phase jump leads with Eq.~(\ref{Eq_ChargeFromRotEigVal}) to the constraint $\nu_1=1\,\text{mod}\,3$, consistent with $\nu_1=-2$. We get phase jumps of $\Delta\varphi_1=4\pi/3$, $\Delta\varphi_2=0$ and $\Delta\varphi_3=2\pi/3$ for the $\nu_1=2$ phase, which is consistent with the constraint $\nu_1=2\,\text{mod}3$ derived from the $\Delta\varphi_1$.
	
	\section{6-fold models\label{appendix6foldmodels}}
	
	The topological phase diagram for the 6-fold model \ref{6foldmodel} of irrep $\bar{R}_7\bar{R}_7$ is invariant under rescaling $\alpha_n\to\alpha_n/r$ with $r>0$. So we can reduce the number of parameters of the 6-fold model to 3 by choosing an $r$ such that $(\alpha_0 + i\alpha_1)/r=e^{i\phi}$. We rename $\alpha_{2/3}/r$ as $\alpha_{2/3}$. The parameters $\phi$, $b=\alpha_2 + i\alpha_3$ correspond to the ones found in \cite{bradlyn2016beyond}.
	
	The characteristic polynomial is of the form
	\begin{align}
		\chi(E)=E^6+A_1E^4+A_2E^2-\text{det}H.
	\end{align}
	We see that the spectrum must be particle hole symmetric. Due to the nodal planes, Chern numbers for odd fillings can not be defined. One way a topological phase transition can happen is by a 6-fold degeneracy. We first show that the only places, where this can happen is on the nodal planes. The determinant of $H$ looks like this
	\begin{align}
		\text{det}H&=B(\alpha_n)k_x^2k_y^2k_z^2.
	\end{align}
	$\text{det}H=0$ is a necessary condition for a 6-fold degeneracy.  When $B\neq 0$, this is only the case on the nodal planes. A 6-fold degeneracy and particle hole symmetry implies $E_n=0$. Therefore $A_1=0\land A_2=0$ on the nodal plane. We parameterize $k_x=\cos(P),k_y=\sin(P),k_z=0$. Then
	\begin{align}
		A_1&=-2(\alpha_2^2+\alpha_3^2+1)\\
		A_2&=(\alpha_2^2+\alpha_3^2+1)^2
	\end{align}
	$A_1=0\land A_2=0$ is not possible. This is also the case for all other nodal planes.
	
	Another way a topological phase transition can happen is by a two-fold degeneracy. At the energy of a two-fold degeneracy, $\chi$ must have a maximum or minimum with $\chi=0$. With $k_x^2+k_y^2+k_z^2=1$ we find that
	\begin{align}
		A_1&=-2(\alpha_2^2+\alpha_3^2+1)\\
		A_2&=(\alpha_2^2+\alpha_3^2+1)^2
	\end{align}
	remains true. This means, $\chi$ has a local maximum at 
	\begin{align}
		E^\text{max}_\pm = \pm\sqrt{\frac{\alpha_2^2+\alpha_3^2+1}{3}}
	\end{align}
	independent of $a_0$ and $\vebm{k}$. $\text{det}H$ is tuning the value of this maximum. Therefore $\text{det}H$ can be tuned such that $\chi(E^\text{max}_\pm)=0$, where $E^\text{max}_\pm$ are the energies of the double degeneracies, in this case of band pairs $(2,3)$ and $(4,5)$. We get the condition
	\begin{align}
		\text{det}H=\frac{4}{27}(\alpha_2^2+\alpha_3^2+1)^3. \label{ZZ}
	\end{align}
	We find that $\text{det}H$ is
	\begin{align}
		\text{det}H=&k_x^2k_y^2k_z^2(2\cos(2\phi)-1+\alpha_2^2+\alpha_3^2)^2\nonumber\\
		&(2\cos(2\phi)+2+4\alpha_2^2+4\alpha_3^2)
	\end{align}
	\begin{figure}[t]
		\centering
		\includegraphics[width=0.49\textwidth]{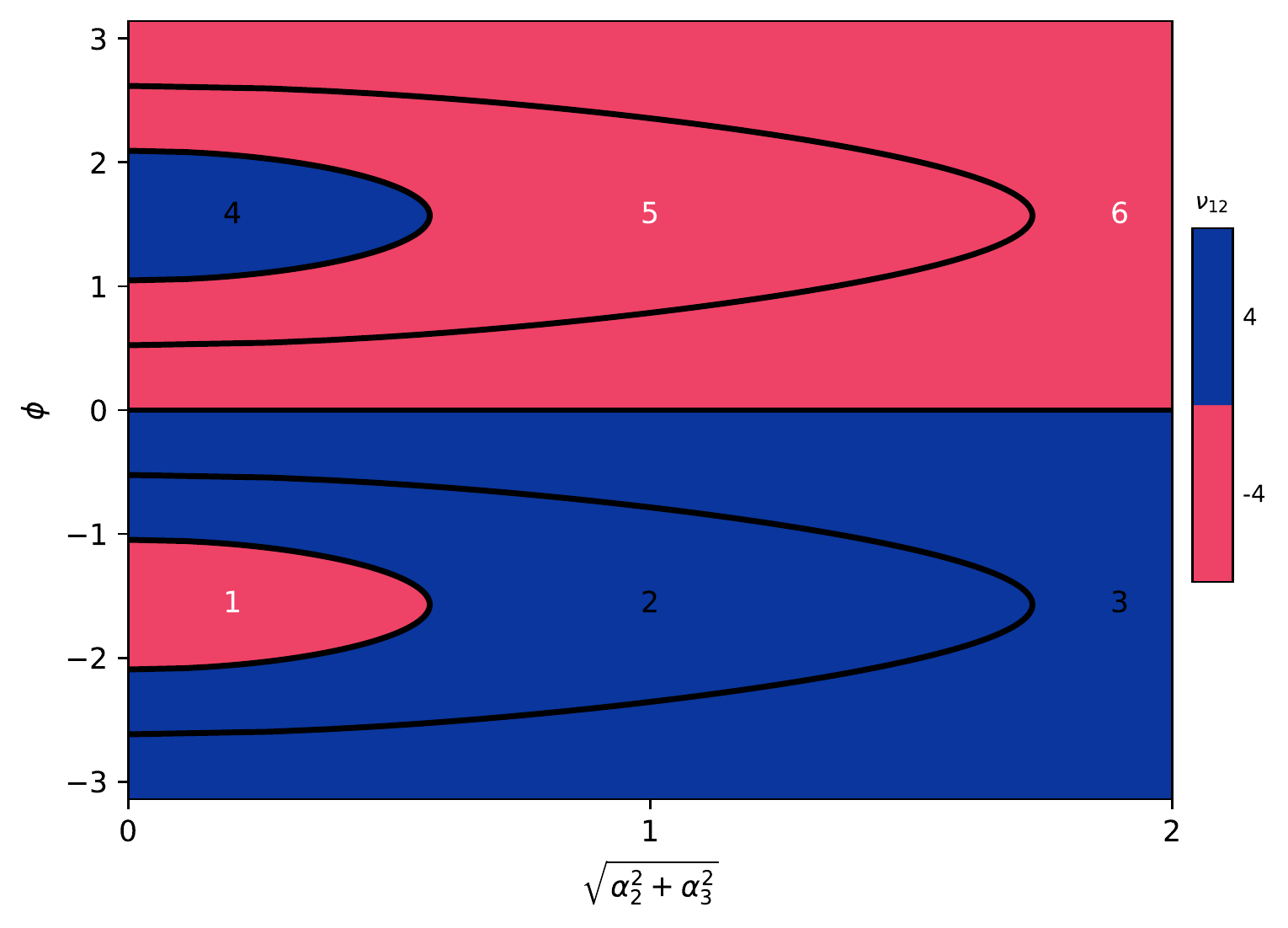}
		\caption{Analytical topological phase diagram for a filling of 2 of the $\bar{R}_7\bar{R}_7$ model. All phases separated by a closing gap are numerated from 1 to 6.\label{appendix6foldtopphase}}
	\end{figure}
	With $F=\alpha_2^2+\alpha_3^2+1$ ($1\leq F$), $S=2\cos(2\phi)-2$ ($0\geq S\geq-4$) and $A^{-1}=27k_x^2k_y^2k_z^2$ ($1\leq A$) we have
	\begin{align}
		(S+F)^2(S+4F)=4AF^3.
	\end{align}
	The only solutions with $1\leq A$ are $A=1\land S=-3F$ or $A=1\land S=0$. $A=1$ implies $k_x=k_y=k_z=\pm\frac{1}{\sqrt{3}}$ again. We get $\cos(2\phi)=\frac{-3\alpha_2^2-3\alpha_3^2-1}{2}$ and $\phi\in\{0,\pi\}$ as surfaced in parameter space separating different topological phases.
	
	Further, $\chi$ has minima at $E^\text{min}=0$ and $E^\text{min}_\pm=\pm(\alpha_2^2+\alpha_3^2+1)$ with $\chi(E^\text{min})=-\text{det}H$. So we once again have the condition $\text{det}H=0$. The only unexplored way this condition can be fulfilled is by $B(\alpha_n)=0$. This leads to another phase separating surface with $\cos(2\phi)=\frac{-\alpha_2^2-\alpha_3^2+1}{2}$, which corresponds to degeneracies of band pairs $(1,2)$, $(3,4)$ and $(5,6)$. Since $\chi$ is independent of $k_n$ at these parameters, this degeneracy occurs at all $k$ points. It turns out, that this gap closing does not lead to a change in Chern number. This can be seen in figure \ref{appendix6foldtopphase}, where the topological phase diagram of this model is show and every phase is numerated from 1 to 6. Additionally, for small off diagonal terms ($\alpha_{2/3}<<1$) the topological phase diagram is equivalent to a double 3-fold point, as expected, since at $b=0$, $H$ is a direct sum of two 3-fold points. At large off diagonal terms, this is no longer the case. This is the only topological phase diagram out of all multifold crossings, where the topological phase depends on the parameters relative magnitude, in this case the ratio $\frac{|\alpha_0+i\alpha_1|}{|\alpha_2+i\alpha_3|}$. The Chern number for the two middle bands is always zero. Following \cite{bradlyn2016beyond}, we can get the SG 212/213 $\bar{R}_7\bar{R}_8$ model by setting $\alpha_0=\frac{\pi}{2}$.
	
	The band 1 symmetry eigenvalue jump of the 3-fold rotation is 0 in phases 3 and 6, $\frac{4\pi}{3}$ in phases 1 and 2 and $\frac{2\pi}{3}$ in phases 4 and 5. The band 2 symmetry eigenvalue jump of the 3-fold rotation is 0 in phases 2 and 5, $\frac{4\pi}{3}$ in phases 1 and 3 and $\frac{2\pi}{3}$ in phases 4 and 6. For band 3, we get symmetry eigenvalue jumps of 0 in phases 1 and 4, $\frac{4\pi}{3}$ in phases 2 and 6 and $\frac{2\pi}{3}$ in phases 3 and 5. For the remaining bands, particle hole symmetry interchanges $\frac{2\pi}{3}\leftrightarrow\frac{4\pi}{3}$. We see that we can still distinguish between phases 1 and 2 (4 and 5) by symmetry eigenvalue jumps.
	
	Bandpairs are degenerate on rotation axis of the two-fold rotation, so Eq.~\ref{Eq_ChargeFromRotEigVal} can not be applied here. Considering Eq.~\ref{Eq_nonabeliansewingjump}, all jumps of $\log\det\mathcal{B}_{C_2}$ are zero for all bandpairs. This puts constraints $\nu_{1,2}=0 \,\text{mod } 2$ and $\nu_{3,4}=0 \,\text{mod } 2$ on the nonabelian Chern numbers. Further, the 3-fold rotation symmetry eigenvalue jumps lead with Eq.~\ref{Eq_nonabeliansymjump} to the following constraints $\nu_{1,2}= 1\,\text{mod }3$ in phases 1, 5 and 6 and $\nu_{1,2}= 2\,\text{mod }3$ in phases 2, 3 and 4. These constraints are consistent with the Chern numbers found in figure \ref{appendix6foldtopphase}. For bands 3 and 4 we get $\nu_{1,2}= 0\,\text{mod }3$, which is also fulfilled.

	\section{Gellmann matrices\label{appendix:gellmanndefinition}}
	We use the following definition of the Gellmann matrices.

	\noindent\begin{minipage}{0.45\linewidth}
		\begin{align}
			\lambda_0&= \begin{pmatrix}
				1 & 0 & 0 \\
				0 & -1 & 0 \\
				0 & 0 & 0
			\end{pmatrix}
		\end{align}
	\end{minipage}
	\begin{minipage}{0.45\linewidth}
		\begin{align}
			\lambda_1&= \begin{pmatrix}
				0 & -i & 0 \\
				i & 0 & 0 \\
				0 & 0 & 0
			\end{pmatrix}
		\end{align}
	\end{minipage}
	\begin{minipage}{0.45\linewidth}
		\begin{align}
			\lambda_2&= \begin{pmatrix}
				0 & 0 & -i \\
				0 & 0 & 0 \\
				i & 0 & 0
			\end{pmatrix}
		\end{align}
	\end{minipage}
	\begin{minipage}{0.45\linewidth}
		\begin{align}
			\lambda_3&= \begin{pmatrix}
				0 & 1 & 0 \\
				1 & 0 & 0 \\
				0 & 0 & 0
			\end{pmatrix}
		\end{align}
	\end{minipage}
	\begin{minipage}{0.45\linewidth}
		\begin{align}
			\lambda_4&= \begin{pmatrix}
				\frac{1}{\sqrt{3}} & 0 & 0 \\
				0 & \frac{1}{\sqrt{3}} & 0 \\
				0 & 0 & -\frac{2}{\sqrt{3}}
			\end{pmatrix}
		\end{align}
	\end{minipage}
	\begin{minipage}{0.45\linewidth}
		\begin{align}
			\lambda_5&= \begin{pmatrix}
				0 & 0 & -i \\
				0 & 0 & 0 \\
				i & 0 & 0
			\end{pmatrix}
		\end{align}
	\end{minipage}
	\begin{minipage}{0.45\linewidth}
		\begin{align}
			\lambda_6&= \begin{pmatrix}
				0 & 0 & 1 \\
				0 & 0 & 0 \\
				1 & 0 & 0
			\end{pmatrix}
		\end{align}
	\end{minipage}
	\begin{minipage}{0.45\linewidth}
		\begin{align}
			\lambda_7&= \begin{pmatrix}
				0 & 0 & 0 \\
				0 & 0 & 1 \\
				0 & 1 & 0
			\end{pmatrix}
		\end{align}
	\end{minipage}
	\begin{minipage}{0.45\linewidth}
		\begin{align}
			\lambda_8&=\mathbb{1}
		\end{align}
	\end{minipage}
	
	\section{Details on the calculation \\ for TaO$_2$ and NbO$_2$}
	\begin{figure*}
		\centering
		\includegraphics[width=\textwidth]{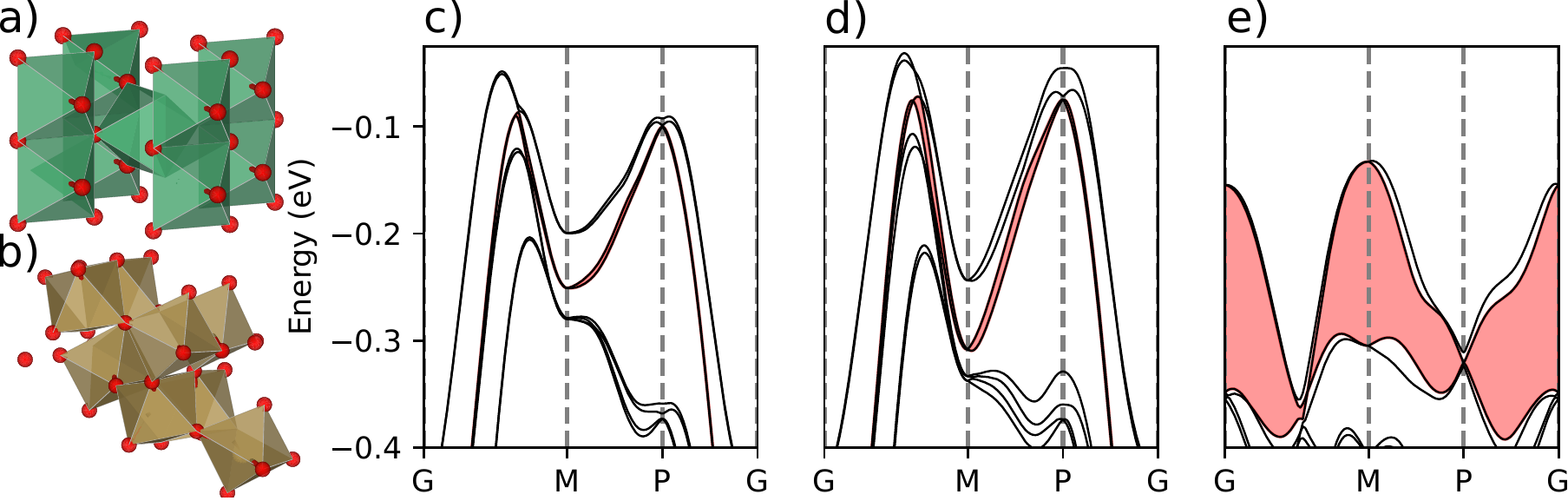}
		\caption{a) Crystal structure of the parent compound NbO$_2$ in SG $\#136$ and b) of TaO$_2$ with increased oxygen-deficiency in SG $\#80$. c) - d) DFT band structures for the reported oxygen-deficient NbO$_2$, the same structure with Nb substituted by Ta and the structure with enhanced oxygen-deficiency modeled by an increased distortion. The topological gaps are colored in red. NbO$_2$ has a small topological gap that can be slightly enhanced by substitution. Distorting the material (by a factor of $1.5$ w.r.t. the distortion of the parent compound in SG $\#136$ to the reported oxygen-deficient structure in SG $\#80$) leads to a band inversion and a significant enhancement of the topological band gap.}
		\label{fig:tao2_bands_appendix}
	\end{figure*}
	
	To enlarge the topological band gap, we distort the reported oxygen-deficient NbO$_2$ structure further, modeling a stronger oxygen-deficiency, as well as substituting Nb with Ta. We assume that the enhanced deficiency leads to a distortion that is larger in magnitude but preserves the ratio of the modes it is composed of. In general, this is a complicated linear combination of many distortion modes. First, we identify which linear combination of modes leads to the above-mentioned structural phase transition by using the ISODISTORT tool.\cite{isodistort2} By comparing the parent and the reported oxygen-deficient compound, we can identify and then exaggerate the distortion by a factor of $1.5$. The band structure of that structure has a much larger topological band gap while preserving the symmetry of the reported oxygen-deficient crystal. The crystallographic axes of the distorted cell are $a_1 = (-4.847, 4.847, 2.967)$, $a_2 = (4.847, -4.847, 2.967)$ and $a_3 = (4.847, 4.847, -2.967)$ in units of Angstrom. The positions of the atoms are \begin{table}[]
		\centering
		\begin{tabular}{c|c c c}
			Atom & $x$ & $y$ & $z$ \\ \hline 
			Ta & 0.276080 & 0.033580 & 0.217160 \\
			Ta & 0.816420 & 0.058920 & 0.782840 \\ 
			Ta & 0.308920 & 0.026080 & 0.742500 \\
			Ta & 0.283580 & 0.566420 & 0.257500 \\
			Ta & 0.206250 & 0.448580 & 0.727330 \\
			Ta & 0.721250 & 0.478920 & 0.272670 \\
			Ta & 0.728920 & 0.956250 & 0.257670 \\
			Ta & 0.698580 & 0.471250 & 0.742330 \\
			O & 0.097210 & 0.840550 & 0.951600 \\ 
			O & 0.888950 & 0.145610 & 0.048400 \\
			O & 0.395610 & 0.847210 & 0.756660 \\
			O & 0.090550 & 0.638950 & 0.243340 \\
			O & 0.618220 & 0.372050 & 0.977430 \\
			O & 0.394620 & 0.640790 & 0.022570 \\
			O & 0.890790 & 0.368220 & 0.746170 \\
			O & 0.622050 & 0.144620 & 0.253830 \\
			O & 0.111550 & 0.357720 & 0.449770 \\
			O & 0.907950 & 0.661780 & 0.550230 \\
			O & 0.911780 & 0.861550 & 0.253830 \\
			O & 0.607720 & 0.657950 & 0.746170 \\
			O & 0.595550 & 0.852220 & 0.484930 \\
			O & 0.367290 & 0.110620 & 0.515070 \\
			O & 0.360620 & 0.345550 & 0.243330 \\
			O & 0.102220 & 0.117290 & 0.756670 
		\end{tabular}
		\caption{Atomic positions of the distorted $\beta$-TaO$_2$ compound. The coordinates are given in lattice vectors $a_1 = (-4.847, 4.847, 2.967)$, $a_2 = (4.847, -4.847, 2.967)$ and $a_3 = (4.847, 4.847, -2.967)$ in units of Angstrom. }
		\label{tab:distorted_cell}
	\end{table}
	
	\section{Details on DFT calculations\label{App_details_calculation}}
	
	The DFT calculations have been performed using the VASP software\cite{vasp1, vasp2, vasp3} and Quantum Espresso\,\cite{giannozzi2009quantum,giannozzi2017advanced} with optimized norm-conserving Vanderbilt pseudopotentials\,\cite{hamann2013optimized} from PseudoDojo\,\cite{van2018pseudodojo} within the PBE approximation\cite{PBE} of the exchange-correlation functional. For wannierization Wannier90\cite{wannier90} has been employment and the surface simulation was carried out using Wanniertools.\cite{wanniertools}
	
	\bibliography{main_arxiv.bib}
\end{document}